\documentclass[aps,prx,showpacs,preprintnumbers,twocolumn,superscriptaddress, reprint, nofootinbib,longbibliography]{revtex4-2}

\usepackage{amsmath,amssymb}

\usepackage{bm}
\usepackage{tipa}
\usepackage{upgreek}
\usepackage{comment}
\usepackage{mathrsfs}
\usepackage{graphicx}
\usepackage{braket}
\usepackage{float}
\usepackage{enumitem}
\usepackage{xcolor}
\usepackage{adjustbox}
\usepackage{mathbbol}
\usepackage{booktabs}
\usepackage{gensymb}
\usepackage[normalem]{ulem}
\usepackage{xcolor}
\usepackage[colorlinks,bookmarks=true,citecolor=blue,linkcolor=red,urlcolor=blue]{hyperref}
\usepackage{hyperref}
\usepackage{cleveref}

\usepackage{pifont}

\renewcommand{\comment}[1]{}
\usepackage{subfigure}
\usepackage{multirow}

\begin{document}

\title{
Fractional Chern insulators and composite Fermi liquids vs. charge order in twisted MoTe$_2$}

\author{Enrique Aguilar-Méndez}
\affiliation{Institute for Theoretical Physics, ETH Z{\"u}rich, 8093 Z{\"u}rich, Switzerland}
\author{Glenn Wagner}
\affiliation{Institute for Theoretical Physics, ETH Z{\"u}rich, 8093 Z{\"u}rich, Switzerland}

\begin{abstract}
The bandstructure of twisted MoTe$_2$ around 3.7$^\circ$ consists of spin-valley locked Chern bands with a bandwidth that is smaller than the interaction scale. As such, the strongly correlated phases familiar from Landau levels can be realized: Quantum Hall states, composite Fermi liquids and charge order. These states have in common that they are all favoured by strong interactions, however changing the quantum geometry and topology of the bands as well as their bandwidth via the displacement field and twist angle tilts the balance between them. We provide a comprehensive exact diagonalization analysis of the competition between these states and in particular elucidate the nature of the charge order at fillings $\nu=-\frac14,-\frac13,-\frac12,-\frac23$. Furthermore, we analyze the effect of band mixing by going beyond a single band per valley. This effect of band mixing is crucial for reconciling the numerical phase diagrams with experimental observations.
\end{abstract}

\maketitle

\section{Introduction}

Twisted bilayer MoTe$_2$ has emerged as a remarkable platform for studying how the interplay between band topology and strong electron correlations gives rise to exotic phases of matter. At small twist angles (around $3.7^{\circ}$), the low-energy bands near charge neutrality become nearly flat, energetically isolated, and topologically non-trivial, carrying a non-zero Chern number. Fractional filling of these bands has been predicted to host a variety of exotic many-body phases including states in which a fractional quantum Hall effect arises without an external magnetic field \cite{Wu2019a,Abouelkomsan2020,Devakul2021,Li2021c}. Such phases are also known as fractional Chern insulators (FCIs)  \cite{Neupert2011,Sheng2011,Regnault2011}.

FCIs have been reported in twisted MoTe$_2$ by multiple experimental probes including both optical and transport measurements \cite{Zeng2023, Cai2023, Park2023, Xu2023, Ji2024, Redekop2024,Kang2024,Xu2025,Park2025a,Wang2025a,Park2025,Chang2026,Sun2026,Li2026,Xu2026}. In transport experiments they manifest as a suppression of the longitudinal resistance $\rho_{xx}$ together with a quantized transverse resistance $\rho_{xy}$ at filling $\nu=-2/3$ \cite{Park2023, Xu2023}. Upon applying a displacement field $D$ there is a substantial enhancement of the charge gap, consistent with a field-driven phase transition \cite{Xu2023}. 

Subsequent experiments have revealed a richer landscape of high-resistance states, the most robust of which occur at fillings $\nu=-2/3,-1/2,-1/3$ and $-1/4$ \cite{Xu2025b}. At fillings $\nu = -2/3$ and $-1/2$ a resistive state is observed at large $D$. At fillings $\nu = -1/3$ and $-1/4$, $\rho_{xx}$ is already large at $D = 0$ and stays elevated. The evolution of this phase diagram with twist angle has also been investigated experimentally~\cite{Sun2026}: as the angle increases, the range of displacement fields over which these resistive states survive shrinks. By $5.19^{\circ}$ high-resistance features persist only at $\nu = -1/3$ and $-1/4$, and only near zero displacement field.

Understanding the nature of these high-resistance states has been an active area of theoretical investigation. Leading candidates are charge ordered states that break the discrete moir\'e translational symmetry, i.e.~charge density waves. In the literature these are often called generalized Wigner crystals (GWCs) to emphasize that they are not atomic-scale CDWs \cite{Hubbard1978,Padhi2021}. The lattice constant of the GWCs is $\sim10$ nm which is comparable to the lattice constants of Wigner crystals in Landau levels. These states are topologically trivial, in contrast with anomalous Hall crystals (AHCs), in which charge order coexists with a quantized Hall response.

Transport and spin-susceptibility measurements in monolayer MoTe$_2$  show evidence of WCs, with antiferromagnetic correlations at zero magnetic field \cite{Zhang2025}. Recent scanning tunneling microscopy measurements have visualized molecular WCs in twisted MoTe$_2$, where several electrons can cluster into one minimum of the moiré potential \cite{Liu2026}. However the experimental work that finds GWCs most similar to the ones considered by us is in the heterobilayer WSe$_2$/WS$_2$, where $\sqrt{3}\times\sqrt{3}$ charge order and stripes have been observed at $\nu=-1/3$ and $\nu=-1/2$ respectively \cite{Li2021}. 

All these observations motivate a systematic study of the phases present in twisted MoTe$_2$ as a function of displacement field at multiple filling factors.

Exact diagonalization (ED) studies at zero displacement have identified some of these candidates and their transitions into FCIs and composite Fermi liquids (CFLs). Increasing the twist angle was found to drive transitions from the CFL at $\nu = -1/4$ into a GWC~\cite{Reddy2023c}, and from the CFL at $\nu =-1/2$ into an AHC ~\cite{Sheng2024a}. At fillings $\nu = -1/3$ and $-2/3$, FCI states were found to destabilize into GWCs when either the twist angle or interaction strength is varied~\cite{Yu2024b, Crepel2023, Reddy2023c, Reddy2023e, He2025, Chen2026c}.

 DMRG calculations at zero displacement field have  confirmed key ED results and revealed GWC phases across a broader range of fillings. DMRG validates the FCI-to-GWC transition at $\nu = -1/3$ with increasing interaction strength \cite{He2025, Chen2026c} and corroborates the CFL-to-AHC transition at $\nu = -1/2$ driven by twist angle \cite{Chen2026c}. Beyond these, GWC phases have been identified at $\nu = -2/5$ \cite{Wang2025} and $-4/3$ \cite{Yu2024b} via ED, and at $\nu = -4/9, -5/9, -4/7, -3/5, -2/3$ and $-4/5$ via DMRG \cite{He2025, Chen2026c,tuo2025fractionalquantumanomaloushall}.

As $D$ increases, the spin gap closes and the ground state cannot be assumed to be spin polarized \cite{Park2023, Anderson2024}, requiring a systematic comparison of ground-state energies across all possible spin polarizations. Only a handful of studies have tackled this regime \cite{Zaklama2024a,He2025}. Using ED, a transition from FCI to GWC was found at $\nu = -1/3$ \cite{Zaklama2024a} for small twist angles  below $3^{\circ}$. For $\nu = -2/3$, DMRG calculations have also identified a displacement field driven FCI to CDW transition \cite{He2025}.

MoTe$_2$ exhibits strong spin–valley locking, so the minimal description of the hole-doped system involves only two bands: the topmost valence band of each spin species. Most calculations adopt this simplification to remain computationally tractable. However, the energy gap separating this first pair of valence bands from the next is smaller than the characteristic interaction strength, so a more accurate treatment should include the next pair as well, yielding a four-band description that allows for ``band mixing". 

In this work, we use ED to explore the phase diagram of twisted MoTe$_2$ at fillings $\nu=-1/4,-1/2,-1/3$, and $-2/3$ as a function of displacement field and one additional control parameter, either twist angle or gate distance. Unlike many previous studies, we do not assume spin polarization. Furthermore, we account for band mixing by allowing first one and then two holes per spin to occupy  the second highest valence band in our calculations.

\begin{figure}
    \centering
    \includegraphics[width=\columnwidth]{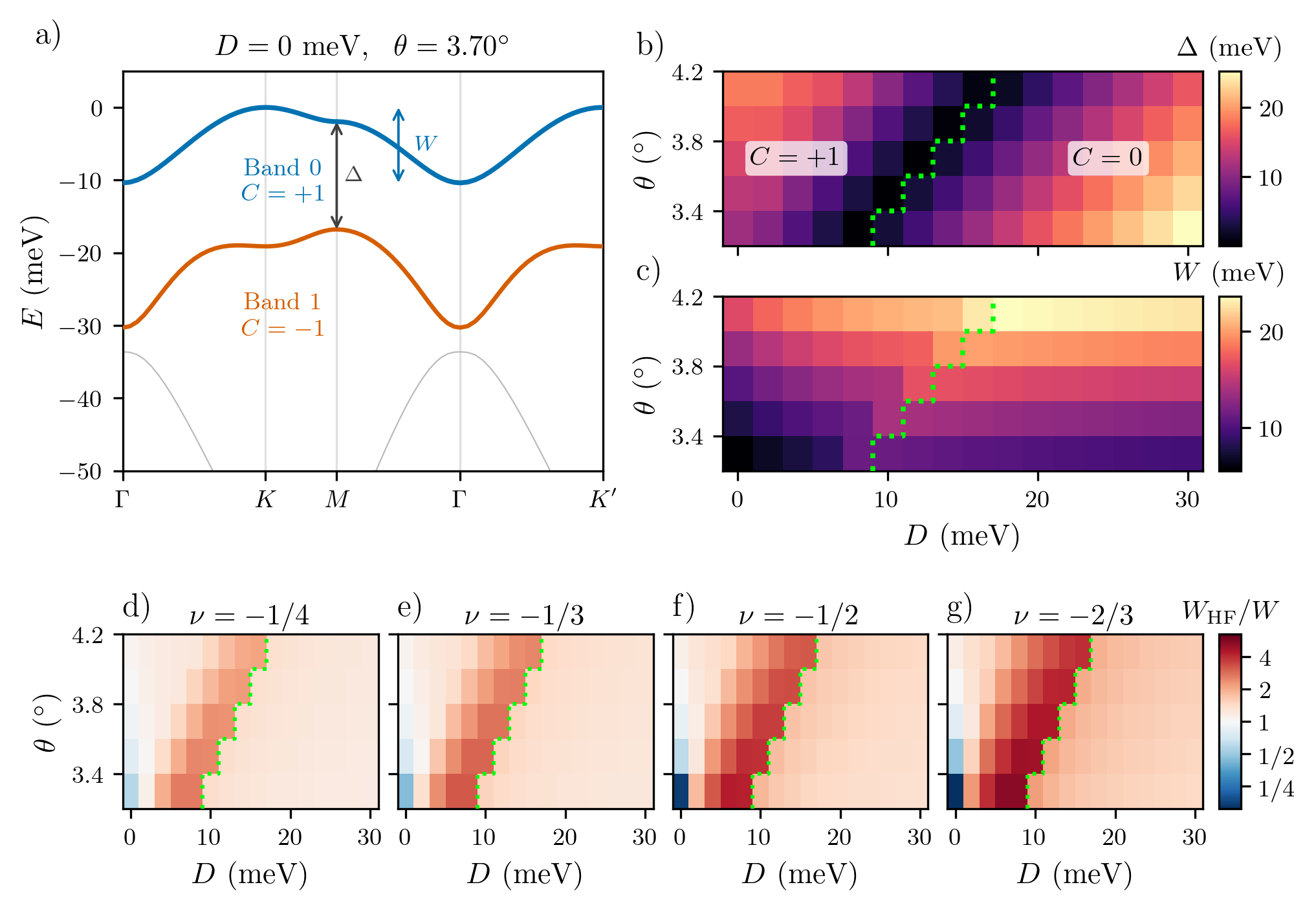}

    \caption{\textbf{Single-particle bands and their Hartree-Fock renormalization.} \textbf{a)} Moir\'e bands in valley $K$. The two highest bands carry Chern number $C=+1$ and $C=-1$.  \textbf{b)} Minimum direct gap $\Delta$ between the two highest bands versus twist angle $\theta$ and displacement field $D$. The dotted line separates the $C=+1$ and $C=0$ region of the highest band and follows the gap closing. \textbf{c)} Band width $W$ of the highest valence band versus $\theta$ and $D$. \textbf{d)-g)} Single shot Hartree-Fock bandwidth $W_{\mathrm{HF}}$ relative to $W$ at $\nu=-1/4,-1/3,-1/2$ and $-2/3$, calculations were done keeping only band 0 and assuming uniform occupation of this band. }
    \label{fig:single-particle}
\end{figure}

\section{Methods}
The starting point of our calculations is a single particle continuum model. Following the same steps as in \cite{Wu2019} we write the Hamiltonian in terms of intralayer $h_{\eta,l}(\bm{r})$ and hopping terms $t_{\eta}(\bm{r})$ that are expanded to first harmonics using the set of DFT parameters obtained in \cite{Jia2023} for twist angle $\theta=3.7^\circ$: $w = -18.8\text{ meV}, V = 16.5\text{ meV}$ and $\psi = -105.9^{\circ}$, $m^*=0.6m_e$. The index $l=b,t$ denotes the bottom and top layers and the index $\eta=\pm1$ the valleys $\pm K $ which are locked to the spins up and down, respectively, due to strong spin orbit coupling. We have the following Hamiltonian: 

\begin{equation}
H_{\eta}=\int d^2\bm{r} 
\begin{pmatrix}
c^{\dagger}_{\eta,b,\bm{r}} & c^{\dagger}_{\eta,t,\bm{r}}
\end{pmatrix}
 \begin{pmatrix}
h_{\eta,b}(\bm{r}) & t_{\eta}(\bm{r}) \\
t_{\eta}^*(\bm{r}) & h_{\eta,t}(\bm{r})
\end{pmatrix}
\begin{pmatrix}
c_{\eta,b,\bm{r}} \\ c_{\eta,t,\bm{r}} 
\end{pmatrix} ,
\end{equation}
where the intralayer term $h_{\eta,l}(\bm{r})$ incorporates the contributions of the kinetic energy, the moir\'e potential and a displacement field $D$ perpendicular to the MoTe$_2$. $t_{\eta}(\bm{r})$ is the intralayer hopping. See appendix \ref{app:hamiltonian} for the detailed expressions for $h_{\eta,l}(\bm{r})$ and $t_{\eta}(\bm{r})$.

We introduce interactions via a Coulomb potential screened by metallic gates located at height $\pm \xi/2$ : 

\begin{equation}
    H_{\text{int}}=\frac{1}{2A}\sum_{\bm{q}}V(\bm{q}):\rho(\bm{q}) \rho(\bm{-q}):\,,
\end{equation}
where $A$ is the area of the system and  $\rho(\mathbf q)$ is the density of holes in the moiré valence bands. Normal ordering is with respect to the hole vacuum.
\begin{equation}
V(q)=\frac{e^2}{2\epsilon_0\epsilon_rq}\tanh{q\frac{\xi}{2}} ,
\label{eq:interaction}
\end{equation}
 where $\epsilon_0$ is the vacuum's permittivity and $\epsilon_r$ is the relative permittivity set to 10 throughout our calculations. 
\begin{equation}
    H=\sum_{\eta}H_{\eta}+H_{\text{int}}
    \label{eqq:full_hamiltonain}
\end{equation}
For the exact diagonalization we project $H$  onto the highest (band 0) and second highest (band 1) valence bands, indicated in panel Fig.~\ref{fig:single-particle}a. To keep the Hilbert space tractable we restrict it to the Fock states with a limited number of holes in band 1. The maximum number of holes in band 1 is $N_\mathrm{max}^1$, which is an approach that has been used previously \cite{Kwan2026,Yu2025moire,Li2025multiband,Goncalves2026spinless}. In our work, $N_\mathrm{max}^1$ is set to be at most 2. See appendix \ref{app:band_projection} for details of the band projection. Total momentum and $S_z$ are conserved and label the sectors we diagonalize. Since the momentum eigenstates are translationally invariant, we introduce a small pinning potential when plotting the real space charge density (see appendix \ref{app:pinned_density}). 

\section{Results}
\subsection{Single-shot Hartree Fock}
Fig.~\ref{fig:single-particle}b shows the direct gap $\Delta$ between bands 0 and 1 of the single particle model. As $D$ increases the gap shrinks until it closes. Where the gap closes, band 0 undergoes a topological transition in which its Chern number changes from $C=+1$ to $C=0$, hence destroying topological phases such as the fractional Chern insulator \cite{Wang2024fractional}. At larger $D$ the gap reopens and grows. Fig.~\ref{fig:single-particle}c shows the width $W$ of band 0. $W$ grows with $D$ up to the transition, and decreases slowly beyond it.
As a first look at how interactions modify the band structure, we perform one shot Hartree-Fock calculations restricted to band 0. We consider a spin-polarized state in which the holes fill this band uniformly at filling $\nu$ (see Appendix~\ref{app:hartree-fock} for details). Figs.~\ref{fig:single-particle}d-g show the resulting bandwidth $W_{\mathrm{HF}}$ relative to the single particle width $W$. The trend is the same at all four fillings. At zero displacement field, interactions narrow the band at small twist angles ($\theta \lesssim 3.7^\circ$) and leave it nearly unchanged at larger ones. As $D$ increases the band broadens, and $W_{\mathrm{HF}}/W$ peaks just before the Chern transition; beyond the transition the band stays moderately wider than the non-interacting one. Both the narrowing and the broadening grow with the filling. 

As shown in Figs.~\ref{CFL_and_band_mixing_one_quarter}, \ref{FCI_and_band_mixing_one_third} and \ref{FCI_and_band_mixing_two_thirds}, we observe a decrease in the ferromagnetic tendency with increasing $D$ across all fillings.  Indeed, it was pointed out in \cite{Chen2026} that the loss of band topology is related to the loss of ferromagnetism. In the following we present results for each of the fillings studied.

\subsection{Exact diagonalization at $\nu=-1/4$}

\begin{figure}
    \centering
    \includegraphics[width=\columnwidth]{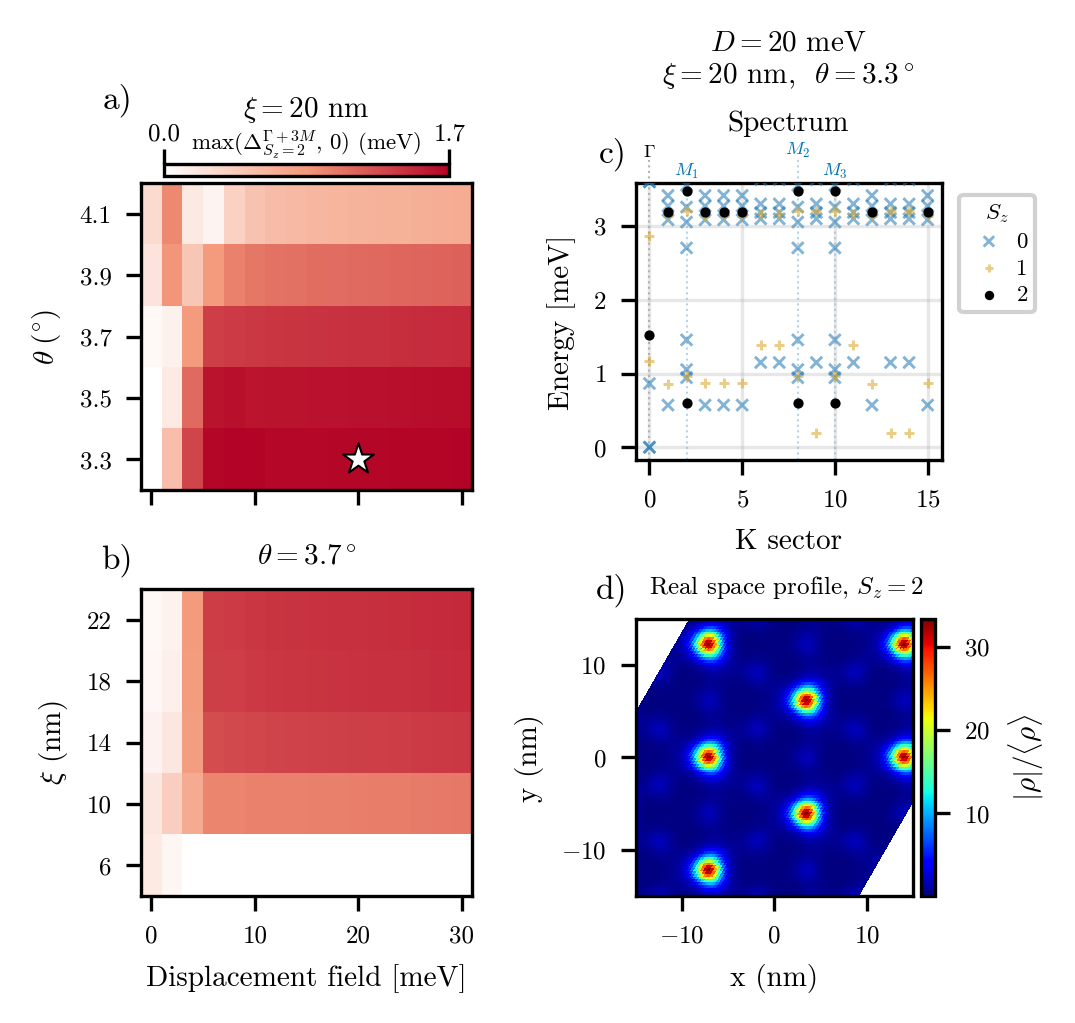}

    \caption{\textbf{ Charge order at $\nu=-1/4$} ($N_s=16$, $N_\mathrm{max}^1=2$).
\textbf{a)} $\max(\Delta^{\Gamma+3M}_{S_z=2},0)$ of the four-state manifold at $\{\Gamma,M_1,M_2,M_3\}$,
versus $D$ and $\theta$ at $\xi=20$ nm. \textbf{b)} same as a) but plotted versus $D$ and $\xi$ at $\theta=3.7^\circ$. 
\textbf{c)} Spectrum at $\theta=3.3^\circ$, $D=20$ meV, $\xi=20$ nm: the $S_z=2$ manifold (black) is
gapped by $1.7$ meV within its sector.
\textbf{d)} charge density: a $2\times2$ superlattice. }
\label{fig:one_quarter}
\end{figure}

In Fig.~\ref{fig:one_quarter} we present the ED results at $\nu=-1/4$. At zero displacement field we observe a CFL only in absence of band mixing (Fig.~\ref{CFL_and_band_mixing_one_quarter}). For $N_\mathrm{max}^1=0$, the ground state is polarized at $D=0$ and the finite-size ground state degeneracy is consistent with the model from Ref.~\cite{Fremling2018trial}. For $N_\mathrm{max}^1\geq1$ as in Fig.~\ref{fig:one_quarter} the CFL is suppressed. For large $D$ we find a robust GWC corresponding to a quadrupling of the unit cell, as pictured in Fig.~\ref{fig:one_quarter}d. 

In the spin polarized sector, this fourfold increase of the unit cell results in a fourfold degeneracy in the ED spectrum of Fig.~\ref{fig:one_quarter}c. However, in the spin-unpolarized sector the degeneracy is larger, due to an emergent $SU(2)$ symmetry arising in the large $D$ limit. The bandstructure of twisted MoTe$_2$ at $D=0$ strongly breaks valley $SU(2)$ symmetry (the two valleys have opposite Chern number) due to the interlayer hopping term $h_{\eta,l}(\bm{r})$. The interlayer hopping will be suppressed in the large $D$ limit restoring the $SU(2)$ symmetry. We can estimate the size of the $SU(2)$ breaking term by performing a perturbative expansion in $w/D$ (see appendix \ref{app:SU2}). For $N_e=4$  we have a degeneracy of $4\times \binom{4}{1}=16$ for $S_z=1$ and  $4\times \binom{4}{2}=24$ for $S_z=0$ coming from the number of possible spin flips. In Fig.~\ref{fig:one_quarter}d the GWC for $S_z=4$ is shown, however the same GWC is obtained for all $S_z$, the only difference is the assignment of spins to each localized orbital
(see appendix \ref{app:other_spin_sectors}). We note that the overall ground state in Fig.~\ref{fig:one_quarter}c is in the $S_z=0$ sector, consistent with the expectation that Anderson superexchange will drive the spins into an antiferromagnetic arrangement. 

Decreasing the twist angle $\theta$ decreases the single particle bandwidth of the system (Fig.~\ref{fig:single-particle}c). This is beneficial for the GWC as can be seen in Fig.~\ref{fig:one_quarter}a, where the gap of the GWC increases as $\theta$ decreases. Fig.~\ref{fig:one_quarter}b demonstrates that decreasing the screening length destroys the GWC, since the GWC requires long-range Coulomb interactions. 

\subsection{Exact diagonalization at $\nu=-1/3$}

\begin{figure}
    \centering
    \includegraphics[width=\columnwidth]{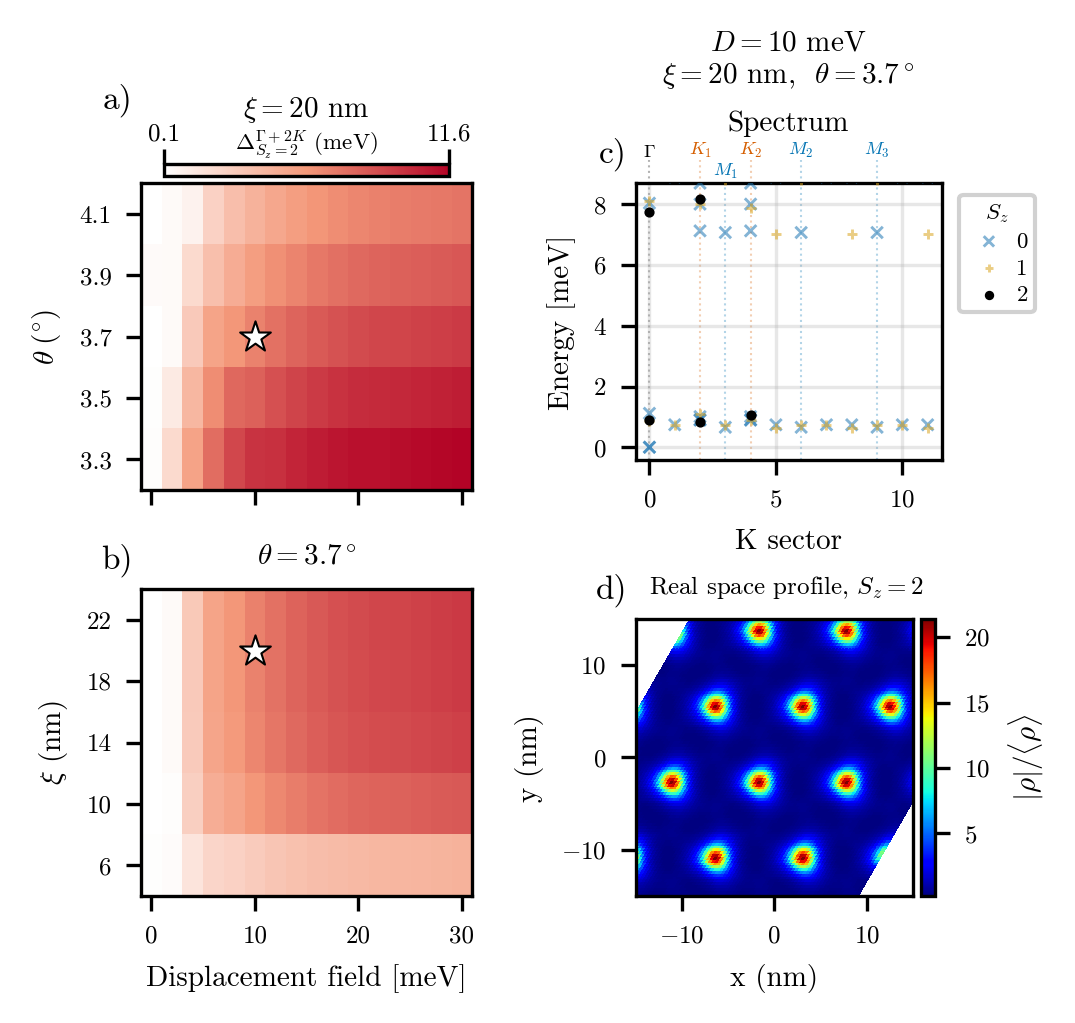}

     \caption{\textbf{Charge order at $\nu=-1/3$} ($N_s=12$, $N_\mathrm{max}^1=2$). \textbf{a)} $\Delta^{\Gamma+2K}_{S_z=2}$ of the three-state manifold at $\{\Gamma,K_1,K_2\}$, versus $D$ 
and $\theta$ at $\xi=20$ nm. \textbf{b)} same as a) but plotted versus $D$ and $\xi$ at $\theta=3.7^\circ$. 
\textbf{c)} Spectrum at $\theta=3.7^\circ$, $D=10$~meV, $\xi=20$ nm: the $S_z=2$ manifold (black) is
gapped by $6.7$ meV within its sector.
\textbf{d)} charge density: a $\sqrt3\times\sqrt3$ modulation.}
\label{fig:one_third}
\end{figure}

Fig.~\ref{fig:one_third} shows the ED results at $\nu=-1/3$. The results in Fig.~\ref{fig:one_third}ab include band mixing with $N_\mathrm{max}^1=2$ and no FCI is observed anywhere in the phase diagram. However, we note that an FCI is found at this filling with $N_\mathrm{max}^1=0$ and the FCI is still present (albeit with a smaller gap) for $N_\mathrm{max}^1=1$ (see Fig.~\ref{FCI_and_band_mixing_one_third}). The destabilizing effect of band mixing on the FCI at $\nu=-1/3$ was first noted in Ref.~\cite{Yu2024b} and is also consistent with experiments, which typically also do not observe any FCI at $\nu=-1/3$. 

The competing GWC is a $\sqrt{3}\times\sqrt{3}$ increase of the unit cell as pictured in Fig.~\ref{fig:one_third}d. For $N_e=4$ the spectrum of the GWC shown in Fig.~\ref{fig:one_third}c has a degeneracy of $3$, $3\times \binom{4}{1}=12$ and  $3\times \binom{4}{2}=18$ for $S_z=2,1,0$ respectively. Fig.~\ref{fig:one_third}ab show that as for $\nu=-1/4$, smaller twist angle and larger gate distance $\xi$ lead to an increase in the neutral gap of the GWC. Although Fig.~\ref{fig:one_third} pictures the GWC for $S_z=2$, the same GWC is obtained for all $S_z$
(see appendix \ref{app:other_spin_sectors}).

\subsection{Exact diagonalization at $\nu=-1/2$}

\begin{figure}
    \centering
    \includegraphics[width=\columnwidth]{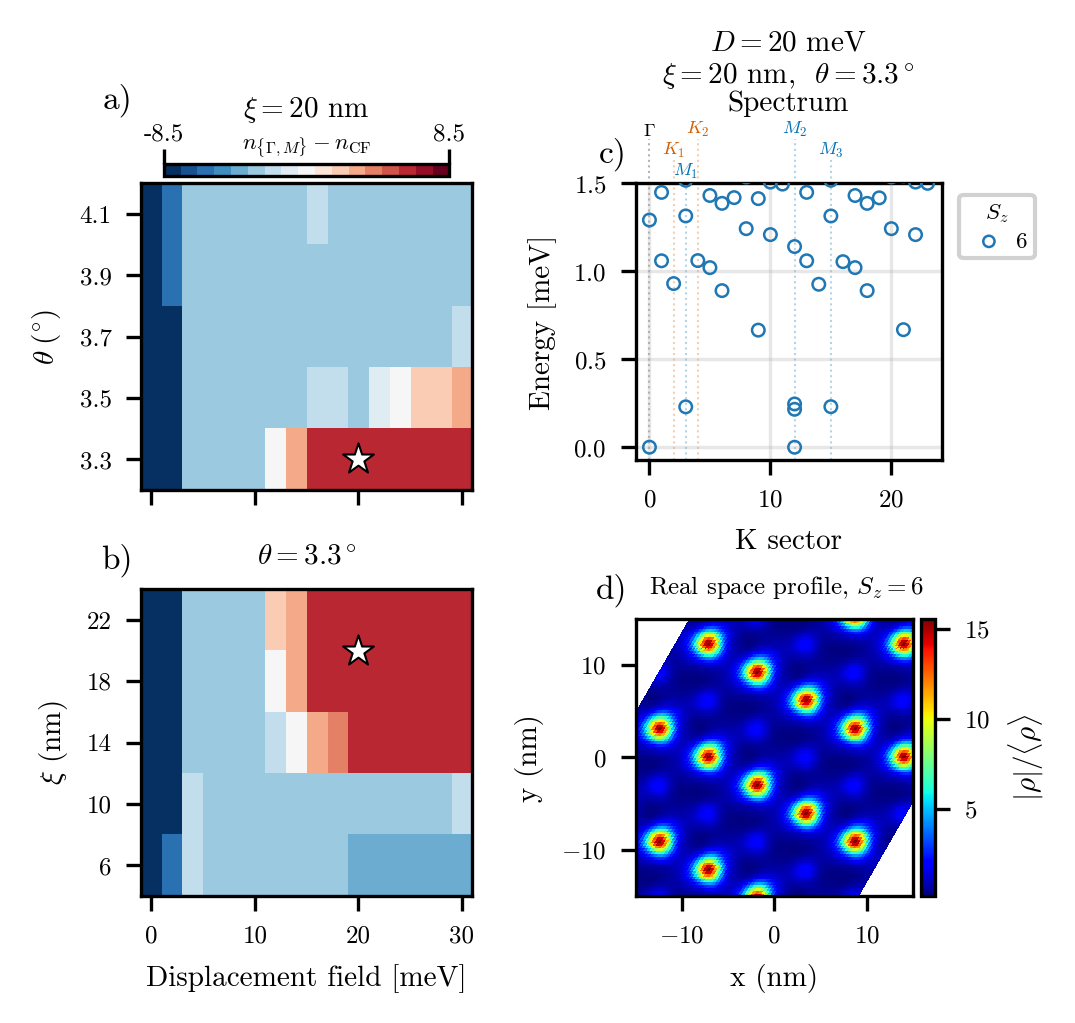}

     \caption{\textbf{CFL versus GWC at $\nu=-1/2$ }($N_s$=24, $S_z=6$, $N_\mathrm{max}^1$=0). \textbf{a)} Order score: among the 8 lowest states the number with momenta in $\{\Gamma,M_1,M_2,M_3\}$ minus the number with CFL momenta, plotted versus $D$ and $\theta$ at $\xi=20$ nm. A score of -8 signals a CFL, positive values signal charge order at the M points. \textbf{b)} same as a) but plotted versus $D$ and $\xi$ at $\theta=3.3^\circ$. \textbf{c)} Spectrum at $\theta = 3.3 ^\circ$, $D=20$ meV and $\xi=20$ nm, the six lowest states lie at $\Gamma$ and $M$ (score $+6$). \textbf{d)} charge density: maxima on alternate rows of moir\'e sites, a stripe. }   
\label{fig:one_half}
\end{figure}

For the results at $\nu=-1/2$ in Fig.~\ref{fig:one_half}ab rather than the many-body gap we plot a parameter which captures the momentum sectors of the ground state manifold. The parameter changes sign depending on whether the system is in the CFL or in the GWC phase. Previous work has found evidence for a zero-field CFL in twisted MoTe$_2$ \cite{Dong2023,Goldman2023} and there is also experimental evidence for it \cite{Anderson2024}. At $D=0$ we indeed find ground states in the momenta predicted by composite fermion theory \cite{Fremling2018trial}. The GWC we find in this case at large $D$ is just a doubling of the unit cell along one axis, as shown in Fig.~\ref{fig:one_half}d. This spontaneously breaks the $C_3$ symmetry. Since the triangular lattice is non bipartite, it is not possible to form a GWC at $\nu=-1/2$ which respects the $C_3$ symmetry. The spin polarized GWC whose spectrum we plot in Fig.~\ref{fig:one_half}c is sixfold degenerate ($2\times 3$ coming from the doubling of the unit cell and the $C_3$ symmetry breaking). Fig.~\ref{fig:one_half}ab shows that the GWC is strongly dependent on $\theta$ and $\xi$ as previously seen at the other fillings. The GWC favours small $\theta$ and large $\xi$. On the other hand, while the CFL is also less robust for large $\theta$ (due to the large bandwidth), it is less sensitive to $\xi$. 

In contrast to \cite{Chen2026c} where a transition from a CFL to a  AHC is observed at $D=0$ as $\epsilon_r$ is tuned, here the charge ordered state we observe cannot be an AHC since at the values of $D$ where it appears, the band it emerges from has already undergone the topological transition and is topologically trivial.

\subsection{Exact diagonalization at $\nu=-2/3$}

At $\nu=-2/3$ we observe a robust FCI. The GWC at $\nu=-2/3$ shown in Fig.~\ref{fig:one_half}d triples the unit cell. In the ED spectrum in Fig.~\ref{fig:one_half}c for $N_e=8$ we obtain degeneracies of $3$, $3\times\binom{8}{1}=24$, $3\times\binom{8}{2}=84$, $3\times\binom{8}{3}=168$, and $3\times\binom{8}{4}=210$ in the $S_z=4,3,2,1,0$ sectors, respectively.

On the one hand decreasing $\xi$ makes the interaction more short-range, which should favour the FCI. However,  decreasing $\xi$ also leads to an overall reduction of the interaction strength, such that the FCI gap reduces with decreasing $\xi$.

\begin{figure}
    \centering
    \includegraphics[width=\columnwidth]{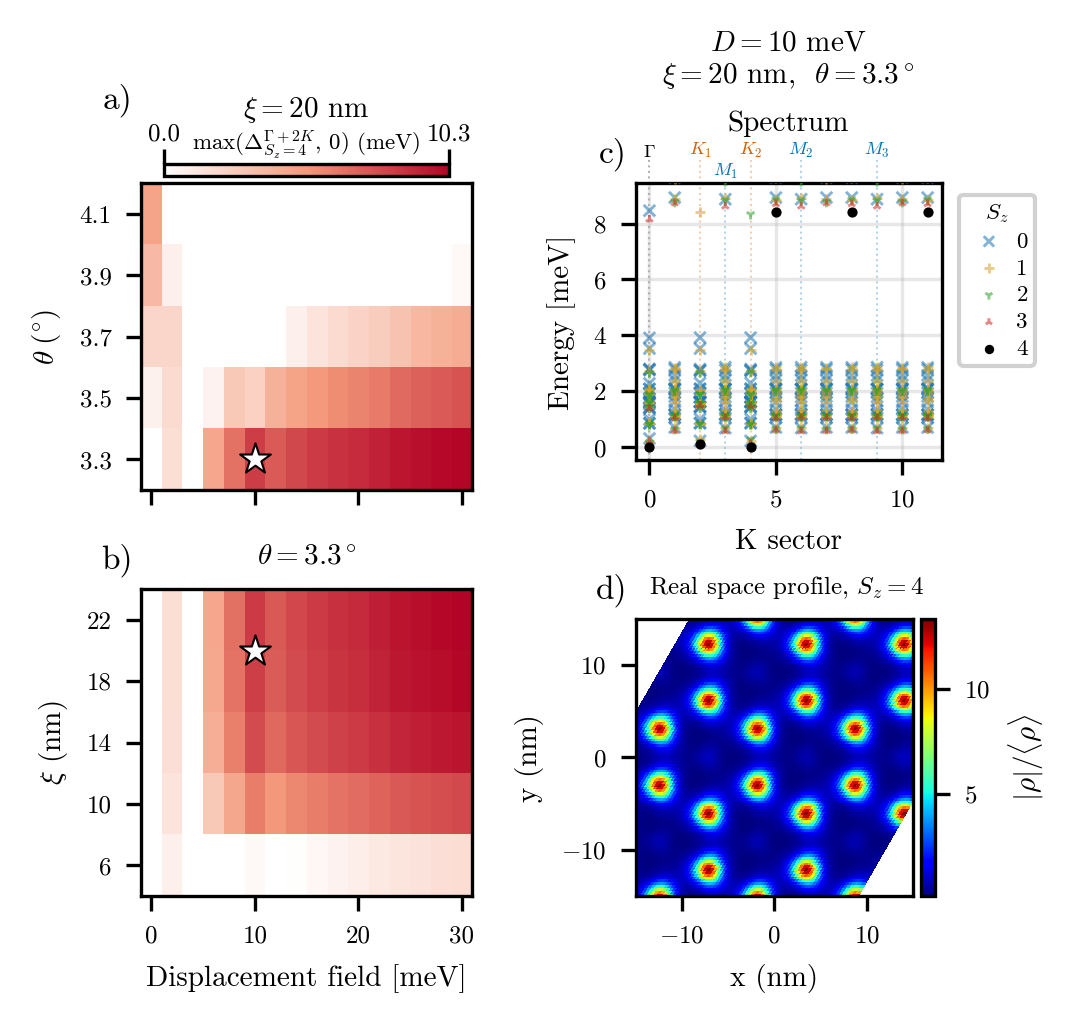}

     \caption{\textbf{Charge order at $\nu=-2/3$} ($N_s=12$, $N_\mathrm{max}^1=2$). \textbf{a)} $\max(\Delta^{\Gamma+2K}_{S_z=4},0)$ of the three-state manifold at $\{\Gamma,K_1,K_2\}$,
versus $D$ and $\theta$ at $\xi=20$ nm . \textbf{b)} same as a) but plotted versus $D$ and $\xi$ at $\theta=3.3^\circ$. \textbf{c)} Spectrum at $\theta=3.3^\circ$, $D=10$ meV, $\xi=20$ nm: the $S_z=4$ manifold (black) is gapped by $8.3$ meV within its sector. \textbf{d)} charge density: a $\sqrt3\times\sqrt3$ modulation.}
    \label{fig:two_thirds}
\end{figure}

Decreasing the twist angle decreases the effective bandwidth of the system which is beneficial for the GWC and the FCI as can be seen in Fig.~\ref{fig:two_thirds}. In Fig.~\ref{fig:two_thirds}a at $3.3^{\circ}$ we see an FCI whose stability is enhanced by increasing D. This is counterintuitive as increasing $D$ increases the band mixing and the bandwidth, making the FCI less competitive. However, we caution that this may be a finite-size effect. 

Figs.~\ref{fig:two_thirds}ab show a rapid disappearance of the FCI with $D$. This is consistent with Fig.~\ref{fig:single-particle}g which shows significant band broadening before the topological transition. A band whose (Hartree-Fock renormalized) bandwidth is wide compared to the interaction scale is less favorable for an FCI.
For this filling ferromagnetism is resilient to band mixing at small $D$ as shown in Fig.~\ref{FCI_and_band_mixing_two_thirds}.

\subsection{Effect of band mixing}

Previous work has emphasized the importance of band mixing in order to correctly model twisted MoTe$_2$. Since the interaction scale is larger than the gap between the single-particle bands, single-band projected ED is not fully justified. In particular, band mixing destroys polarization for fillings $\nu>-1/2$ \cite{moralesdurán2026bandmixingparticleholeasymmetry,Abouelkomsan2024band,Yu2024b}. The intuitive understanding of particle-hole asymmetry arising from Landau level mixing in a Landau level model is as follows \cite{moralesdurán2026bandmixingparticleholeasymmetry}: The WC competing with the quantum Hall state has a different lattice constant for $\nu=-1/3$ vs.~$\nu=-2/3$, which leads to the $\nu=-2/3$ quantum Hall state being more robust.

Band mixing has previously been examined  \cite{Kwan2026,Li2026large,Hou2025,Goncalves2026spinless} and at $\nu=-2/3$, assuming full spin polarization, it was shown to modify the range of twist angles over which FCIs are observed \cite{Xu2024}. When spin polarization is not assumed, band mixing was found to improve the agreement between the computed ground-state magnetization at $\nu=-1/3$ and $\nu=-4/3$ and experimental observations \cite{Yu2024b}. In addition,  DMRG calculations reported that band mixing substantially stabilizes a GWC at $\nu=-1/3$ \cite{He2025}. Notably, all of these studies were performed at zero displacement field.

\begin{figure}
    \centering
    \includegraphics[width=\columnwidth]{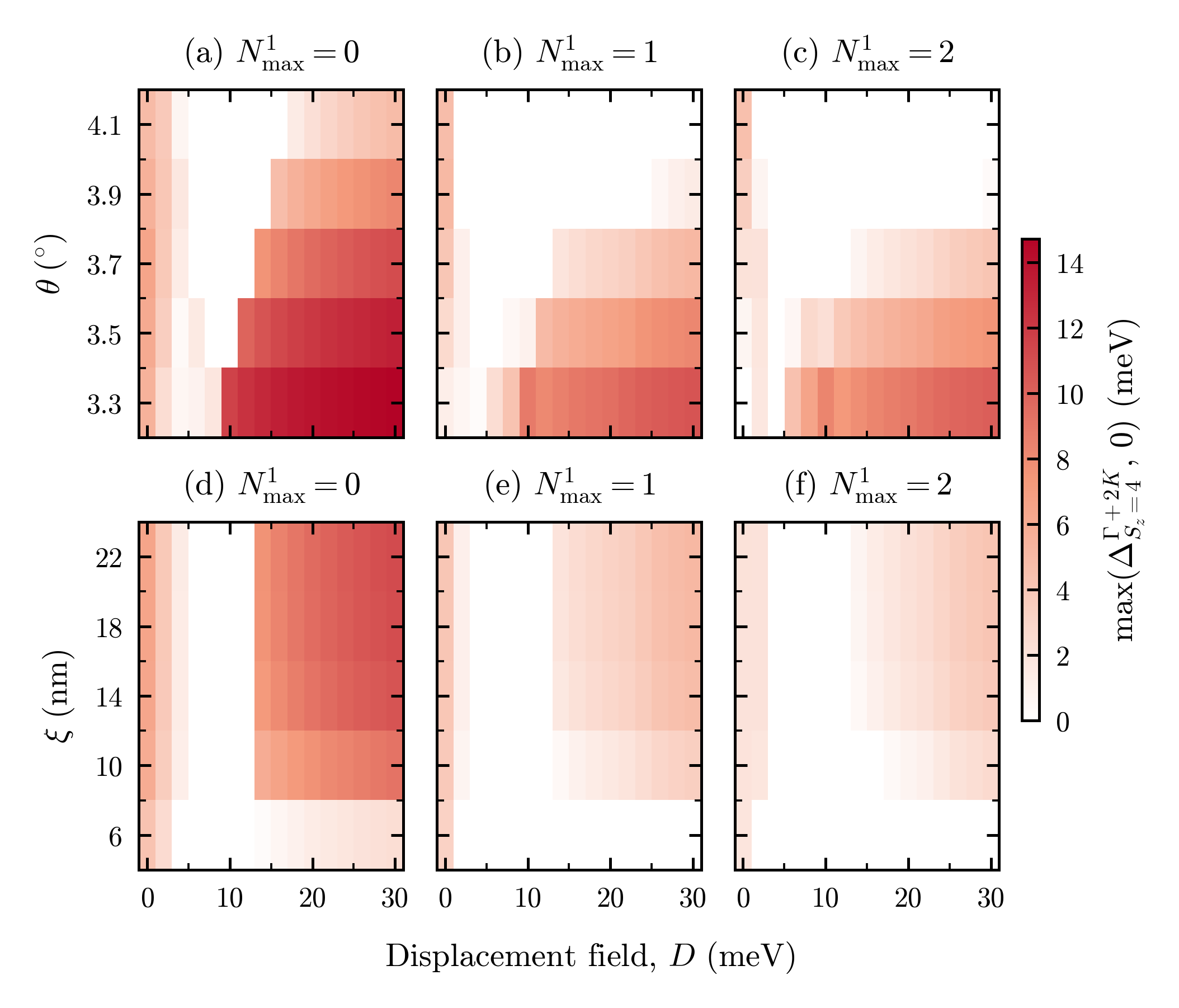}

    \caption{ \textbf{Band mixing at $\nu=-2/3$} ($N_s=12$). \textbf{a)-c)} The gap of Fig.~\ref{fig:two_thirds} for $N_\mathrm{max}^1=0$, $1$, $2$, versus $D$ and $\theta$ at $\xi=20$ nm. \textbf{d)-f)} same as previous panels but plotted versus $D$ and $\xi$ at $\theta=3.7^\circ$.  Band mixing lowers the gap and shrinks the gapped region; the change from $N_\mathrm{max}^1=1$ to $2$ is small. }
    \label{fig:band_mixing}
\end{figure}

Fig.~\ref{fig:band_mixing} shows the effect of the band mixing (tuned via increasing $N_\mathrm{max}^1$) on the FCI and GWC at $\nu=-2/3$. The gaps of both the FCI and GWC decrease as $N_\mathrm{max}^1$ increases which may arise from the fact that the Hilbert space dimension is increasing which means more available competing states that can cause the gap to shrink. 

The effect of $N_\mathrm{max}^1$ on the FCI is largest at small twist angles, which is due to the single-particle gap between band 0 and band 1 being larger in that regime. On the other hand, the band mixing does not sensitively depend on $\xi$ since this leaves the single-particle gap unaffected.

\section{Discussion}

Experiments in twisted MoTe$_2$ observe resisitive states at high displacement fields which are thought to be charge ordered states that are often termed GWCs. The resistive states are observed in the large $D$ regime, where the system is largely layer polarized. However the resistive states only occur at commensurate fillings $\nu$ indicating that the moiré lattice still plays a crucial role. 

In this work we studied the competition between the FCI or CFL and the charge ordered GWCs in twisted MoTe$_2$, in particular with respect to tuning the displacement field $D$. The topmost valence band of the non-interacting bandstructure undergoes a topological phase transition around $D\sim10$ meV from a $|C|=1$ to a $|C|=0$ band. However the phases related to the topological band (FCI and CFL) are destabilized for displacement fields far below $D\sim10$ meV. We observe that while the non-interacting bandwidth is relatively featureless across the transition (Fig.~\ref{fig:single-particle}b) the Hartree-Fock renormalized bandstructure becomes significantly more dispersive as $D$ increases (Figs.~\ref{fig:single-particle}d-g). This destroys phases such as the CFL and FCI which rely on flat (HF-renormalized) bands.  

Our calculations show that the GWCs are more robust at smaller filling fractions, something which is consistent with WCs typically being low-density phases. Strictly speaking, the GWCs phases are not true WCs, since they break a discrete rather than a continuous translation symmetry. However, the energetic stabilization mechanism is the same, the GWCs require low densities and long-range interactions (large $\xi$). 

Our exact diagonalization calculation predicts the charge profile of the GWCs at the different fillings. In future, it would be interesting to confirm this experimentally. The most unambiguous way to experimentally characterise the GWCs in twisted MoTe$_2$ would be using scanning tunneling microscopy (STM). STM requires single gate screened samples and hence there is no ability to tune the displacement field. However since the GWC exists in a large range of displacement fields it may still be possible to observe (unlike the FCI which exists only in a small window of displacement fields). 

Since we observe the GWCs to be strongly suppressed by small $\xi$, devices can be constructed in which the GWCs should be weak or non-existent. However, this would not necessarily lead to a stronger FCI, since the main mechanism by which the FCI is destroyed when increasing $D$ is the large (HF-renormalized) bandwidth, rather than a direct competition with the GWC.

\section{Acknowledgements}
Exact diagonalization calculations were performed using DiagHam \cite{Regnault2001DiagHam}. G.W.~and E.A.M.~are supported by the Swiss National Science Foundation (SNSF) via Ambizione grant number PZ00P2-216183. 

\bibliographystyle{apsrev4-2}
\bibliography{references}

\begin{thebibliography}{58}%
\makeatletter
\providecommand \@ifxundefined [1]{%
 \@ifx{#1\undefined}
}%
\providecommand \@ifnum [1]{%
 \ifnum #1\expandafter \@firstoftwo
 \else \expandafter \@secondoftwo
 \fi
}%
\providecommand \@ifx [1]{%
 \ifx #1\expandafter \@firstoftwo
 \else \expandafter \@secondoftwo
 \fi
}%
\providecommand \natexlab [1]{#1}%
\providecommand \enquote  [1]{``#1''}%
\providecommand \bibnamefont  [1]{#1}%
\providecommand \bibfnamefont [1]{#1}%
\providecommand \citenamefont [1]{#1}%
\providecommand \href@noop [0]{\@secondoftwo}%
\providecommand \href [0]{\begingroup \@sanitize@url \@href}%
\providecommand \@href[1]{\@@startlink{#1}\@@href}%
\providecommand \@@href[1]{\endgroup#1\@@endlink}%
\providecommand \@sanitize@url [0]{\catcode `\\12\catcode `\$12\catcode `\&12\catcode `\#12\catcode `\^12\catcode `\_12\catcode `\%12\relax}%
\providecommand \@@startlink[1]{}%
\providecommand \@@endlink[0]{}%
\providecommand \url  [0]{\begingroup\@sanitize@url \@url }%
\providecommand \@url [1]{\endgroup\@href {#1}{\urlprefix }}%
\providecommand \urlprefix  [0]{URL }%
\providecommand \Eprint [0]{\href }%
\providecommand \doibase [0]{https://doi.org/}%
\providecommand \selectlanguage [0]{\@gobble}%
\providecommand \bibinfo  [0]{\@secondoftwo}%
\providecommand \bibfield  [0]{\@secondoftwo}%
\providecommand \translation [1]{[#1]}%
\providecommand \BibitemOpen [0]{}%
\providecommand \bibitemStop [0]{}%
\providecommand \bibitemNoStop [0]{.\EOS\space}%
\providecommand \EOS [0]{\spacefactor3000\relax}%
\providecommand \BibitemShut  [1]{\csname bibitem#1\endcsname}%
\let\auto@bib@innerbib\@empty
\bibitem [{\citenamefont {Wu}\ \emph {et~al.}(2019{\natexlab{a}})\citenamefont {Wu}, \citenamefont {Lovorn}, \citenamefont {Tutuc}, \citenamefont {Martin},\ and\ \citenamefont {MacDonald}}]{Wu2019a}%
  \BibitemOpen
  \bibfield  {author} {\bibinfo {author} {\bibfnamefont {F.}~\bibnamefont {Wu}}, \bibinfo {author} {\bibfnamefont {T.}~\bibnamefont {Lovorn}}, \bibinfo {author} {\bibfnamefont {E.}~\bibnamefont {Tutuc}}, \bibinfo {author} {\bibfnamefont {I.}~\bibnamefont {Martin}},\ and\ \bibinfo {author} {\bibfnamefont {A.~H.}\ \bibnamefont {MacDonald}},\ }\href {https://doi.org/10.1103/PhysRevLett.122.086402} {\bibfield  {journal} {\bibinfo  {journal} {Physical Review Letters}\ }\textbf {\bibinfo {volume} {122}},\ \bibinfo {pages} {086402} (\bibinfo {year} {2019}{\natexlab{a}})}\BibitemShut {NoStop}%
\bibitem [{\citenamefont {Abouelkomsan}\ \emph {et~al.}(2020)\citenamefont {Abouelkomsan}, \citenamefont {Liu},\ and\ \citenamefont {Bergholtz}}]{Abouelkomsan2020}%
  \BibitemOpen
  \bibfield  {author} {\bibinfo {author} {\bibfnamefont {A.}~\bibnamefont {Abouelkomsan}}, \bibinfo {author} {\bibfnamefont {Z.}~\bibnamefont {Liu}},\ and\ \bibinfo {author} {\bibfnamefont {E.~J.}\ \bibnamefont {Bergholtz}},\ }\href {https://doi.org/10.1103/PhysRevLett.124.106803} {\bibfield  {journal} {\bibinfo  {journal} {Physical Review Letters}\ }\textbf {\bibinfo {volume} {124}},\ \bibinfo {pages} {106803} (\bibinfo {year} {2020})}\BibitemShut {NoStop}%
\bibitem [{\citenamefont {Devakul}\ \emph {et~al.}(2021)\citenamefont {Devakul}, \citenamefont {Cr{\'e}pel}, \citenamefont {Zhang},\ and\ \citenamefont {Fu}}]{Devakul2021}%
  \BibitemOpen
  \bibfield  {author} {\bibinfo {author} {\bibfnamefont {T.}~\bibnamefont {Devakul}}, \bibinfo {author} {\bibfnamefont {V.}~\bibnamefont {Cr{\'e}pel}}, \bibinfo {author} {\bibfnamefont {Y.}~\bibnamefont {Zhang}},\ and\ \bibinfo {author} {\bibfnamefont {L.}~\bibnamefont {Fu}},\ }\href {https://doi.org/10.1038/s41467-021-27042-9} {\bibfield  {journal} {\bibinfo  {journal} {Nature Communications}\ }\textbf {\bibinfo {volume} {12}},\ \bibinfo {pages} {6730} (\bibinfo {year} {2021})}\BibitemShut {NoStop}%
\bibitem [{\citenamefont {Li}\ \emph {et~al.}(2021{\natexlab{a}})\citenamefont {Li}, \citenamefont {Kumar}, \citenamefont {Sun},\ and\ \citenamefont {Lin}}]{Li2021c}%
  \BibitemOpen
  \bibfield  {author} {\bibinfo {author} {\bibfnamefont {H.}~\bibnamefont {Li}}, \bibinfo {author} {\bibfnamefont {U.}~\bibnamefont {Kumar}}, \bibinfo {author} {\bibfnamefont {K.}~\bibnamefont {Sun}},\ and\ \bibinfo {author} {\bibfnamefont {S.-Z.}\ \bibnamefont {Lin}},\ }\href {https://doi.org/10.1103/PhysRevResearch.3.L032070} {\bibfield  {journal} {\bibinfo  {journal} {Physical Review Research}\ }\textbf {\bibinfo {volume} {3}},\ \bibinfo {pages} {L032070} (\bibinfo {year} {2021}{\natexlab{a}})}\BibitemShut {NoStop}%
\bibitem [{\citenamefont {Neupert}\ \emph {et~al.}(2011)\citenamefont {Neupert}, \citenamefont {Santos}, \citenamefont {Chamon},\ and\ \citenamefont {Mudry}}]{Neupert2011}%
  \BibitemOpen
  \bibfield  {author} {\bibinfo {author} {\bibfnamefont {T.}~\bibnamefont {Neupert}}, \bibinfo {author} {\bibfnamefont {L.}~\bibnamefont {Santos}}, \bibinfo {author} {\bibfnamefont {C.}~\bibnamefont {Chamon}},\ and\ \bibinfo {author} {\bibfnamefont {C.}~\bibnamefont {Mudry}},\ }\href {https://doi.org/10.1103/PhysRevLett.106.236804} {\bibfield  {journal} {\bibinfo  {journal} {Physical Review Letters}\ }\textbf {\bibinfo {volume} {106}},\ \bibinfo {pages} {236804} (\bibinfo {year} {2011})}\BibitemShut {NoStop}%
\bibitem [{\citenamefont {Sheng}\ \emph {et~al.}(2011)\citenamefont {Sheng}, \citenamefont {Gu}, \citenamefont {Sun},\ and\ \citenamefont {Sheng}}]{Sheng2011}%
  \BibitemOpen
  \bibfield  {author} {\bibinfo {author} {\bibfnamefont {D.~N.}\ \bibnamefont {Sheng}}, \bibinfo {author} {\bibfnamefont {Z.-C.}\ \bibnamefont {Gu}}, \bibinfo {author} {\bibfnamefont {K.}~\bibnamefont {Sun}},\ and\ \bibinfo {author} {\bibfnamefont {L.}~\bibnamefont {Sheng}},\ }\href {https://doi.org/10.1038/ncomms1380} {\bibfield  {journal} {\bibinfo  {journal} {Nature Communications}\ }\textbf {\bibinfo {volume} {2}},\ \bibinfo {pages} {389} (\bibinfo {year} {2011})}\BibitemShut {NoStop}%
\bibitem [{\citenamefont {Regnault}\ and\ \citenamefont {Bernevig}(2011)}]{Regnault2011}%
  \BibitemOpen
  \bibfield  {author} {\bibinfo {author} {\bibfnamefont {N.}~\bibnamefont {Regnault}}\ and\ \bibinfo {author} {\bibfnamefont {B.~A.}\ \bibnamefont {Bernevig}},\ }\href {https://doi.org/10.1103/PhysRevX.1.021014} {\bibfield  {journal} {\bibinfo  {journal} {Physical Review X}\ }\textbf {\bibinfo {volume} {1}},\ \bibinfo {pages} {021014} (\bibinfo {year} {2011})}\BibitemShut {NoStop}%
\bibitem [{\citenamefont {Zeng}\ \emph {et~al.}(2023)\citenamefont {Zeng}, \citenamefont {Xia}, \citenamefont {Kang}, \citenamefont {Zhu}, \citenamefont {Kn{\"u}ppel}, \citenamefont {Vaswani}, \citenamefont {Watanabe}, \citenamefont {Taniguchi}, \citenamefont {Mak},\ and\ \citenamefont {Shan}}]{Zeng2023}%
  \BibitemOpen
  \bibfield  {author} {\bibinfo {author} {\bibfnamefont {Y.}~\bibnamefont {Zeng}}, \bibinfo {author} {\bibfnamefont {Z.}~\bibnamefont {Xia}}, \bibinfo {author} {\bibfnamefont {K.}~\bibnamefont {Kang}}, \bibinfo {author} {\bibfnamefont {J.}~\bibnamefont {Zhu}}, \bibinfo {author} {\bibfnamefont {P.}~\bibnamefont {Kn{\"u}ppel}}, \bibinfo {author} {\bibfnamefont {C.}~\bibnamefont {Vaswani}}, \bibinfo {author} {\bibfnamefont {K.}~\bibnamefont {Watanabe}}, \bibinfo {author} {\bibfnamefont {T.}~\bibnamefont {Taniguchi}}, \bibinfo {author} {\bibfnamefont {K.~F.}\ \bibnamefont {Mak}},\ and\ \bibinfo {author} {\bibfnamefont {J.}~\bibnamefont {Shan}},\ }\href {https://doi.org/10.1038/s41586-023-06452-3} {\bibfield  {journal} {\bibinfo  {journal} {Nature}\ }\textbf {\bibinfo {volume} {622}},\ \bibinfo {pages} {69} (\bibinfo {year} {2023})}\BibitemShut {NoStop}%
\bibitem [{\citenamefont {Cai}\ \emph {et~al.}(2023)\citenamefont {Cai}, \citenamefont {Anderson}, \citenamefont {Wang}, \citenamefont {Zhang}, \citenamefont {Liu}, \citenamefont {Holtzmann}, \citenamefont {Zhang}, \citenamefont {Fan}, \citenamefont {Taniguchi}, \citenamefont {Watanabe}, \citenamefont {Ran}, \citenamefont {Cao}, \citenamefont {Fu}, \citenamefont {Xiao}, \citenamefont {Yao},\ and\ \citenamefont {Xu}}]{Cai2023}%
  \BibitemOpen
  \bibfield  {author} {\bibinfo {author} {\bibfnamefont {J.}~\bibnamefont {Cai}}, \bibinfo {author} {\bibfnamefont {E.}~\bibnamefont {Anderson}}, \bibinfo {author} {\bibfnamefont {C.}~\bibnamefont {Wang}}, \bibinfo {author} {\bibfnamefont {X.}~\bibnamefont {Zhang}}, \bibinfo {author} {\bibfnamefont {X.}~\bibnamefont {Liu}}, \bibinfo {author} {\bibfnamefont {W.}~\bibnamefont {Holtzmann}}, \bibinfo {author} {\bibfnamefont {Y.}~\bibnamefont {Zhang}}, \bibinfo {author} {\bibfnamefont {F.}~\bibnamefont {Fan}}, \bibinfo {author} {\bibfnamefont {T.}~\bibnamefont {Taniguchi}}, \bibinfo {author} {\bibfnamefont {K.}~\bibnamefont {Watanabe}}, \bibinfo {author} {\bibfnamefont {Y.}~\bibnamefont {Ran}}, \bibinfo {author} {\bibfnamefont {T.}~\bibnamefont {Cao}}, \bibinfo {author} {\bibfnamefont {L.}~\bibnamefont {Fu}}, \bibinfo {author} {\bibfnamefont {D.}~\bibnamefont {Xiao}}, \bibinfo {author} {\bibfnamefont {W.}~\bibnamefont {Yao}},\ and\ \bibinfo {author} {\bibfnamefont {X.}~\bibnamefont {Xu}},\ }\href
  {https://doi.org/10.1038/s41586-023-06289-w} {\bibfield  {journal} {\bibinfo  {journal} {Nature}\ }\textbf {\bibinfo {volume} {622}},\ \bibinfo {pages} {63} (\bibinfo {year} {2023})}\BibitemShut {NoStop}%
\bibitem [{\citenamefont {Park}\ \emph {et~al.}(2023)\citenamefont {Park}, \citenamefont {Cai}, \citenamefont {Anderson}, \citenamefont {Zhang}, \citenamefont {Zhu}, \citenamefont {Liu}, \citenamefont {Wang}, \citenamefont {Holtzmann}, \citenamefont {Hu}, \citenamefont {Liu}, \citenamefont {Taniguchi}, \citenamefont {Watanabe}, \citenamefont {Chu}, \citenamefont {Cao}, \citenamefont {Fu}, \citenamefont {Yao}, \citenamefont {Chang}, \citenamefont {Cobden}, \citenamefont {Xiao},\ and\ \citenamefont {Xu}}]{Park2023}%
  \BibitemOpen
  \bibfield  {author} {\bibinfo {author} {\bibfnamefont {H.}~\bibnamefont {Park}}, \bibinfo {author} {\bibfnamefont {J.}~\bibnamefont {Cai}}, \bibinfo {author} {\bibfnamefont {E.}~\bibnamefont {Anderson}}, \bibinfo {author} {\bibfnamefont {Y.}~\bibnamefont {Zhang}}, \bibinfo {author} {\bibfnamefont {J.}~\bibnamefont {Zhu}}, \bibinfo {author} {\bibfnamefont {X.}~\bibnamefont {Liu}}, \bibinfo {author} {\bibfnamefont {C.}~\bibnamefont {Wang}}, \bibinfo {author} {\bibfnamefont {W.}~\bibnamefont {Holtzmann}}, \bibinfo {author} {\bibfnamefont {C.}~\bibnamefont {Hu}}, \bibinfo {author} {\bibfnamefont {Z.}~\bibnamefont {Liu}}, \bibinfo {author} {\bibfnamefont {T.}~\bibnamefont {Taniguchi}}, \bibinfo {author} {\bibfnamefont {K.}~\bibnamefont {Watanabe}}, \bibinfo {author} {\bibfnamefont {J.-H.}\ \bibnamefont {Chu}}, \bibinfo {author} {\bibfnamefont {T.}~\bibnamefont {Cao}}, \bibinfo {author} {\bibfnamefont {L.}~\bibnamefont {Fu}}, \bibinfo {author} {\bibfnamefont {W.}~\bibnamefont {Yao}}, \bibinfo {author}
  {\bibfnamefont {C.-Z.}\ \bibnamefont {Chang}}, \bibinfo {author} {\bibfnamefont {D.}~\bibnamefont {Cobden}}, \bibinfo {author} {\bibfnamefont {D.}~\bibnamefont {Xiao}},\ and\ \bibinfo {author} {\bibfnamefont {X.}~\bibnamefont {Xu}},\ }\href {https://doi.org/10.1038/s41586-023-06536-0} {\bibfield  {journal} {\bibinfo  {journal} {Nature}\ }\textbf {\bibinfo {volume} {622}},\ \bibinfo {pages} {74} (\bibinfo {year} {2023})}\BibitemShut {NoStop}%
\bibitem [{\citenamefont {Xu}\ \emph {et~al.}(2023)\citenamefont {Xu}, \citenamefont {Sun}, \citenamefont {Jia}, \citenamefont {Liu}, \citenamefont {Xu}, \citenamefont {Li}, \citenamefont {Gu}, \citenamefont {Watanabe}, \citenamefont {Taniguchi}, \citenamefont {Tong}, \citenamefont {Jia}, \citenamefont {Shi}, \citenamefont {Jiang}, \citenamefont {Zhang}, \citenamefont {Liu},\ and\ \citenamefont {Li}}]{Xu2023}%
  \BibitemOpen
  \bibfield  {author} {\bibinfo {author} {\bibfnamefont {F.}~\bibnamefont {Xu}}, \bibinfo {author} {\bibfnamefont {Z.}~\bibnamefont {Sun}}, \bibinfo {author} {\bibfnamefont {T.}~\bibnamefont {Jia}}, \bibinfo {author} {\bibfnamefont {C.}~\bibnamefont {Liu}}, \bibinfo {author} {\bibfnamefont {C.}~\bibnamefont {Xu}}, \bibinfo {author} {\bibfnamefont {C.}~\bibnamefont {Li}}, \bibinfo {author} {\bibfnamefont {Y.}~\bibnamefont {Gu}}, \bibinfo {author} {\bibfnamefont {K.}~\bibnamefont {Watanabe}}, \bibinfo {author} {\bibfnamefont {T.}~\bibnamefont {Taniguchi}}, \bibinfo {author} {\bibfnamefont {B.}~\bibnamefont {Tong}}, \bibinfo {author} {\bibfnamefont {J.}~\bibnamefont {Jia}}, \bibinfo {author} {\bibfnamefont {Z.}~\bibnamefont {Shi}}, \bibinfo {author} {\bibfnamefont {S.}~\bibnamefont {Jiang}}, \bibinfo {author} {\bibfnamefont {Y.}~\bibnamefont {Zhang}}, \bibinfo {author} {\bibfnamefont {X.}~\bibnamefont {Liu}},\ and\ \bibinfo {author} {\bibfnamefont {T.}~\bibnamefont {Li}},\ }\href
  {https://doi.org/10.1103/PhysRevX.13.031037} {\bibfield  {journal} {\bibinfo  {journal} {Physical Review X}\ }\textbf {\bibinfo {volume} {13}},\ \bibinfo {pages} {031037} (\bibinfo {year} {2023})}\BibitemShut {NoStop}%
\bibitem [{\citenamefont {Ji}\ \emph {et~al.}(2024)\citenamefont {Ji}, \citenamefont {Park}, \citenamefont {Barber}, \citenamefont {Hu}, \citenamefont {Watanabe}, \citenamefont {Taniguchi}, \citenamefont {Chu}, \citenamefont {Xu},\ and\ \citenamefont {Shen}}]{Ji2024}%
  \BibitemOpen
  \bibfield  {author} {\bibinfo {author} {\bibfnamefont {Z.}~\bibnamefont {Ji}}, \bibinfo {author} {\bibfnamefont {H.}~\bibnamefont {Park}}, \bibinfo {author} {\bibfnamefont {M.~E.}\ \bibnamefont {Barber}}, \bibinfo {author} {\bibfnamefont {C.}~\bibnamefont {Hu}}, \bibinfo {author} {\bibfnamefont {K.}~\bibnamefont {Watanabe}}, \bibinfo {author} {\bibfnamefont {T.}~\bibnamefont {Taniguchi}}, \bibinfo {author} {\bibfnamefont {J.-H.}\ \bibnamefont {Chu}}, \bibinfo {author} {\bibfnamefont {X.}~\bibnamefont {Xu}},\ and\ \bibinfo {author} {\bibfnamefont {Z.-X.}\ \bibnamefont {Shen}},\ }\href {https://doi.org/10.1038/s41586-024-08092-7} {\bibfield  {journal} {\bibinfo  {journal} {Nature}\ }\textbf {\bibinfo {volume} {635}},\ \bibinfo {pages} {578} (\bibinfo {year} {2024})}\BibitemShut {NoStop}%
\bibitem [{\citenamefont {Redekop}\ \emph {et~al.}(2024)\citenamefont {Redekop}, \citenamefont {Zhang}, \citenamefont {Park}, \citenamefont {Cai}, \citenamefont {Anderson}, \citenamefont {Sheekey}, \citenamefont {Arp}, \citenamefont {Babikyan}, \citenamefont {Salters}, \citenamefont {Watanabe}, \citenamefont {Taniguchi}, \citenamefont {Huber}, \citenamefont {Xu},\ and\ \citenamefont {Young}}]{Redekop2024}%
  \BibitemOpen
  \bibfield  {author} {\bibinfo {author} {\bibfnamefont {E.}~\bibnamefont {Redekop}}, \bibinfo {author} {\bibfnamefont {C.}~\bibnamefont {Zhang}}, \bibinfo {author} {\bibfnamefont {H.}~\bibnamefont {Park}}, \bibinfo {author} {\bibfnamefont {J.}~\bibnamefont {Cai}}, \bibinfo {author} {\bibfnamefont {E.}~\bibnamefont {Anderson}}, \bibinfo {author} {\bibfnamefont {O.}~\bibnamefont {Sheekey}}, \bibinfo {author} {\bibfnamefont {T.}~\bibnamefont {Arp}}, \bibinfo {author} {\bibfnamefont {G.}~\bibnamefont {Babikyan}}, \bibinfo {author} {\bibfnamefont {S.}~\bibnamefont {Salters}}, \bibinfo {author} {\bibfnamefont {K.}~\bibnamefont {Watanabe}}, \bibinfo {author} {\bibfnamefont {T.}~\bibnamefont {Taniguchi}}, \bibinfo {author} {\bibfnamefont {M.~E.}\ \bibnamefont {Huber}}, \bibinfo {author} {\bibfnamefont {X.}~\bibnamefont {Xu}},\ and\ \bibinfo {author} {\bibfnamefont {A.~F.}\ \bibnamefont {Young}},\ }\href {https://doi.org/10.1038/s41586-024-08153-x} {\bibfield  {journal} {\bibinfo  {journal} {Nature}\ }\textbf
  {\bibinfo {volume} {635}},\ \bibinfo {pages} {584} (\bibinfo {year} {2024})}\BibitemShut {NoStop}%
\bibitem [{\citenamefont {Kang}\ \emph {et~al.}(2024)\citenamefont {Kang}, \citenamefont {Shen}, \citenamefont {Qiu}, \citenamefont {Zeng}, \citenamefont {Xia}, \citenamefont {Watanabe}, \citenamefont {Taniguchi}, \citenamefont {Shan},\ and\ \citenamefont {Mak}}]{Kang2024}%
  \BibitemOpen
  \bibfield  {author} {\bibinfo {author} {\bibfnamefont {K.}~\bibnamefont {Kang}}, \bibinfo {author} {\bibfnamefont {B.}~\bibnamefont {Shen}}, \bibinfo {author} {\bibfnamefont {Y.}~\bibnamefont {Qiu}}, \bibinfo {author} {\bibfnamefont {Y.}~\bibnamefont {Zeng}}, \bibinfo {author} {\bibfnamefont {Z.}~\bibnamefont {Xia}}, \bibinfo {author} {\bibfnamefont {K.}~\bibnamefont {Watanabe}}, \bibinfo {author} {\bibfnamefont {T.}~\bibnamefont {Taniguchi}}, \bibinfo {author} {\bibfnamefont {J.}~\bibnamefont {Shan}},\ and\ \bibinfo {author} {\bibfnamefont {K.~F.}\ \bibnamefont {Mak}},\ }\href {https://doi.org/10.1038/s41586-024-07214-5} {\bibfield  {journal} {\bibinfo  {journal} {Nature}\ }\textbf {\bibinfo {volume} {628}},\ \bibinfo {pages} {522} (\bibinfo {year} {2024})},\ \Eprint {https://arxiv.org/abs/2402.03294} {arXiv:2402.03294 [cond-mat]} \BibitemShut {NoStop}%
\bibitem [{\citenamefont {Xu}\ \emph {et~al.}(2025{\natexlab{a}})\citenamefont {Xu}, \citenamefont {Chang}, \citenamefont {Xiao}, \citenamefont {Zhang}, \citenamefont {Liu}, \citenamefont {Sun}, \citenamefont {Mao}, \citenamefont {Peshcherenko}, \citenamefont {Li}, \citenamefont {Watanabe}, \citenamefont {Taniguchi}, \citenamefont {Tong}, \citenamefont {Lu}, \citenamefont {Jia}, \citenamefont {Qian}, \citenamefont {Shi}, \citenamefont {Zhang}, \citenamefont {Liu}, \citenamefont {Jiang},\ and\ \citenamefont {Li}}]{Xu2025}%
  \BibitemOpen
  \bibfield  {author} {\bibinfo {author} {\bibfnamefont {F.}~\bibnamefont {Xu}}, \bibinfo {author} {\bibfnamefont {X.}~\bibnamefont {Chang}}, \bibinfo {author} {\bibfnamefont {J.}~\bibnamefont {Xiao}}, \bibinfo {author} {\bibfnamefont {Y.}~\bibnamefont {Zhang}}, \bibinfo {author} {\bibfnamefont {F.}~\bibnamefont {Liu}}, \bibinfo {author} {\bibfnamefont {Z.}~\bibnamefont {Sun}}, \bibinfo {author} {\bibfnamefont {N.}~\bibnamefont {Mao}}, \bibinfo {author} {\bibfnamefont {N.}~\bibnamefont {Peshcherenko}}, \bibinfo {author} {\bibfnamefont {J.}~\bibnamefont {Li}}, \bibinfo {author} {\bibfnamefont {K.}~\bibnamefont {Watanabe}}, \bibinfo {author} {\bibfnamefont {T.}~\bibnamefont {Taniguchi}}, \bibinfo {author} {\bibfnamefont {B.}~\bibnamefont {Tong}}, \bibinfo {author} {\bibfnamefont {L.}~\bibnamefont {Lu}}, \bibinfo {author} {\bibfnamefont {J.}~\bibnamefont {Jia}}, \bibinfo {author} {\bibfnamefont {D.}~\bibnamefont {Qian}}, \bibinfo {author} {\bibfnamefont {Z.}~\bibnamefont {Shi}}, \bibinfo {author} {\bibfnamefont
  {Y.}~\bibnamefont {Zhang}}, \bibinfo {author} {\bibfnamefont {X.}~\bibnamefont {Liu}}, \bibinfo {author} {\bibfnamefont {S.}~\bibnamefont {Jiang}},\ and\ \bibinfo {author} {\bibfnamefont {T.}~\bibnamefont {Li}},\ }\href {https://doi.org/10.1038/s41567-025-02803-1} {\bibfield  {journal} {\bibinfo  {journal} {Nature Physics}\ }\textbf {\bibinfo {volume} {21}},\ \bibinfo {pages} {542} (\bibinfo {year} {2025}{\natexlab{a}})},\ \Eprint {https://arxiv.org/abs/2406.09687} {arXiv:2406.09687 [cond-mat]} \BibitemShut {NoStop}%
\bibitem [{\citenamefont {Park}\ \emph {et~al.}(2025{\natexlab{a}})\citenamefont {Park}, \citenamefont {Cai}, \citenamefont {Anderson}, \citenamefont {Zhang}, \citenamefont {Liu}, \citenamefont {Holtzmann}, \citenamefont {Li}, \citenamefont {Wang}, \citenamefont {Hu}, \citenamefont {Zhao}, \citenamefont {Taniguchi}, \citenamefont {Watanabe}, \citenamefont {Yang}, \citenamefont {Cobden}, \citenamefont {Chu}, \citenamefont {Regnault}, \citenamefont {Bernevig}, \citenamefont {Fu}, \citenamefont {Cao}, \citenamefont {Xiao},\ and\ \citenamefont {Xu}}]{Park2025a}%
  \BibitemOpen
  \bibfield  {author} {\bibinfo {author} {\bibfnamefont {H.}~\bibnamefont {Park}}, \bibinfo {author} {\bibfnamefont {J.}~\bibnamefont {Cai}}, \bibinfo {author} {\bibfnamefont {E.}~\bibnamefont {Anderson}}, \bibinfo {author} {\bibfnamefont {X.-W.}\ \bibnamefont {Zhang}}, \bibinfo {author} {\bibfnamefont {X.}~\bibnamefont {Liu}}, \bibinfo {author} {\bibfnamefont {W.}~\bibnamefont {Holtzmann}}, \bibinfo {author} {\bibfnamefont {W.}~\bibnamefont {Li}}, \bibinfo {author} {\bibfnamefont {C.}~\bibnamefont {Wang}}, \bibinfo {author} {\bibfnamefont {C.}~\bibnamefont {Hu}}, \bibinfo {author} {\bibfnamefont {Y.}~\bibnamefont {Zhao}}, \bibinfo {author} {\bibfnamefont {T.}~\bibnamefont {Taniguchi}}, \bibinfo {author} {\bibfnamefont {K.}~\bibnamefont {Watanabe}}, \bibinfo {author} {\bibfnamefont {J.}~\bibnamefont {Yang}}, \bibinfo {author} {\bibfnamefont {D.}~\bibnamefont {Cobden}}, \bibinfo {author} {\bibfnamefont {J.-H.}\ \bibnamefont {Chu}}, \bibinfo {author} {\bibfnamefont {N.}~\bibnamefont {Regnault}}, \bibinfo
  {author} {\bibfnamefont {B.~A.}\ \bibnamefont {Bernevig}}, \bibinfo {author} {\bibfnamefont {L.}~\bibnamefont {Fu}}, \bibinfo {author} {\bibfnamefont {T.}~\bibnamefont {Cao}}, \bibinfo {author} {\bibfnamefont {D.}~\bibnamefont {Xiao}},\ and\ \bibinfo {author} {\bibfnamefont {X.}~\bibnamefont {Xu}},\ }\href {https://doi.org/10.1038/s41567-025-02804-0} {\bibfield  {journal} {\bibinfo  {journal} {Nature Physics}\ }\textbf {\bibinfo {volume} {21}},\ \bibinfo {pages} {549} (\bibinfo {year} {2025}{\natexlab{a}})},\ \Eprint {https://arxiv.org/abs/2406.09591} {arXiv:2406.09591 [cond-mat]} \BibitemShut {NoStop}%
\bibitem [{\citenamefont {Wang}\ \emph {et~al.}(2025{\natexlab{a}})\citenamefont {Wang}, \citenamefont {Choe}, \citenamefont {Anderson}, \citenamefont {Li}, \citenamefont {Ingham}, \citenamefont {Arsenault}, \citenamefont {Li}, \citenamefont {Hu}, \citenamefont {Taniguchi}, \citenamefont {Watanabe}, \citenamefont {Roy}, \citenamefont {Basov}, \citenamefont {Xiao}, \citenamefont {Queiroz}, \citenamefont {Hone}, \citenamefont {Xu},\ and\ \citenamefont {Zhu}}]{Wang2025a}%
  \BibitemOpen
  \bibfield  {author} {\bibinfo {author} {\bibfnamefont {Y.}~\bibnamefont {Wang}}, \bibinfo {author} {\bibfnamefont {J.}~\bibnamefont {Choe}}, \bibinfo {author} {\bibfnamefont {E.}~\bibnamefont {Anderson}}, \bibinfo {author} {\bibfnamefont {W.}~\bibnamefont {Li}}, \bibinfo {author} {\bibfnamefont {J.}~\bibnamefont {Ingham}}, \bibinfo {author} {\bibfnamefont {E.~A.}\ \bibnamefont {Arsenault}}, \bibinfo {author} {\bibfnamefont {Y.}~\bibnamefont {Li}}, \bibinfo {author} {\bibfnamefont {X.}~\bibnamefont {Hu}}, \bibinfo {author} {\bibfnamefont {T.}~\bibnamefont {Taniguchi}}, \bibinfo {author} {\bibfnamefont {K.}~\bibnamefont {Watanabe}}, \bibinfo {author} {\bibfnamefont {X.}~\bibnamefont {Roy}}, \bibinfo {author} {\bibfnamefont {D.}~\bibnamefont {Basov}}, \bibinfo {author} {\bibfnamefont {D.}~\bibnamefont {Xiao}}, \bibinfo {author} {\bibfnamefont {R.}~\bibnamefont {Queiroz}}, \bibinfo {author} {\bibfnamefont {J.~C.}\ \bibnamefont {Hone}}, \bibinfo {author} {\bibfnamefont {X.}~\bibnamefont {Xu}},\ and\ \bibinfo
  {author} {\bibfnamefont {X.-Y.}\ \bibnamefont {Zhu}},\ }\href {https://doi.org/10.1038/s41586-025-08954-8} {\bibfield  {journal} {\bibinfo  {journal} {Nature}\ ,\ \bibinfo {pages} {1149}} (\bibinfo {year} {2025}{\natexlab{a}})},\ \Eprint {https://arxiv.org/abs/2502.21153} {arXiv:2502.21153 [cond-mat]} \BibitemShut {NoStop}%
\bibitem [{\citenamefont {Park}\ \emph {et~al.}(2025{\natexlab{b}})\citenamefont {Park}, \citenamefont {Li}, \citenamefont {Hu}, \citenamefont {Beach}, \citenamefont {Gon{\c c}alves}, \citenamefont {{Mendez-Valderrama}}, \citenamefont {{Herzog-Arbeitman}}, \citenamefont {Taniguchi}, \citenamefont {Watanabe}, \citenamefont {Cobden}, \citenamefont {Fu}, \citenamefont {Bernevig}, \citenamefont {Regnault}, \citenamefont {Chu}, \citenamefont {Xiao},\ and\ \citenamefont {Xu}}]{Park2025}%
  \BibitemOpen
  \bibfield  {author} {\bibinfo {author} {\bibfnamefont {H.}~\bibnamefont {Park}}, \bibinfo {author} {\bibfnamefont {W.}~\bibnamefont {Li}}, \bibinfo {author} {\bibfnamefont {C.}~\bibnamefont {Hu}}, \bibinfo {author} {\bibfnamefont {C.}~\bibnamefont {Beach}}, \bibinfo {author} {\bibfnamefont {M.}~\bibnamefont {Gon{\c c}alves}}, \bibinfo {author} {\bibfnamefont {J.~F.}\ \bibnamefont {{Mendez-Valderrama}}}, \bibinfo {author} {\bibfnamefont {J.}~\bibnamefont {{Herzog-Arbeitman}}}, \bibinfo {author} {\bibfnamefont {T.}~\bibnamefont {Taniguchi}}, \bibinfo {author} {\bibfnamefont {K.}~\bibnamefont {Watanabe}}, \bibinfo {author} {\bibfnamefont {D.}~\bibnamefont {Cobden}}, \bibinfo {author} {\bibfnamefont {L.}~\bibnamefont {Fu}}, \bibinfo {author} {\bibfnamefont {B.~A.}\ \bibnamefont {Bernevig}}, \bibinfo {author} {\bibfnamefont {N.}~\bibnamefont {Regnault}}, \bibinfo {author} {\bibfnamefont {J.-H.}\ \bibnamefont {Chu}}, \bibinfo {author} {\bibfnamefont {D.}~\bibnamefont {Xiao}},\ and\ \bibinfo {author}
  {\bibfnamefont {X.}~\bibnamefont {Xu}},\ }\href {https://doi.org/10.48550/arXiv.2503.10989} {\bibinfo {title} {Observation of {{High-Temperature Dissipationless Fractional Chern Insulator}}}} (\bibinfo {year} {2025}{\natexlab{b}}),\ \Eprint {https://arxiv.org/abs/2503.10989} {arXiv:2503.10989 [cond-mat.mes-hall]} \BibitemShut {NoStop}%
\bibitem [{\citenamefont {Chang}\ \emph {et~al.}(2026)\citenamefont {Chang}, \citenamefont {Liu}, \citenamefont {Xu}, \citenamefont {Xu}, \citenamefont {Xiao}, \citenamefont {Sun}, \citenamefont {Jiao}, \citenamefont {Zhang}, \citenamefont {Wang}, \citenamefont {Shen}, \citenamefont {He}, \citenamefont {Watanabe}, \citenamefont {Taniguchi}, \citenamefont {Zhong}, \citenamefont {Jia}, \citenamefont {Shi}, \citenamefont {Liu}, \citenamefont {Zhang}, \citenamefont {Qian}, \citenamefont {Li},\ and\ \citenamefont {Jiang}}]{Chang2026}%
  \BibitemOpen
  \bibfield  {author} {\bibinfo {author} {\bibfnamefont {X.}~\bibnamefont {Chang}}, \bibinfo {author} {\bibfnamefont {F.}~\bibnamefont {Liu}}, \bibinfo {author} {\bibfnamefont {F.}~\bibnamefont {Xu}}, \bibinfo {author} {\bibfnamefont {C.}~\bibnamefont {Xu}}, \bibinfo {author} {\bibfnamefont {J.}~\bibnamefont {Xiao}}, \bibinfo {author} {\bibfnamefont {Z.}~\bibnamefont {Sun}}, \bibinfo {author} {\bibfnamefont {P.}~\bibnamefont {Jiao}}, \bibinfo {author} {\bibfnamefont {Y.}~\bibnamefont {Zhang}}, \bibinfo {author} {\bibfnamefont {S.}~\bibnamefont {Wang}}, \bibinfo {author} {\bibfnamefont {B.}~\bibnamefont {Shen}}, \bibinfo {author} {\bibfnamefont {R.}~\bibnamefont {He}}, \bibinfo {author} {\bibfnamefont {K.}~\bibnamefont {Watanabe}}, \bibinfo {author} {\bibfnamefont {T.}~\bibnamefont {Taniguchi}}, \bibinfo {author} {\bibfnamefont {R.}~\bibnamefont {Zhong}}, \bibinfo {author} {\bibfnamefont {J.}~\bibnamefont {Jia}}, \bibinfo {author} {\bibfnamefont {Z.}~\bibnamefont {Shi}}, \bibinfo {author} {\bibfnamefont
  {X.}~\bibnamefont {Liu}}, \bibinfo {author} {\bibfnamefont {Y.}~\bibnamefont {Zhang}}, \bibinfo {author} {\bibfnamefont {D.}~\bibnamefont {Qian}}, \bibinfo {author} {\bibfnamefont {T.}~\bibnamefont {Li}},\ and\ \bibinfo {author} {\bibfnamefont {S.}~\bibnamefont {Jiang}},\ }\href {https://doi.org/10.1038/s41467-026-71479-9} {\bibfield  {journal} {\bibinfo  {journal} {Nature Communications}\ }\textbf {\bibinfo {volume} {17}},\ \bibinfo {pages} {4874} (\bibinfo {year} {2026})},\ \Eprint {https://arxiv.org/abs/2503.13213} {arXiv:2503.13213 [cond-mat]} \BibitemShut {NoStop}%
\bibitem [{\citenamefont {Sun}\ \emph {et~al.}(2026)\citenamefont {Sun}, \citenamefont {Xu}, \citenamefont {Li}, \citenamefont {Jiang}, \citenamefont {Gao}, \citenamefont {Xu}, \citenamefont {Jia}, \citenamefont {Cheng}, \citenamefont {Zhang}, \citenamefont {Tian}, \citenamefont {Watanabe}, \citenamefont {Taniguchi}, \citenamefont {Jia}, \citenamefont {Jiang}, \citenamefont {Zhang}, \citenamefont {Zhang}, \citenamefont {Lei}, \citenamefont {Liu},\ and\ \citenamefont {Li}}]{Sun2026}%
  \BibitemOpen
  \bibfield  {author} {\bibinfo {author} {\bibfnamefont {Z.}~\bibnamefont {Sun}}, \bibinfo {author} {\bibfnamefont {F.}~\bibnamefont {Xu}}, \bibinfo {author} {\bibfnamefont {J.}~\bibnamefont {Li}}, \bibinfo {author} {\bibfnamefont {Y.}~\bibnamefont {Jiang}}, \bibinfo {author} {\bibfnamefont {J.}~\bibnamefont {Gao}}, \bibinfo {author} {\bibfnamefont {C.}~\bibnamefont {Xu}}, \bibinfo {author} {\bibfnamefont {T.}~\bibnamefont {Jia}}, \bibinfo {author} {\bibfnamefont {K.}~\bibnamefont {Cheng}}, \bibinfo {author} {\bibfnamefont {J.}~\bibnamefont {Zhang}}, \bibinfo {author} {\bibfnamefont {W.}~\bibnamefont {Tian}}, \bibinfo {author} {\bibfnamefont {K.}~\bibnamefont {Watanabe}}, \bibinfo {author} {\bibfnamefont {T.}~\bibnamefont {Taniguchi}}, \bibinfo {author} {\bibfnamefont {J.}~\bibnamefont {Jia}}, \bibinfo {author} {\bibfnamefont {S.}~\bibnamefont {Jiang}}, \bibinfo {author} {\bibfnamefont {Y.}~\bibnamefont {Zhang}}, \bibinfo {author} {\bibfnamefont {Y.}~\bibnamefont {Zhang}}, \bibinfo {author} {\bibfnamefont
  {S.}~\bibnamefont {Lei}}, \bibinfo {author} {\bibfnamefont {X.}~\bibnamefont {Liu}},\ and\ \bibinfo {author} {\bibfnamefont {T.}~\bibnamefont {Li}},\ }\href {https://arxiv.org/abs/2603.16412v1} {\bibinfo {title} {Twist-angle evolution from valley-polarized fractional topological phases to valley-degenerate superconductivity in twisted bilayer {{MoTe2}}}} (\bibinfo {year} {2026})\BibitemShut {NoStop}%
\bibitem [{\citenamefont {Li}\ \emph {et~al.}(2026{\natexlab{a}})\citenamefont {Li}, \citenamefont {Wang~Beach}, \citenamefont {Hu}, \citenamefont {Taniguchi}, \citenamefont {Watanabe}, \citenamefont {Chu}, \citenamefont {Imamo{\u g}lu}, \citenamefont {Cao}, \citenamefont {Xiao},\ and\ \citenamefont {Xu}}]{Li2026}%
  \BibitemOpen
  \bibfield  {author} {\bibinfo {author} {\bibfnamefont {W.}~\bibnamefont {Li}}, \bibinfo {author} {\bibfnamefont {C.}~\bibnamefont {Wang~Beach}}, \bibinfo {author} {\bibfnamefont {C.}~\bibnamefont {Hu}}, \bibinfo {author} {\bibfnamefont {T.}~\bibnamefont {Taniguchi}}, \bibinfo {author} {\bibfnamefont {K.}~\bibnamefont {Watanabe}}, \bibinfo {author} {\bibfnamefont {J.-H.}\ \bibnamefont {Chu}}, \bibinfo {author} {\bibfnamefont {A.}~\bibnamefont {Imamo{\u g}lu}}, \bibinfo {author} {\bibfnamefont {T.}~\bibnamefont {Cao}}, \bibinfo {author} {\bibfnamefont {D.}~\bibnamefont {Xiao}},\ and\ \bibinfo {author} {\bibfnamefont {X.}~\bibnamefont {Xu}},\ }\href {https://doi.org/10.1038/s41586-026-10101-w} {\bibfield  {journal} {\bibinfo  {journal} {Nature}\ }\textbf {\bibinfo {volume} {651}},\ \bibinfo {pages} {48} (\bibinfo {year} {2026}{\natexlab{a}})}\BibitemShut {NoStop}%
\bibitem [{\citenamefont {Xu}\ \emph {et~al.}(2026)\citenamefont {Xu}, \citenamefont {Sun}, \citenamefont {Li}, \citenamefont {Zheng}, \citenamefont {Xu}, \citenamefont {Gao}, \citenamefont {Jia}, \citenamefont {Su}, \citenamefont {Watanabe}, \citenamefont {Taniguchi}, \citenamefont {Tong}, \citenamefont {Lu}, \citenamefont {Jia}, \citenamefont {Shi}, \citenamefont {Jiang}, \citenamefont {Lin}, \citenamefont {Zhang}, \citenamefont {Zhang}, \citenamefont {Lei}, \citenamefont {Liu},\ and\ \citenamefont {Li}}]{Xu2026}%
  \BibitemOpen
  \bibfield  {author} {\bibinfo {author} {\bibfnamefont {F.}~\bibnamefont {Xu}}, \bibinfo {author} {\bibfnamefont {Z.}~\bibnamefont {Sun}}, \bibinfo {author} {\bibfnamefont {J.}~\bibnamefont {Li}}, \bibinfo {author} {\bibfnamefont {C.}~\bibnamefont {Zheng}}, \bibinfo {author} {\bibfnamefont {C.}~\bibnamefont {Xu}}, \bibinfo {author} {\bibfnamefont {J.}~\bibnamefont {Gao}}, \bibinfo {author} {\bibfnamefont {T.}~\bibnamefont {Jia}}, \bibinfo {author} {\bibfnamefont {Y.}~\bibnamefont {Su}}, \bibinfo {author} {\bibfnamefont {K.}~\bibnamefont {Watanabe}}, \bibinfo {author} {\bibfnamefont {T.}~\bibnamefont {Taniguchi}}, \bibinfo {author} {\bibfnamefont {B.}~\bibnamefont {Tong}}, \bibinfo {author} {\bibfnamefont {L.}~\bibnamefont {Lu}}, \bibinfo {author} {\bibfnamefont {J.}~\bibnamefont {Jia}}, \bibinfo {author} {\bibfnamefont {Z.}~\bibnamefont {Shi}}, \bibinfo {author} {\bibfnamefont {S.}~\bibnamefont {Jiang}}, \bibinfo {author} {\bibfnamefont {J.}~\bibnamefont {Lin}}, \bibinfo {author} {\bibfnamefont
  {Y.}~\bibnamefont {Zhang}}, \bibinfo {author} {\bibfnamefont {Y.}~\bibnamefont {Zhang}}, \bibinfo {author} {\bibfnamefont {S.}~\bibnamefont {Lei}}, \bibinfo {author} {\bibfnamefont {X.}~\bibnamefont {Liu}},\ and\ \bibinfo {author} {\bibfnamefont {T.}~\bibnamefont {Li}},\ }\href {https://doi.org/10.48550/arXiv.2504.06972} {\bibinfo {title} {Signatures of unconventional superconductivity near reentrant and fractional quantum anomalous {{Hall}} insulators}} (\bibinfo {year} {2026}),\ \Eprint {https://arxiv.org/abs/2504.06972} {arXiv:2504.06972 [cond-mat.mes-hall]} \BibitemShut {NoStop}%
\bibitem [{\citenamefont {Xu}\ \emph {et~al.}(2025{\natexlab{b}})\citenamefont {Xu}, \citenamefont {Sun}, \citenamefont {Li}, \citenamefont {Zheng}, \citenamefont {Xu}, \citenamefont {Gao}, \citenamefont {Jia}, \citenamefont {Watanabe}, \citenamefont {Taniguchi}, \citenamefont {Tong}, \citenamefont {Lu}, \citenamefont {Jia}, \citenamefont {Shi}, \citenamefont {Jiang}, \citenamefont {Zhang}, \citenamefont {Zhang}, \citenamefont {Lei}, \citenamefont {Liu},\ and\ \citenamefont {Li}}]{Xu2025b}%
  \BibitemOpen
  \bibfield  {author} {\bibinfo {author} {\bibfnamefont {F.}~\bibnamefont {Xu}}, \bibinfo {author} {\bibfnamefont {Z.}~\bibnamefont {Sun}}, \bibinfo {author} {\bibfnamefont {J.}~\bibnamefont {Li}}, \bibinfo {author} {\bibfnamefont {C.}~\bibnamefont {Zheng}}, \bibinfo {author} {\bibfnamefont {C.}~\bibnamefont {Xu}}, \bibinfo {author} {\bibfnamefont {J.}~\bibnamefont {Gao}}, \bibinfo {author} {\bibfnamefont {T.}~\bibnamefont {Jia}}, \bibinfo {author} {\bibfnamefont {K.}~\bibnamefont {Watanabe}}, \bibinfo {author} {\bibfnamefont {T.}~\bibnamefont {Taniguchi}}, \bibinfo {author} {\bibfnamefont {B.}~\bibnamefont {Tong}}, \bibinfo {author} {\bibfnamefont {L.}~\bibnamefont {Lu}}, \bibinfo {author} {\bibfnamefont {J.}~\bibnamefont {Jia}}, \bibinfo {author} {\bibfnamefont {Z.}~\bibnamefont {Shi}}, \bibinfo {author} {\bibfnamefont {S.}~\bibnamefont {Jiang}}, \bibinfo {author} {\bibfnamefont {Y.}~\bibnamefont {Zhang}}, \bibinfo {author} {\bibfnamefont {Y.}~\bibnamefont {Zhang}}, \bibinfo {author} {\bibfnamefont
  {S.}~\bibnamefont {Lei}}, \bibinfo {author} {\bibfnamefont {X.}~\bibnamefont {Liu}},\ and\ \bibinfo {author} {\bibfnamefont {T.}~\bibnamefont {Li}},\ }\href {https://doi.org/10.48550/arXiv.2504.06972} {\bibinfo {title} {Signatures of unconventional superconductivity near reentrant and fractional quantum anomalous {{Hall}} insulators}} (\bibinfo {year} {2025}{\natexlab{b}}),\ \Eprint {https://arxiv.org/abs/2504.06972} {arXiv:2504.06972 [cond-mat]} \BibitemShut {NoStop}%
\bibitem [{\citenamefont {Hubbard}(1978)}]{Hubbard1978}%
  \BibitemOpen
  \bibfield  {author} {\bibinfo {author} {\bibfnamefont {J.}~\bibnamefont {Hubbard}},\ }\href {https://doi.org/10.1103/PhysRevB.17.494} {\bibfield  {journal} {\bibinfo  {journal} {Phys. Rev. B}\ }\textbf {\bibinfo {volume} {17}},\ \bibinfo {pages} {494} (\bibinfo {year} {1978})}\BibitemShut {NoStop}%
\bibitem [{\citenamefont {Padhi}\ \emph {et~al.}(2021)\citenamefont {Padhi}, \citenamefont {Chitra},\ and\ \citenamefont {Phillips}}]{Padhi2021}%
  \BibitemOpen
  \bibfield  {author} {\bibinfo {author} {\bibfnamefont {B.}~\bibnamefont {Padhi}}, \bibinfo {author} {\bibfnamefont {R.}~\bibnamefont {Chitra}},\ and\ \bibinfo {author} {\bibfnamefont {P.~W.}\ \bibnamefont {Phillips}},\ }\href {https://doi.org/10.1103/PhysRevB.103.125146} {\bibfield  {journal} {\bibinfo  {journal} {Phys. Rev. B}\ }\textbf {\bibinfo {volume} {103}},\ \bibinfo {pages} {125146} (\bibinfo {year} {2021})}\BibitemShut {NoStop}%
\bibitem [{\citenamefont {Zhang}\ \emph {et~al.}(2025)\citenamefont {Zhang}, \citenamefont {Wang}, \citenamefont {Jiang}, \citenamefont {Liu}, \citenamefont {Watanabe}, \citenamefont {Taniguchi}, \citenamefont {Liu}, \citenamefont {Lei}, \citenamefont {Li},\ and\ \citenamefont {Xu}}]{Zhang2025}%
  \BibitemOpen
  \bibfield  {author} {\bibinfo {author} {\bibfnamefont {M.}~\bibnamefont {Zhang}}, \bibinfo {author} {\bibfnamefont {Z.}~\bibnamefont {Wang}}, \bibinfo {author} {\bibfnamefont {Y.}~\bibnamefont {Jiang}}, \bibinfo {author} {\bibfnamefont {Y.}~\bibnamefont {Liu}}, \bibinfo {author} {\bibfnamefont {K.}~\bibnamefont {Watanabe}}, \bibinfo {author} {\bibfnamefont {T.}~\bibnamefont {Taniguchi}}, \bibinfo {author} {\bibfnamefont {S.}~\bibnamefont {Liu}}, \bibinfo {author} {\bibfnamefont {S.}~\bibnamefont {Lei}}, \bibinfo {author} {\bibfnamefont {Y.}~\bibnamefont {Li}},\ and\ \bibinfo {author} {\bibfnamefont {Y.}~\bibnamefont {Xu}},\ }\href {https://doi.org/10.48550/arXiv.2506.20392} {\bibinfo {title} {Transport {{Evidence}} for {{Wigner Crystals}} in {{Monolayer MoTe2}}}} (\bibinfo {year} {2025}),\ \Eprint {https://arxiv.org/abs/2506.20392} {arXiv:2506.20392 [cond-mat.str-el]} \BibitemShut {NoStop}%
\bibitem [{\citenamefont {Liu}\ \emph {et~al.}(2026)\citenamefont {Liu}, \citenamefont {Gu}, \citenamefont {Bao}, \citenamefont {Mao}, \citenamefont {Jiang}, \citenamefont {Liu}, \citenamefont {Guan}, \citenamefont {Li}, \citenamefont {Zheng}, \citenamefont {Liu}, \citenamefont {Watanabe}, \citenamefont {Taniguchi}, \citenamefont {Duan}, \citenamefont {Jia}, \citenamefont {Liu}, \citenamefont {Li}, \citenamefont {Zhang}, \citenamefont {Li},\ and\ \citenamefont {Wang}}]{Liu2026}%
  \BibitemOpen
  \bibfield  {author} {\bibinfo {author} {\bibfnamefont {Y.}~\bibnamefont {Liu}}, \bibinfo {author} {\bibfnamefont {Y.}~\bibnamefont {Gu}}, \bibinfo {author} {\bibfnamefont {T.}~\bibnamefont {Bao}}, \bibinfo {author} {\bibfnamefont {N.}~\bibnamefont {Mao}}, \bibinfo {author} {\bibfnamefont {S.}~\bibnamefont {Jiang}}, \bibinfo {author} {\bibfnamefont {L.}~\bibnamefont {Liu}}, \bibinfo {author} {\bibfnamefont {D.}~\bibnamefont {Guan}}, \bibinfo {author} {\bibfnamefont {Y.}~\bibnamefont {Li}}, \bibinfo {author} {\bibfnamefont {H.}~\bibnamefont {Zheng}}, \bibinfo {author} {\bibfnamefont {C.}~\bibnamefont {Liu}}, \bibinfo {author} {\bibfnamefont {K.}~\bibnamefont {Watanabe}}, \bibinfo {author} {\bibfnamefont {T.}~\bibnamefont {Taniguchi}}, \bibinfo {author} {\bibfnamefont {W.}~\bibnamefont {Duan}}, \bibinfo {author} {\bibfnamefont {J.}~\bibnamefont {Jia}}, \bibinfo {author} {\bibfnamefont {X.}~\bibnamefont {Liu}}, \bibinfo {author} {\bibfnamefont {C.}~\bibnamefont {Li}}, \bibinfo {author} {\bibfnamefont
  {Y.}~\bibnamefont {Zhang}}, \bibinfo {author} {\bibfnamefont {T.}~\bibnamefont {Li}},\ and\ \bibinfo {author} {\bibfnamefont {S.}~\bibnamefont {Wang}},\ }\href {https://doi.org/10.1093/nsr/nwag014} {\bibfield  {journal} {\bibinfo  {journal} {National Science Review}\ }\textbf {\bibinfo {volume} {13}},\ \bibinfo {pages} {nwag014} (\bibinfo {year} {2026})}\BibitemShut {NoStop}%
\bibitem [{\citenamefont {Li}\ \emph {et~al.}(2021{\natexlab{b}})\citenamefont {Li}, \citenamefont {Li}, \citenamefont {Regan}, \citenamefont {Wang}, \citenamefont {Zhao}, \citenamefont {Kahn}, \citenamefont {Yumigeta}, \citenamefont {Blei}, \citenamefont {Taniguchi}, \citenamefont {Watanabe}, \citenamefont {Tongay}, \citenamefont {Zettl}, \citenamefont {Crommie},\ and\ \citenamefont {Wang}}]{Li2021}%
  \BibitemOpen
  \bibfield  {author} {\bibinfo {author} {\bibfnamefont {H.}~\bibnamefont {Li}}, \bibinfo {author} {\bibfnamefont {S.}~\bibnamefont {Li}}, \bibinfo {author} {\bibfnamefont {E.~C.}\ \bibnamefont {Regan}}, \bibinfo {author} {\bibfnamefont {D.}~\bibnamefont {Wang}}, \bibinfo {author} {\bibfnamefont {W.}~\bibnamefont {Zhao}}, \bibinfo {author} {\bibfnamefont {S.}~\bibnamefont {Kahn}}, \bibinfo {author} {\bibfnamefont {K.}~\bibnamefont {Yumigeta}}, \bibinfo {author} {\bibfnamefont {M.}~\bibnamefont {Blei}}, \bibinfo {author} {\bibfnamefont {T.}~\bibnamefont {Taniguchi}}, \bibinfo {author} {\bibfnamefont {K.}~\bibnamefont {Watanabe}}, \bibinfo {author} {\bibfnamefont {S.}~\bibnamefont {Tongay}}, \bibinfo {author} {\bibfnamefont {A.}~\bibnamefont {Zettl}}, \bibinfo {author} {\bibfnamefont {M.~F.}\ \bibnamefont {Crommie}},\ and\ \bibinfo {author} {\bibfnamefont {F.}~\bibnamefont {Wang}},\ }\href {https://doi.org/10.1038/s41586-021-03874-9} {\bibfield  {journal} {\bibinfo  {journal} {Nature}\ }\textbf {\bibinfo
  {volume} {597}},\ \bibinfo {pages} {650} (\bibinfo {year} {2021}{\natexlab{b}})}\BibitemShut {NoStop}%
\bibitem [{\citenamefont {Reddy}\ and\ \citenamefont {Fu}(2023)}]{Reddy2023c}%
  \BibitemOpen
  \bibfield  {author} {\bibinfo {author} {\bibfnamefont {A.~P.}\ \bibnamefont {Reddy}}\ and\ \bibinfo {author} {\bibfnamefont {L.}~\bibnamefont {Fu}},\ }\href {https://doi.org/10.48550/arXiv.2308.10406} {\bibinfo {title} {Toward a global phase diagram of the fractional quantum anomalous {{Hall}} effect}} (\bibinfo {year} {2023}),\ \Eprint {https://arxiv.org/abs/2308.10406} {arXiv:2308.10406 [cond-mat]} \BibitemShut {NoStop}%
\bibitem [{\citenamefont {Sheng}\ \emph {et~al.}(2024)\citenamefont {Sheng}, \citenamefont {Reddy}, \citenamefont {Abouelkomsan}, \citenamefont {Bergholtz},\ and\ \citenamefont {Fu}}]{Sheng2024a}%
  \BibitemOpen
  \bibfield  {author} {\bibinfo {author} {\bibfnamefont {D.~N.}\ \bibnamefont {Sheng}}, \bibinfo {author} {\bibfnamefont {A.~P.}\ \bibnamefont {Reddy}}, \bibinfo {author} {\bibfnamefont {A.}~\bibnamefont {Abouelkomsan}}, \bibinfo {author} {\bibfnamefont {E.~J.}\ \bibnamefont {Bergholtz}},\ and\ \bibinfo {author} {\bibfnamefont {L.}~\bibnamefont {Fu}},\ }\href {https://doi.org/10.1103/PhysRevLett.133.066601} {\bibfield  {journal} {\bibinfo  {journal} {Physical Review Letters}\ }\textbf {\bibinfo {volume} {133}},\ \bibinfo {pages} {066601} (\bibinfo {year} {2024})}\BibitemShut {NoStop}%
\bibitem [{\citenamefont {Yu}\ \emph {et~al.}(2024)\citenamefont {Yu}, \citenamefont {{Herzog-Arbeitman}}, \citenamefont {Wang}, \citenamefont {Vafek}, \citenamefont {Bernevig},\ and\ \citenamefont {Regnault}}]{Yu2024b}%
  \BibitemOpen
  \bibfield  {author} {\bibinfo {author} {\bibfnamefont {J.}~\bibnamefont {Yu}}, \bibinfo {author} {\bibfnamefont {J.}~\bibnamefont {{Herzog-Arbeitman}}}, \bibinfo {author} {\bibfnamefont {M.}~\bibnamefont {Wang}}, \bibinfo {author} {\bibfnamefont {O.}~\bibnamefont {Vafek}}, \bibinfo {author} {\bibfnamefont {B.~A.}\ \bibnamefont {Bernevig}},\ and\ \bibinfo {author} {\bibfnamefont {N.}~\bibnamefont {Regnault}},\ }\href {https://doi.org/10.1103/PhysRevB.109.045147} {\bibfield  {journal} {\bibinfo  {journal} {Physical Review B}\ }\textbf {\bibinfo {volume} {109}},\ \bibinfo {pages} {045147} (\bibinfo {year} {2024})}\BibitemShut {NoStop}%
\bibitem [{\citenamefont {Cr{\'e}pel}\ and\ \citenamefont {Fu}(2023)}]{Crepel2023}%
  \BibitemOpen
  \bibfield  {author} {\bibinfo {author} {\bibfnamefont {V.}~\bibnamefont {Cr{\'e}pel}}\ and\ \bibinfo {author} {\bibfnamefont {L.}~\bibnamefont {Fu}},\ }\href {https://doi.org/10.1103/PhysRevB.107.L201109} {\bibfield  {journal} {\bibinfo  {journal} {Physical Review B}\ }\textbf {\bibinfo {volume} {107}},\ \bibinfo {pages} {L201109} (\bibinfo {year} {2023})},\ \Eprint {https://arxiv.org/abs/2207.08895} {arXiv:2207.08895 [cond-mat]} \BibitemShut {NoStop}%
\bibitem [{\citenamefont {Reddy}\ \emph {et~al.}(2023)\citenamefont {Reddy}, \citenamefont {Alsallom}, \citenamefont {Zhang}, \citenamefont {Devakul},\ and\ \citenamefont {Fu}}]{Reddy2023e}%
  \BibitemOpen
  \bibfield  {author} {\bibinfo {author} {\bibfnamefont {A.~P.}\ \bibnamefont {Reddy}}, \bibinfo {author} {\bibfnamefont {F.}~\bibnamefont {Alsallom}}, \bibinfo {author} {\bibfnamefont {Y.}~\bibnamefont {Zhang}}, \bibinfo {author} {\bibfnamefont {T.}~\bibnamefont {Devakul}},\ and\ \bibinfo {author} {\bibfnamefont {L.}~\bibnamefont {Fu}},\ }\href {https://doi.org/10.1103/PhysRevB.108.085117} {\bibfield  {journal} {\bibinfo  {journal} {Physical Review B}\ }\textbf {\bibinfo {volume} {108}},\ \bibinfo {pages} {085117} (\bibinfo {year} {2023})}\BibitemShut {NoStop}%
\bibitem [{\citenamefont {He}\ \emph {et~al.}(2025)\citenamefont {He}, \citenamefont {Simon},\ and\ \citenamefont {Parameswaran}}]{He2025}%
  \BibitemOpen
  \bibfield  {author} {\bibinfo {author} {\bibfnamefont {Y.}~\bibnamefont {He}}, \bibinfo {author} {\bibfnamefont {S.~H.}\ \bibnamefont {Simon}},\ and\ \bibinfo {author} {\bibfnamefont {S.~A.}\ \bibnamefont {Parameswaran}},\ }\href {https://doi.org/10.48550/arXiv.2505.06354} {\bibinfo {title} {Fractional {{Chern Insulators}} and {{Competing States}} in a {{Twisted MoTe}}\$\_2\$ {{Lattice Model}}}} (\bibinfo {year} {2025}),\ \Eprint {https://arxiv.org/abs/2505.06354} {arXiv:2505.06354 [cond-mat]} \BibitemShut {NoStop}%
\bibitem [{\citenamefont {Chen}\ \emph {et~al.}(2026{\natexlab{a}})\citenamefont {Chen}, \citenamefont {Li}, \citenamefont {Wang},\ and\ \citenamefont {Li}}]{Chen2026c}%
  \BibitemOpen
  \bibfield  {author} {\bibinfo {author} {\bibfnamefont {J.}~\bibnamefont {Chen}}, \bibinfo {author} {\bibfnamefont {Q.}~\bibnamefont {Li}}, \bibinfo {author} {\bibfnamefont {X.}~\bibnamefont {Wang}},\ and\ \bibinfo {author} {\bibfnamefont {W.}~\bibnamefont {Li}},\ }\href {https://doi.org/10.1016/j.scib.2026.01.014} {\bibfield  {journal} {\bibinfo  {journal} {Science Bulletin}\ }\textbf {\bibinfo {volume} {71}},\ \bibinfo {pages} {1034} (\bibinfo {year} {2026}{\natexlab{a}})}\BibitemShut {NoStop}%
\bibitem [{\citenamefont {Wang}\ \emph {et~al.}(2025{\natexlab{b}})\citenamefont {Wang}, \citenamefont {Yu}, \citenamefont {Sharma},\ and\ \citenamefont {Liu}}]{Wang2025}%
  \BibitemOpen
  \bibfield  {author} {\bibinfo {author} {\bibfnamefont {B.}~\bibnamefont {Wang}}, \bibinfo {author} {\bibfnamefont {J.}~\bibnamefont {Yu}}, \bibinfo {author} {\bibfnamefont {P.}~\bibnamefont {Sharma}},\ and\ \bibinfo {author} {\bibfnamefont {C.-C.}\ \bibnamefont {Liu}},\ }\href {https://doi.org/10.48550/arXiv.2504.11177} {\bibinfo {title} {Pressure-{{Tunable Generalized Wigner Crystal}} and {{Fractional Chern Insulator}} in twisted {{MoTe}}\$\_2\$}} (\bibinfo {year} {2025}{\natexlab{b}}),\ \Eprint {https://arxiv.org/abs/2504.11177} {arXiv:2504.11177 [cond-mat]} \BibitemShut {NoStop}%
\bibitem [{\citenamefont {Tuo}\ \emph {et~al.}(2025)\citenamefont {Tuo}, \citenamefont {Li},\ and\ \citenamefont {Yao}}]{tuo2025fractionalquantumanomaloushall}%
  \BibitemOpen
  \bibfield  {author} {\bibinfo {author} {\bibfnamefont {C.}~\bibnamefont {Tuo}}, \bibinfo {author} {\bibfnamefont {M.-R.}\ \bibnamefont {Li}},\ and\ \bibinfo {author} {\bibfnamefont {H.}~\bibnamefont {Yao}},\ }\href {https://arxiv.org/abs/2512.23608} {\bibinfo {title} {Fractional quantum anomalous hall and anyon density-wave halo in a minimal interacting lattice model of twisted bilayer mote$_2$}} (\bibinfo {year} {2025}),\ \Eprint {https://arxiv.org/abs/2512.23608} {arXiv:2512.23608 [cond-mat.str-el]} \BibitemShut {NoStop}%
\bibitem [{\citenamefont {Anderson}\ \emph {et~al.}(2024)\citenamefont {Anderson}, \citenamefont {Cai}, \citenamefont {Reddy}, \citenamefont {Park}, \citenamefont {Holtzmann}, \citenamefont {Davis}, \citenamefont {Taniguchi}, \citenamefont {Watanabe}, \citenamefont {Smolenski}, \citenamefont {Imamo{\u g}lu}, \citenamefont {Cao}, \citenamefont {Xiao}, \citenamefont {Fu}, \citenamefont {Yao},\ and\ \citenamefont {Xu}}]{Anderson2024}%
  \BibitemOpen
  \bibfield  {author} {\bibinfo {author} {\bibfnamefont {E.}~\bibnamefont {Anderson}}, \bibinfo {author} {\bibfnamefont {J.}~\bibnamefont {Cai}}, \bibinfo {author} {\bibfnamefont {A.~P.}\ \bibnamefont {Reddy}}, \bibinfo {author} {\bibfnamefont {H.}~\bibnamefont {Park}}, \bibinfo {author} {\bibfnamefont {W.}~\bibnamefont {Holtzmann}}, \bibinfo {author} {\bibfnamefont {K.}~\bibnamefont {Davis}}, \bibinfo {author} {\bibfnamefont {T.}~\bibnamefont {Taniguchi}}, \bibinfo {author} {\bibfnamefont {K.}~\bibnamefont {Watanabe}}, \bibinfo {author} {\bibfnamefont {T.}~\bibnamefont {Smolenski}}, \bibinfo {author} {\bibfnamefont {A.}~\bibnamefont {Imamo{\u g}lu}}, \bibinfo {author} {\bibfnamefont {T.}~\bibnamefont {Cao}}, \bibinfo {author} {\bibfnamefont {D.}~\bibnamefont {Xiao}}, \bibinfo {author} {\bibfnamefont {L.}~\bibnamefont {Fu}}, \bibinfo {author} {\bibfnamefont {W.}~\bibnamefont {Yao}},\ and\ \bibinfo {author} {\bibfnamefont {X.}~\bibnamefont {Xu}},\ }\href {https://doi.org/10.1038/s41586-024-08134-0} {\bibfield
  {journal} {\bibinfo  {journal} {Nature}\ }\textbf {\bibinfo {volume} {635}},\ \bibinfo {pages} {590} (\bibinfo {year} {2024})}\BibitemShut {NoStop}%
\bibitem [{\citenamefont {Zaklama}\ \emph {et~al.}(2024)\citenamefont {Zaklama}, \citenamefont {Luo},\ and\ \citenamefont {Fu}}]{Zaklama2024a}%
  \BibitemOpen
  \bibfield  {author} {\bibinfo {author} {\bibfnamefont {T.}~\bibnamefont {Zaklama}}, \bibinfo {author} {\bibfnamefont {D.}~\bibnamefont {Luo}},\ and\ \bibinfo {author} {\bibfnamefont {L.}~\bibnamefont {Fu}},\ }\href {https://doi.org/10.48550/arXiv.2411.03496} {\bibinfo {title} {Structure factor and topological bound of twisted bilayer semiconductors at fractional fillings}} (\bibinfo {year} {2024}),\ \Eprint {https://arxiv.org/abs/2411.03496} {arXiv:2411.03496 [cond-mat]} \BibitemShut {NoStop}%
\bibitem [{\citenamefont {Wu}\ \emph {et~al.}(2019{\natexlab{b}})\citenamefont {Wu}, \citenamefont {Lovorn}, \citenamefont {Tutuc}, \citenamefont {Martin},\ and\ \citenamefont {MacDonald}}]{Wu2019}%
  \BibitemOpen
  \bibfield  {author} {\bibinfo {author} {\bibfnamefont {F.}~\bibnamefont {Wu}}, \bibinfo {author} {\bibfnamefont {T.}~\bibnamefont {Lovorn}}, \bibinfo {author} {\bibfnamefont {E.}~\bibnamefont {Tutuc}}, \bibinfo {author} {\bibfnamefont {I.}~\bibnamefont {Martin}},\ and\ \bibinfo {author} {\bibfnamefont {A.~H.}\ \bibnamefont {MacDonald}},\ }\href {https://doi.org/10.1103/PhysRevLett.122.086402} {\bibfield  {journal} {\bibinfo  {journal} {Physical Review Letters}\ }\textbf {\bibinfo {volume} {122}},\ \bibinfo {pages} {086402} (\bibinfo {year} {2019}{\natexlab{b}})}\BibitemShut {NoStop}%
\bibitem [{\citenamefont {Jia}\ \emph {et~al.}(2023)\citenamefont {Jia}, \citenamefont {Yu}, \citenamefont {Liu}, \citenamefont {{Herzog-Arbeitman}}, \citenamefont {Qi}, \citenamefont {Regnault}, \citenamefont {Weng}, \citenamefont {Bernevig},\ and\ \citenamefont {Wu}}]{Jia2023}%
  \BibitemOpen
  \bibfield  {author} {\bibinfo {author} {\bibfnamefont {Y.}~\bibnamefont {Jia}}, \bibinfo {author} {\bibfnamefont {J.}~\bibnamefont {Yu}}, \bibinfo {author} {\bibfnamefont {J.}~\bibnamefont {Liu}}, \bibinfo {author} {\bibfnamefont {J.}~\bibnamefont {{Herzog-Arbeitman}}}, \bibinfo {author} {\bibfnamefont {Z.}~\bibnamefont {Qi}}, \bibinfo {author} {\bibfnamefont {N.}~\bibnamefont {Regnault}}, \bibinfo {author} {\bibfnamefont {H.}~\bibnamefont {Weng}}, \bibinfo {author} {\bibfnamefont {B.~A.}\ \bibnamefont {Bernevig}},\ and\ \bibinfo {author} {\bibfnamefont {Q.}~\bibnamefont {Wu}},\ }\href {https://doi.org/10.48550/arXiv.2311.04958} {\bibinfo {title} {Moir\'e {{Fractional Chern Insulators I}}: {{First-principles}} calculations and {{Continuum Models}} of {{Twisted Bilayer MoTe}}\$\_2\$}} (\bibinfo {year} {2023}),\ \Eprint {https://arxiv.org/abs/2311.04958} {arXiv:2311.04958 [cond-mat]} \BibitemShut {NoStop}%
\bibitem [{\citenamefont {Kwan}\ \emph {et~al.}(2026)\citenamefont {Kwan}, \citenamefont {Wagner}, \citenamefont {Yu}, \citenamefont {Dagnino}, \citenamefont {Jiang}, \citenamefont {Xu}, \citenamefont {Bernevig}, \citenamefont {Neupert},\ and\ \citenamefont {Regnault}}]{Kwan2026}%
  \BibitemOpen
  \bibfield  {author} {\bibinfo {author} {\bibfnamefont {Y.~H.}\ \bibnamefont {Kwan}}, \bibinfo {author} {\bibfnamefont {G.}~\bibnamefont {Wagner}}, \bibinfo {author} {\bibfnamefont {J.}~\bibnamefont {Yu}}, \bibinfo {author} {\bibfnamefont {A.~K.}\ \bibnamefont {Dagnino}}, \bibinfo {author} {\bibfnamefont {Y.}~\bibnamefont {Jiang}}, \bibinfo {author} {\bibfnamefont {X.}~\bibnamefont {Xu}}, \bibinfo {author} {\bibfnamefont {B.~A.}\ \bibnamefont {Bernevig}}, \bibinfo {author} {\bibfnamefont {T.}~\bibnamefont {Neupert}},\ and\ \bibinfo {author} {\bibfnamefont {N.}~\bibnamefont {Regnault}},\ }\href {https://doi.org/10.1038/s42005-025-02483-6} {\bibfield  {journal} {\bibinfo  {journal} {Communications Physics}\ }\textbf {\bibinfo {volume} {9}},\ \bibinfo {pages} {52} (\bibinfo {year} {2026})},\ \Eprint {https://arxiv.org/abs/2407.02560} {arXiv:2407.02560 [cond-mat]} \BibitemShut {NoStop}%
\bibitem [{\citenamefont {Yu}\ \emph {et~al.}(2025)\citenamefont {Yu}, \citenamefont {Herzog-Arbeitman}, \citenamefont {Kwan}, \citenamefont {Regnault},\ and\ \citenamefont {Bernevig}}]{Yu2025moire}%
  \BibitemOpen
  \bibfield  {author} {\bibinfo {author} {\bibfnamefont {J.}~\bibnamefont {Yu}}, \bibinfo {author} {\bibfnamefont {J.}~\bibnamefont {Herzog-Arbeitman}}, \bibinfo {author} {\bibfnamefont {Y.~H.}\ \bibnamefont {Kwan}}, \bibinfo {author} {\bibfnamefont {N.}~\bibnamefont {Regnault}},\ and\ \bibinfo {author} {\bibfnamefont {B.~A.}\ \bibnamefont {Bernevig}},\ }\href {https://doi.org/10.1103/PhysRevB.112.075110} {\bibfield  {journal} {\bibinfo  {journal} {Phys. Rev. B}\ }\textbf {\bibinfo {volume} {112}},\ \bibinfo {pages} {075110} (\bibinfo {year} {2025})}\BibitemShut {NoStop}%
\bibitem [{\citenamefont {Li}\ \emph {et~al.}(2025)\citenamefont {Li}, \citenamefont {Bernevig},\ and\ \citenamefont {Regnault}}]{Li2025multiband}%
  \BibitemOpen
  \bibfield  {author} {\bibinfo {author} {\bibfnamefont {H.}~\bibnamefont {Li}}, \bibinfo {author} {\bibfnamefont {B.~A.}\ \bibnamefont {Bernevig}},\ and\ \bibinfo {author} {\bibfnamefont {N.}~\bibnamefont {Regnault}},\ }\href {https://doi.org/10.1103/zlqg-sj86} {\bibfield  {journal} {\bibinfo  {journal} {Phys. Rev. B}\ }\textbf {\bibinfo {volume} {112}},\ \bibinfo {pages} {075130} (\bibinfo {year} {2025})}\BibitemShut {NoStop}%
\bibitem [{\citenamefont {Gon{\ifmmode \mbox{\c{c}}\else \c{c}\fi{}}alves}\ \emph {et~al.}(2026)\citenamefont {Gon{\ifmmode \mbox{\c{c}}\else \c{c}\fi{}}alves}, \citenamefont {Mendez-Valderrama}, \citenamefont {Herzog-Arbeitman}, \citenamefont {Yu}, \citenamefont {Xu}, \citenamefont {Xiao}, \citenamefont {Bernevig},\ and\ \citenamefont {Regnault}}]{Goncalves2026spinless}%
  \BibitemOpen
  \bibfield  {author} {\bibinfo {author} {\bibfnamefont {M.}~\bibnamefont {Gon{\ifmmode \mbox{\c{c}}\else \c{c}\fi{}}alves}}, \bibinfo {author} {\bibfnamefont {J.~F.}\ \bibnamefont {Mendez-Valderrama}}, \bibinfo {author} {\bibfnamefont {J.}~\bibnamefont {Herzog-Arbeitman}}, \bibinfo {author} {\bibfnamefont {J.}~\bibnamefont {Yu}}, \bibinfo {author} {\bibfnamefont {X.}~\bibnamefont {Xu}}, \bibinfo {author} {\bibfnamefont {D.}~\bibnamefont {Xiao}}, \bibinfo {author} {\bibfnamefont {B.~A.}\ \bibnamefont {Bernevig}},\ and\ \bibinfo {author} {\bibfnamefont {N.}~\bibnamefont {Regnault}},\ }\href {https://doi.org/10.1103/gmkp-bxf2} {\bibfield  {journal} {\bibinfo  {journal} {Phys. Rev. Lett.}\ }\textbf {\bibinfo {volume} {136}},\ \bibinfo {pages} {196503} (\bibinfo {year} {2026})}\BibitemShut {NoStop}%
\bibitem [{\citenamefont {Wang}\ \emph {et~al.}(2024)\citenamefont {Wang}, \citenamefont {Zhang}, \citenamefont {Liu}, \citenamefont {He}, \citenamefont {Xu}, \citenamefont {Ran}, \citenamefont {Cao},\ and\ \citenamefont {Xiao}}]{Wang2024fractional}%
  \BibitemOpen
  \bibfield  {author} {\bibinfo {author} {\bibfnamefont {C.}~\bibnamefont {Wang}}, \bibinfo {author} {\bibfnamefont {X.-W.}\ \bibnamefont {Zhang}}, \bibinfo {author} {\bibfnamefont {X.}~\bibnamefont {Liu}}, \bibinfo {author} {\bibfnamefont {Y.}~\bibnamefont {He}}, \bibinfo {author} {\bibfnamefont {X.}~\bibnamefont {Xu}}, \bibinfo {author} {\bibfnamefont {Y.}~\bibnamefont {Ran}}, \bibinfo {author} {\bibfnamefont {T.}~\bibnamefont {Cao}},\ and\ \bibinfo {author} {\bibfnamefont {D.}~\bibnamefont {Xiao}},\ }\href {https://doi.org/10.1103/PhysRevLett.132.036501} {\bibfield  {journal} {\bibinfo  {journal} {Phys. Rev. Lett.}\ }\textbf {\bibinfo {volume} {132}},\ \bibinfo {pages} {036501} (\bibinfo {year} {2024})}\BibitemShut {NoStop}%
\bibitem [{\citenamefont {Chen}\ \emph {et~al.}(2026{\natexlab{b}})\citenamefont {Chen}, \citenamefont {Ghorashi}, \citenamefont {Cano},\ and\ \citenamefont {Cr{\'e}pel}}]{Chen2026}%
  \BibitemOpen
  \bibfield  {author} {\bibinfo {author} {\bibfnamefont {L.}~\bibnamefont {Chen}}, \bibinfo {author} {\bibfnamefont {S.~A.~A.}\ \bibnamefont {Ghorashi}}, \bibinfo {author} {\bibfnamefont {J.}~\bibnamefont {Cano}},\ and\ \bibinfo {author} {\bibfnamefont {V.}~\bibnamefont {Cr{\'e}pel}},\ }\href {https://doi.org/10.1103/qbg3-3yxv} {\bibfield  {journal} {\bibinfo  {journal} {Phys. Rev. Lett.}\ }\textbf {\bibinfo {volume} {136}},\ \bibinfo {pages} {186602} (\bibinfo {year} {2026}{\natexlab{b}})}\BibitemShut {NoStop}%
\bibitem [{\citenamefont {Fremling}\ \emph {et~al.}(2018)\citenamefont {Fremling}, \citenamefont {Moran}, \citenamefont {Slingerland},\ and\ \citenamefont {Simon}}]{Fremling2018trial}%
  \BibitemOpen
  \bibfield  {author} {\bibinfo {author} {\bibfnamefont {M.}~\bibnamefont {Fremling}}, \bibinfo {author} {\bibfnamefont {N.}~\bibnamefont {Moran}}, \bibinfo {author} {\bibfnamefont {J.~K.}\ \bibnamefont {Slingerland}},\ and\ \bibinfo {author} {\bibfnamefont {S.~H.}\ \bibnamefont {Simon}},\ }\href {https://doi.org/10.1103/PhysRevB.97.035149} {\bibfield  {journal} {\bibinfo  {journal} {Phys. Rev. B}\ }\textbf {\bibinfo {volume} {97}},\ \bibinfo {pages} {035149} (\bibinfo {year} {2018})}\BibitemShut {NoStop}%
\bibitem [{\citenamefont {Dong}\ \emph {et~al.}(2023)\citenamefont {Dong}, \citenamefont {Wang}, \citenamefont {Ledwith}, \citenamefont {Vishwanath},\ and\ \citenamefont {Parker}}]{Dong2023}%
  \BibitemOpen
  \bibfield  {author} {\bibinfo {author} {\bibfnamefont {J.}~\bibnamefont {Dong}}, \bibinfo {author} {\bibfnamefont {J.}~\bibnamefont {Wang}}, \bibinfo {author} {\bibfnamefont {P.~J.}\ \bibnamefont {Ledwith}}, \bibinfo {author} {\bibfnamefont {A.}~\bibnamefont {Vishwanath}},\ and\ \bibinfo {author} {\bibfnamefont {D.~E.}\ \bibnamefont {Parker}},\ }\href {https://doi.org/10.1103/PhysRevLett.131.136502} {\bibfield  {journal} {\bibinfo  {journal} {Phys. Rev. Lett.}\ }\textbf {\bibinfo {volume} {131}},\ \bibinfo {pages} {136502} (\bibinfo {year} {2023})}\BibitemShut {NoStop}%
\bibitem [{\citenamefont {Goldman}\ \emph {et~al.}(2023)\citenamefont {Goldman}, \citenamefont {Reddy}, \citenamefont {Paul},\ and\ \citenamefont {Fu}}]{Goldman2023}%
  \BibitemOpen
  \bibfield  {author} {\bibinfo {author} {\bibfnamefont {H.}~\bibnamefont {Goldman}}, \bibinfo {author} {\bibfnamefont {A.~P.}\ \bibnamefont {Reddy}}, \bibinfo {author} {\bibfnamefont {N.}~\bibnamefont {Paul}},\ and\ \bibinfo {author} {\bibfnamefont {L.}~\bibnamefont {Fu}},\ }\href {https://doi.org/10.1103/PhysRevLett.131.136501} {\bibfield  {journal} {\bibinfo  {journal} {Phys. Rev. Lett.}\ }\textbf {\bibinfo {volume} {131}},\ \bibinfo {pages} {136501} (\bibinfo {year} {2023})}\BibitemShut {NoStop}%
\bibitem [{\citenamefont {Morales-Durán}\ \emph {et~al.}(2026)\citenamefont {Morales-Durán}, \citenamefont {Shi}, \citenamefont {Voinea}, \citenamefont {Potasz},\ and\ \citenamefont {Cano}}]{moralesdurán2026bandmixingparticleholeasymmetry}%
  \BibitemOpen
  \bibfield  {author} {\bibinfo {author} {\bibfnamefont {N.}~\bibnamefont {Morales-Durán}}, \bibinfo {author} {\bibfnamefont {J.}~\bibnamefont {Shi}}, \bibinfo {author} {\bibfnamefont {C.}~\bibnamefont {Voinea}}, \bibinfo {author} {\bibfnamefont {P.}~\bibnamefont {Potasz}},\ and\ \bibinfo {author} {\bibfnamefont {J.}~\bibnamefont {Cano}},\ }\href {https://arxiv.org/abs/2604.16847} {\bibinfo {title} {Band mixing and particle-hole asymmetry in moir\'e fractional chern insulators}} (\bibinfo {year} {2026}),\ \Eprint {https://arxiv.org/abs/2604.16847} {arXiv:2604.16847 [cond-mat.str-el]} \BibitemShut {NoStop}%
\bibitem [{\citenamefont {Abouelkomsan}\ \emph {et~al.}(2024)\citenamefont {Abouelkomsan}, \citenamefont {Reddy}, \citenamefont {Fu},\ and\ \citenamefont {Bergholtz}}]{Abouelkomsan2024band}%
  \BibitemOpen
  \bibfield  {author} {\bibinfo {author} {\bibfnamefont {A.}~\bibnamefont {Abouelkomsan}}, \bibinfo {author} {\bibfnamefont {A.~P.}\ \bibnamefont {Reddy}}, \bibinfo {author} {\bibfnamefont {L.}~\bibnamefont {Fu}},\ and\ \bibinfo {author} {\bibfnamefont {E.~J.}\ \bibnamefont {Bergholtz}},\ }\href {https://doi.org/10.1103/PhysRevB.109.L121107} {\bibfield  {journal} {\bibinfo  {journal} {Phys. Rev. B}\ }\textbf {\bibinfo {volume} {109}},\ \bibinfo {pages} {L121107} (\bibinfo {year} {2024})}\BibitemShut {NoStop}%
\bibitem [{\citenamefont {Li}\ \emph {et~al.}(2026{\natexlab{b}})\citenamefont {Li}, \citenamefont {Yu}, \citenamefont {Xu}, \citenamefont {Bernevig},\ and\ \citenamefont {Regnault}}]{Li2026large}%
  \BibitemOpen
  \bibfield  {author} {\bibinfo {author} {\bibfnamefont {H.}~\bibnamefont {Li}}, \bibinfo {author} {\bibfnamefont {J.}~\bibnamefont {Yu}}, \bibinfo {author} {\bibfnamefont {X.}~\bibnamefont {Xu}}, \bibinfo {author} {\bibfnamefont {B.~A.}\ \bibnamefont {Bernevig}},\ and\ \bibinfo {author} {\bibfnamefont {N.}~\bibnamefont {Regnault}},\ }\href {https://doi.org/10.48550/arXiv.2608.23675} {\bibinfo {title} {Large scale theoretical investigation of the phase diagram of twisted bilayer {{MoTe$_2$}} at fractional fillings: Agreements and contradictions with current experiments}} (\bibinfo {year} {2026}{\natexlab{b}}),\ \Eprint {https://arxiv.org/abs/2608.23675} {arXiv:2608.23675 [cond-mat]} \BibitemShut {NoStop}%
\bibitem [{\citenamefont {Hou}\ and\ \citenamefont {Nevidomskyy}(2025)}]{Hou2025}%
  \BibitemOpen
  \bibfield  {author} {\bibinfo {author} {\bibfnamefont {R.}~\bibnamefont {Hou}}\ and\ \bibinfo {author} {\bibfnamefont {A.~H.}\ \bibnamefont {Nevidomskyy}},\ }\href {https://doi.org/10.48550/arXiv.2511.16641} {\bibinfo {title} {Stabilizing fractional {{Chern}} states in twisted {{MoTe$_2$}}: Multi-band correlations via non-perturbative renormalization group}} (\bibinfo {year} {2025}),\ \Eprint {https://arxiv.org/abs/2511.16641} {arXiv:2511.16641 [cond-mat]} \BibitemShut {NoStop}%
\bibitem [{\citenamefont {Xu}\ \emph {et~al.}(2024)\citenamefont {Xu}, \citenamefont {Li}, \citenamefont {Xu}, \citenamefont {Bi},\ and\ \citenamefont {Zhang}}]{Xu2024}%
  \BibitemOpen
  \bibfield  {author} {\bibinfo {author} {\bibfnamefont {C.}~\bibnamefont {Xu}}, \bibinfo {author} {\bibfnamefont {J.}~\bibnamefont {Li}}, \bibinfo {author} {\bibfnamefont {Y.}~\bibnamefont {Xu}}, \bibinfo {author} {\bibfnamefont {Z.}~\bibnamefont {Bi}},\ and\ \bibinfo {author} {\bibfnamefont {Y.}~\bibnamefont {Zhang}},\ }\href {https://doi.org/10.1073/pnas.2316749121} {\bibfield  {journal} {\bibinfo  {journal} {Proceedings of the National Academy of Sciences}\ }\textbf {\bibinfo {volume} {121}},\ \bibinfo {pages} {e2316749121} (\bibinfo {year} {2024})}\BibitemShut {NoStop}%
\bibitem [{\citenamefont {Regnault}(2001)}]{Regnault2001DiagHam}%
  \BibitemOpen
  \bibfield  {author} {\bibinfo {author} {\bibfnamefont {N.}~\bibnamefont {Regnault}},\ }\href@noop {} {\bibinfo {title} {{DiagHam}}},\ \bibinfo {howpublished} {\url{http://www.nick-ux.org/diagham}} (\bibinfo {year} {2001}),\ \bibinfo {note} {code repository}\BibitemShut {NoStop}%
\bibitem [{\citenamefont {L{\"a}uchli}\ \emph {et~al.}(2013)\citenamefont {L{\"a}uchli}, \citenamefont {Liu}, \citenamefont {Bergholtz},\ and\ \citenamefont {Moessner}}]{Lauchli2013}%
  \BibitemOpen
  \bibfield  {author} {\bibinfo {author} {\bibfnamefont {A.~M.}\ \bibnamefont {L{\"a}uchli}}, \bibinfo {author} {\bibfnamefont {Z.}~\bibnamefont {Liu}}, \bibinfo {author} {\bibfnamefont {E.~J.}\ \bibnamefont {Bergholtz}},\ and\ \bibinfo {author} {\bibfnamefont {R.}~\bibnamefont {Moessner}},\ }\href {https://doi.org/10.1103/PhysRevLett.111.126802} {\bibfield  {journal} {\bibinfo  {journal} {Physical Review Letters}\ }\textbf {\bibinfo {volume} {111}},\ \bibinfo {pages} {126802} (\bibinfo {year} {2013})}\BibitemShut {NoStop}%
\bibitem [{\citenamefont {Repellin}\ \emph {et~al.}(2014)\citenamefont {Repellin}, \citenamefont {Bernevig},\ and\ \citenamefont {Regnault}}]{Repellin2014}%
  \BibitemOpen
  \bibfield  {author} {\bibinfo {author} {\bibfnamefont {C.}~\bibnamefont {Repellin}}, \bibinfo {author} {\bibfnamefont {B.~A.}\ \bibnamefont {Bernevig}},\ and\ \bibinfo {author} {\bibfnamefont {N.}~\bibnamefont {Regnault}},\ }\href {https://doi.org/10.1103/PhysRevB.90.245401} {\bibfield  {journal} {\bibinfo  {journal} {Physical Review B}\ }\textbf {\bibinfo {volume} {90}},\ \bibinfo {pages} {245401} (\bibinfo {year} {2014})}\BibitemShut {NoStop}%
\end{thebibliography}%

\clearpage
\newpage

\begin{appendix}
\onecolumngrid
	\begin{center}
		\textbf{\large --- Supplementary Material ---\\Fractional Chern insulators and composite Fermi liquids vs. charge order in twisted MoTe$_2$}\\
		\medskip
		\text{Enrique Aguilar-Méndez and Glenn Wagner}
	\end{center}
	
		\setcounter{equation}{0}
	\setcounter{figure}{0}
	\setcounter{table}{0}
	\makeatletter
    \@removefromreset{equation}{section}
    \makeatother
	\renewcommand{\theequation}{S\arabic{equation}}
	\renewcommand{\thefigure}{S\arabic{figure}}
    \renewcommand{\thetable}{S\arabic{table}}
	\renewcommand{\bibnumfmt}[1]{[S#1]}

\section{Single particle Hamiltonian}
\label{app:hamiltonian}

 The interlayer hopping of the single particle Hamiltonian is given by:
\begin{equation}
t_{\eta}(\bm{r})=w\sum_{i=1,2,3}e^{-i\eta\bm{q}_i\cdot\bm{r}}. 
\end{equation}

 The intralayer terms are given by:
\begin{equation}
h_{\eta,l}(\bm{r})=\frac{\hbar^2\nabla^2}{2m^*}+V_{\eta,l}(\bm{r})+(-1)^l\frac{D}{2}.     
\end{equation}
 
 Here $D$ is a displacement field perpendicular to the MoTe$_2$ layers, $m^*$ 
 is the effective mass of monolayer MoTe$_2$ set to $0.6$ times the electron mass as in \cite{Jia2023} and $V_{\eta,l}(\bm{r})$ is the moir\'e potential given by: 

\begin{equation}
V_{\eta,l}(\bm{r})=Ve^{-(-1)^li\psi}\sum_{i=1,2,3}e^{i\bm{g}_i\cdot \bm{r}}+Ve^{(-1)^li\psi}\sum_{i=1,2,3}e^{-i\bm{g}_i\cdot \bm{r}} ,
\label{potential}
\end{equation}

where $\bm{g}_i= C_3^{i-1}\bm{b}_1$, $\bm{b}_1=\bm{q}_3-\bm{q}_2$, $\bm{q}_3=C_3^2\bm{q}_1$,$\bm{q}_2=C_3 \bm{q}_1$, $\bm{q}_1=K_b-K_t$. $C_3$ is a counter clock-wise rotation matrix by $2\pi/3$ and $K_b$, $K_t$ are the positions of the $K$ points in momentum space for the bottom and top layer. $K_b$ and  $K_t$ differ due to the rotation angle  $\theta$ between the layers. Let us call $a_0=0.352$ nm the lattice constant of MoTe$_2$ . If we start from an aligned AA stacking and rotate the top (bottom) layer by $\theta/2$ ($-\theta/2$) the positions of the K points in momentum space are:

\begin{equation}
    K_{b/t}=\frac{4 \pi}{3a_0} \begin{pmatrix}\cos(\theta/2) \\ \pm \sin(\theta/2). \end{pmatrix}
\end{equation}

\section{$SU(2)$ symmetric limit of tMoTe$_2$}
\label{app:SU2}

Time reversal relates the two valley Hamiltonians via
$H_{-\eta}(\bm r) = H_{\eta}^*(\bm r)$, so an exact valley symmetry
would require $H_\eta$ to be independent of $\eta$, which in turn
requires $H_\eta$ to be real. Inspecting the Hamiltonian term by term,
the kinetic energy, the displacement field $D$, and the intralayer
moir\'e potential are all real and carry no
explicit $\eta$ dependence. The interlayer tunneling
$t_\eta(\bm r)=w\sum_i e^{-i\eta\bm q_i\cdot\bm r}$ is the only term
that explicitly breaks valley symmetry, since
$t_{-\eta}(\bm r)=t_\eta^*(\bm r)\neq t_\eta(\bm r)$.

At large $D$ the two layers are pushed apart in energy and the
low-energy physics is governed by the lower layer (say layer $t$),
with layer $b$ pushed up by $D$. Valley-symmetry breaking in the
effective single-layer theory then enters only through virtual
interlayer tunneling, which can be treated via a Schrieffer--Wolff
expansion in $w/D$.

For $D$ large compared to the moir\'e kinetic bandwidth
$W\sim\hbar^2 q^2/m^*$, the energy denominator will be $D$. The second order contribution will then be $|t_\eta(\bm r)|^2/D$. However, $|t_\eta(\bm r)|^2=|t_{-\eta}(\bm r)|^2$, hence this
leading correction is \emph{valley-independent}: it renormalizes the
single-layer moir\'e potential identically for $\eta=\pm1$.

We therefore need to go to next higher order and include the order $W/D$ corrections to the energy denominator. This generates a correction
\begin{equation}
\delta H^{\rm val. break} \sim \frac{w^2}{D}\cdot\frac{W}{D}
\propto \frac{1}{D^2}.
\end{equation}

\section{Single-particle basis and band projection}
\label{app:band_projection}
We discretize the continuum model of twisted MoTe$_2$ on an $N_x\times N_y$ momentum mesh. Diagonalizing it yields a set of eigenvalues and eigenvectors $ \epsilon_{\mathbf{k},\sigma}^{(m)}, \quad u_{\mathbf{k},\sigma,\mathbf{G},f}^{(m)},$ 
where $m$ labels the band ($m=0$ is the topmost valence band), $\sigma$ is the valley/spin index, $\mathbf{G}$ is a reciprocal lattice vector (with an appropriate cutoff) and $f$ is the layer index. We work in the hole basis, in which the vacuum is the state with the valence bands completely filled with electrons. Accordingly, the single-particle term enters the many-body Hamiltonian with $\epsilon_{\mathbf{k},\sigma}^{(m)}\to-\epsilon_{\mathbf{k},\sigma}^{(m)}$ and the eigenvectors are complex conjugated, $u_{\mathbf{k},\sigma,\mathbf{G},f}^{(m)}\to u_{\mathbf{k},\sigma,\mathbf{G},f}^{(m)*}$. Denoting the conjugated eigenvectors by $u$ from now on, the band and plane-wave hole creation operators are related by

\begin{equation}
    c_{\mathbf{k}+\mathbf{G},\sigma,f}^\dagger = \sum_{m}u_{\mathbf{k},\sigma,\mathbf{G},f}^{(m)*}\, d_{\mathbf{k},\sigma,m}^\dagger .
\end{equation}

In terms of the plane-wave operators, the hole density at momentum $\mathbf q+\mathbf Q$, with $\mathbf q$ in the first Brillouin zone and $\mathbf Q$ a reciprocal lattice vector, is:

\begin{equation}
  \rho(\mathbf q+\mathbf Q)=\sum_{\mathbf k,\mathbf G,\sigma,f} c^\dagger_{\mathbf k+\mathbf G+\mathbf q+\mathbf Q,\sigma,f}\,c_{\mathbf k+\mathbf G,\sigma,f}  
\end{equation}

For a crystal momentum $\mathbf k$ we define $\lfloor \mathbf k \rfloor$ as its reduction to the first Brillouin zone and $\lceil \mathbf k \rceil = \mathbf k - \lfloor \mathbf k \rfloor$ as the reciprocal lattice vector required to perform this reduction. Keeping the bands $m\in\{0,1\}$, the density becomes
\begin{equation}
    \rho(\mathbf q+\mathbf Q)=\sum_{\mathbf k,\sigma,m,m'}
    F^{\sigma}_{mm'}(\mathbf k,\mathbf q,\mathbf Q)\,
    d^\dagger_{\lfloor\mathbf k+\mathbf q\rfloor,\sigma,m}\,d_{\mathbf k,\sigma,m'},
    \label{eq:rho_q}
\end{equation}
with the form factors
\begin{equation}
    F^{\sigma}_{mm'}(\mathbf k,\mathbf q,\mathbf Q)=\sum_{\mathbf G,f}
    u^{(m)*}_{\lfloor\mathbf k+\mathbf q\rfloor,\sigma,
    \mathbf G+\mathbf Q+\lceil\mathbf k+\mathbf q\rceil,f}\,
    u^{(m')}_{\mathbf k,\sigma,\mathbf G,f}.
    \label{eq:form_factor}
\end{equation}

\section{Hartree-Fock renormalization}
\label{app:hartree-fock}

Using the operators $d_{\mathbf{k},\sigma,m}$ of Eq.~\ref{eqq:full_hamiltonain}, the interaction term is:
\begin{equation}
    H^\textrm{int}=\frac{1}{2}\sum_{r,s,r',s'}V_{r,s,r',s'}\,d^{\dagger}_r d^{\dagger}_s d_{s'} d_{r'},
    \label{eq:interacting_hamiltonian}
\end{equation}
where $r=(\mathbf k_r,\sigma_r,m_r)$, and likewise $s$, $r'$, $s'$, combine momentum, spin and band. $V_{r,s,r',s'}$, 
contains the area $A$ of the system, the potential of Eq.~\ref{eq:interaction} and the form factors. Using Wick's theorem on $H^\textrm{int}$ gives the Hartree-Fock Hamiltonian
\begin{equation}
    H^\textrm{HF}_{s,s'}=\epsilon_s\,\delta_{s,s'}+\sum_{r,r'}\big(V_{r,s,r',s'}-V_{r,s,s',r'}\big)P_{r,r'},
    \label{eq:hami_hartree-fock}
\end{equation}
with $\epsilon_s$ the bare band energy and $P_{r,r'}=\langle d^\dagger_r d_{r'}\rangle$ the one-body density matrix of the occupied state. In our one shot calculation $P$ is fixed to a fully polarized state in which the holes occupy band 0 of spin $\uparrow$ uniformly, with probability $\nu$ at every momentum:
\begin{equation}
    P_{r,r'}=\nu\,\delta_{\mathbf k_r,\mathbf k_{r'}}\,\delta_{\sigma_r,\uparrow}\,\delta_{\sigma_{r'},\uparrow}\,
    \delta_{m_r,0}\,\delta_{m_{r'},0}.
    \label{eqq:projector}
\end{equation}
Inserting Eq.~\ref{eqq:projector} in Eq.~\ref{eq:hami_hartree-fock}, the sums over $r,r'$ collapse to $r=r'=(\mathbf k',\uparrow,0)$, and momentum and spin conservation make $H^\textrm{HF}$ diagonal in momentum and spin. For the occupied spin and considering only band 0, with $s=(\mathbf k,\uparrow,0)$ and $s'=(\mathbf k,\uparrow,0)$, and omitting the spin and band labels,
\begin{equation}
    H^\textrm{HF}(\mathbf k)=\epsilon_{\mathbf k}\,
    +\nu\sum_{\mathbf k'}\Big(V_{\mathbf k',\mathbf k,\mathbf k',\mathbf k}
    -V_{\mathbf k',\mathbf k,\mathbf k,\mathbf k'}\Big).
    \label{eq:hami_hartree-fock_projected}
\end{equation}
The Hartree-Fock bandwidth is $W_\textrm{HF}=\max_{\mathbf k}H^\textrm{HF}(\mathbf k)-\min_{\mathbf k}H^\textrm{HF}(\mathbf k)$.

\section{charge density calculation}
\label{app:pinned_density}

Exact diagonalization of the interacting Hamiltonian yields a many-body spectrum
$ E_{\mathbf{k},S}^{(n)}, \quad |\psi_{\mathbf{k},S}^{(n)}\rangle,$
where $\mathbf{k}$ is the total crystal momentum, $S$ labels the spin sector and $n$ labels the eigenstates within a given $(\mathbf{k},S)$ sector. We fix $S$ from the outset, order the states of that sector by energy, $ E_{\mathbf{p}_1,S}^{(n_1)}\le E_{\mathbf{p}_2,S}^{(n_2)}\le\dots, $ and keep the $N$ lowest of them, $\mathcal{B}=\left\{(\mathbf{p}_1,n_1) \dots,(\mathbf{p}_N,n_N)\right\},$
allowing several level indices $n$ to occur with the same momentum $\mathbf{p}$. The number $N$ is chosen so that the cut falls in a gap of the chosen sector. Note that the retained states are the lowest ones of that sector, not necessarily of the full spectrum.

The real-space creation operator for a hole is
\begin{equation}
    \psi^\dagger_{f,\sigma}(\mathbf{r})=\frac{1}{\sqrt{A}}\sum_{\mathbf{k},\mathbf{G}}e^{i(\mathbf{k}+\mathbf{G})\cdot\mathbf{r}}c_{\mathbf{k}+\mathbf{G},\sigma,f}^\dagger,
\end{equation}
where $A$ is the area of the system, so that the hole density operator in layer $f$ is
\begin{equation}
    \rho_f(\mathbf{r})=\sum_{\sigma}\psi_{f,\sigma}^\dagger(\mathbf{r})\psi_{f,\sigma}(\mathbf{r})
    =\frac1A\sum_{\mathbf{k},\mathbf{G},\mathbf{k}',\mathbf{G}',\sigma,m,m'}e^{i(\mathbf{k}+\mathbf{G})\cdot \mathbf{r}}e^{-i(\mathbf{k}'+\mathbf{G}')\cdot \mathbf{r}}\, u^{(m)*}_{\mathbf{k},\sigma,\mathbf{G},f}u^{(m')}_{\mathbf{k}',\sigma,\mathbf{G}',f}d_{\mathbf{k},\sigma,m}^\dagger d_{\mathbf{k}',\sigma,m'}.
    \label{eq:rho_r}
\end{equation}
and the total density is $\rho(\mathbf{r})=\sum_f\rho_f(\mathbf{r})$. An impurity potential $U(\mathbf{r})$ that acts on layer $f_0$  couples to this density,

\begin{align}
    H_{\mathrm{imp}}&=\int \mathrm{d}^2{\mathbf{r}}\  U(\mathbf{r})\rho_{f_0}(\mathbf{r})
    \\
    &=\sum_{\mathbf{k},\mathbf{G},\mathbf{k}',\mathbf{G}',\sigma,m,m'}
    \tilde U(\mathbf{k}-\mathbf{k}',\mathbf{G}-\mathbf{G}')\,
    u^{(m)*}_{\mathbf{k},\sigma,\mathbf{G},f_0}u^{(m')}_{\mathbf{k}',\sigma,\mathbf{G}',f_0}\,
    d_{\mathbf{k},\sigma,m}^\dagger d_{\mathbf{k}',\sigma,m'},
    \label{eq:Himp} 
\end{align}
where 

\begin{equation}
    \tilde U(\mathbf{k}-\mathbf{k}',\mathbf{G}-\mathbf{G}')=\frac1A\int \mathrm{d}^2{\mathbf{r}}\ e^{i(\mathbf{k}+\mathbf{G})\cdot \mathbf{r}}e^{-i(\mathbf{k}'+\mathbf{G}')\cdot \mathbf{r}}U(\mathbf{r}).
\end{equation}

For a delta function impurity at position $\mathbf{w}$ on layer $f_0$, $U(\mathbf{r})=U_0A\,\delta(\mathbf{r}-\mathbf{w})$, and hence $\tilde U(\mathbf{k}-\mathbf{k}',\mathbf{G}-\mathbf{G}')=U_0\,e^{i(\mathbf{k}-\mathbf{k}'+\mathbf{G}-\mathbf{G}')\cdot\mathbf{w}}$, where $U_0$ has units of energy. Since $d_{\mathbf{k},\sigma,m}^\dagger d_{\mathbf{k}',\sigma,m'}$ changes the total momentum by $\mathbf{k}-\mathbf{k}'$, its only nonvanishing matrix elements between many-body eigenstates are
\begin{equation}
    S^{(nn')}(\mathbf{p},\mathbf{k},\mathbf{k}',\sigma,m,m')=\langle \psi_{\mathbf{p},S}^{(n)}|d_{\mathbf{k},\sigma,m}^\dagger d_{\mathbf{k}',\sigma,m'} |\psi_{\lfloor \mathbf{p}+\mathbf{k}'-\mathbf{k}\rfloor,S}^{(n')}\rangle.
    \label{eq:Smatrix}
\end{equation}
Projected onto $\mathcal{B}$, the total Hamiltonian (interacting Hamiltonian plus $H_{\mathrm{imp}}$) has the matrix elements
\begin{align}
\hspace{-1.1cm}
&\langle \psi_{\mathbf{p},S}^{(n)}|H_{\mathrm{tot}}|\psi_{\mathbf{p}',S}^{(n')}\rangle=E_{\mathbf{p},S}^{(n)} \delta_{\mathbf{p}\mathbf{p}'}\delta_{nn'}\nonumber\\
&+\sum_{\mathbf{k},\mathbf{G},\mathbf{k}',\mathbf{G}',\sigma,m,m'}
\tilde U(\mathbf{k}-\mathbf{k}',\mathbf{G}-\mathbf{G}')\,
u^{(m)*}_{\mathbf{k},\sigma,\mathbf{G},f_0}u^{(m')}_{\mathbf{k}',\sigma,\mathbf{G}',f_0}\,
S^{(nn')}(\mathbf{p},\mathbf{k},\mathbf{k}',\sigma,m,m')\,
\delta_{\lfloor \mathbf{p}+\mathbf{k}'-\mathbf{k}\rfloor,\mathbf{p}'},
\label{eq:Htot}
\end{align}
with $(\mathbf{p},n)$ and $(\mathbf{p}',n')$ both running over $\mathcal{B}$. Diagonalizing this 
matrix yields the ground state
\begin{equation}
    |\mathrm{GS}\rangle=\sum_{(\mathbf{k},n)\in\mathcal{B}} \alpha_{\mathbf{k}}^{(n)}|\psi_{\mathbf{k},S}^{(n)}\rangle,
\end{equation}
whose density follows from Eq.~\eqref{eq:rho_r} as
$\langle \mathrm{GS}|\rho(\mathbf{r})|\mathrm{GS}\rangle
=\sum_f\langle \mathrm{GS}|\rho_f(\mathbf{r})|\mathrm{GS}\rangle$.

\newpage
\section{Additional exact diagonalization results}

\begin{figure}[H]
\centering
\includegraphics[width=\textwidth]{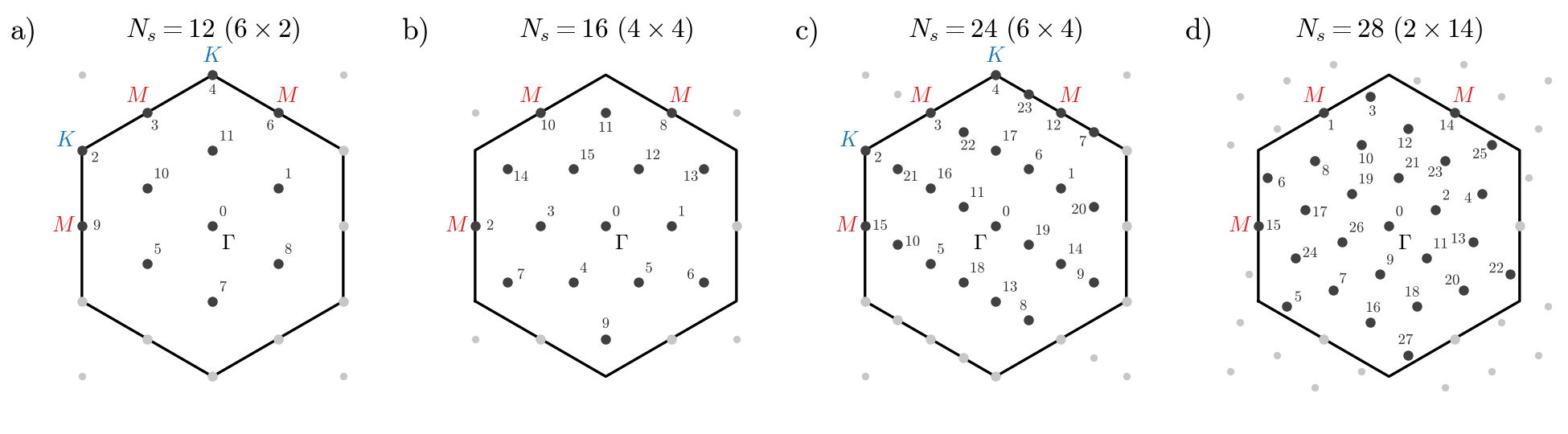}
\caption{\textbf{The clusters of the exact diagonalization.} \textbf{a)}-\textbf{d)} clusters with 12, 16, 24 and 28 sites, respectively.  The hexagon is the first moir\'e Brillouin zone and the dark dots are the  $N_s$  momentum sectors on the horizontal axis of the spectra. $N_x\times N_y$ is the shape of the grid before applying titled boundary conditions \cite{Lauchli2013,Repellin2014}. The gray dots are images of the cluster momenta under reciprocal lattice vectors. The three nonequivalent $M$ points lie on every mesh, the $K$ points only on $N_s$ = 12 and 24.  The $N_s$ = 24 mesh is not invariant under $C_3$ rotations; the other three are.}
\label{fig:clusters}
\end{figure}

All the exact diagonalization results shown here were carried out for holes in the highest moir\'e valence band. Continuum parameters were taken as $w=-18.8$ meV, $V=16.5$ meV, $\psi=-105.9^\circ$, $m^*=0.6\,m_e$ \cite{Jia2023}. The gate-screened Coulomb interaction was fixed at strength $U=10/\varepsilon=2$ and screening length $\xi$. $\theta$ is the twist angle, $D$ the interlayer potential, and $N_\mathrm{max}^1$ the maximum number of holes in the second moir\'e band. The clusters used in the calculations throughout this paper are shown in Fig.~\ref{fig:clusters}.

\subsection{Phase diagram maps, ED spectra and density profiles}

\begin{figure}[H]
    \centering
    \includegraphics[width=0.8\columnwidth]{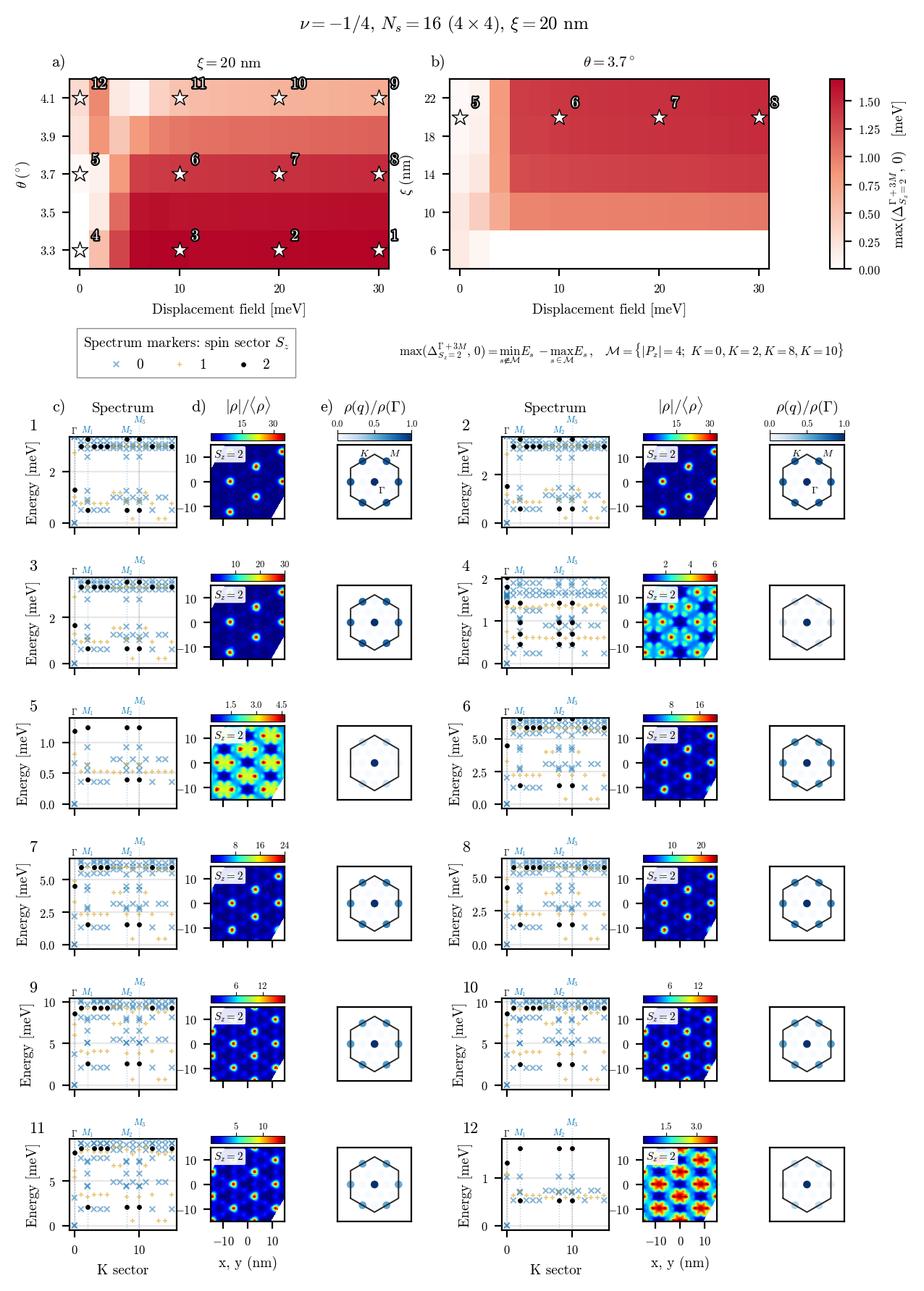}

    \caption{$\bm{\nu=-1/4}$ on $N_s=16$ ($4\times4$) , $N_\mathrm{max}^1=2$, $\xi=20$~nm. \textbf{a)} $\max(\Delta^{\Gamma+3M}_{S_z=2},0)$ of the fully polarized four-state manifold at $\{\Gamma,M_1,M_2,M_3\}$ against $D$ and $\theta$ at $\xi=20$~nm. \textbf{b)} same as a) but plotted against $D$ and $\xi$ at $\theta=3.7^\circ$. \textbf{Blocks 1-12}: data points at $\theta=3.3,3.7,4.1^\circ$ at $D=0,10,20,30$~meV. In each block the panels show: \textbf{c)} the spectrum, \textbf{d)} the charge density and \textbf{e)} its Fourier transform. Panels  d) and e) are built from that same $S_z=2$ four-state manifold, a 2$\times$2 superlattice is visible in the density wherever the gap is open.}
    \label{fig:one_quarter_full_page}
\end{figure}

\begin{figure}[H]
    \centering
    \includegraphics[width=0.8\columnwidth]{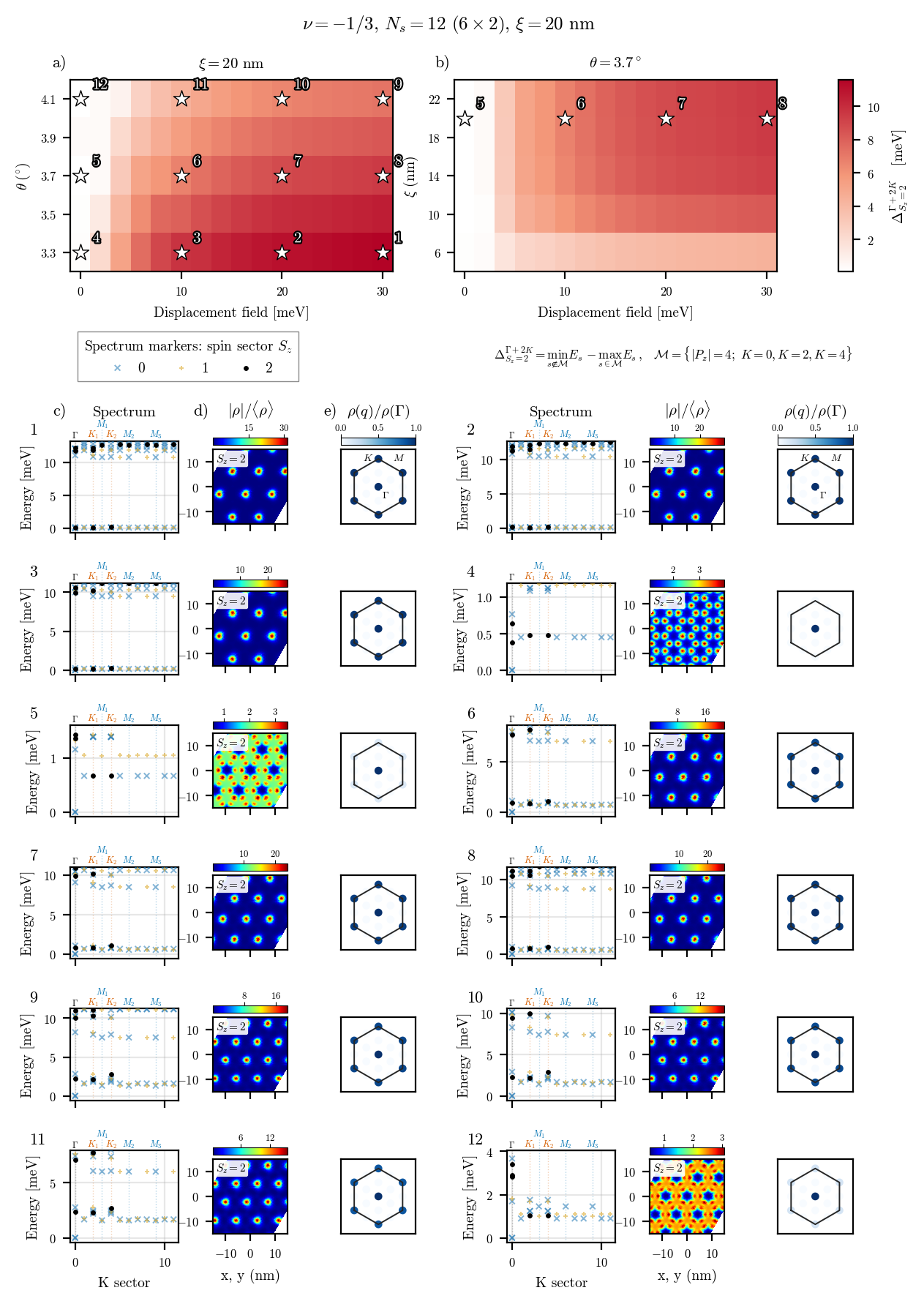}

    \caption{$\bm{\nu=-1/3}$ on $N_s=12$ ($6\times2$), $N_\mathrm{max}^1=2$, $\xi=20$~nm. \textbf{a)} $\Delta^{\Gamma+2K}_{S_z=2}$ of the fully polarized three-state manifold at $\{\Gamma,K_1,K_2\}$ against $D$ and $\theta$ at $\xi=20$~nm. \textbf{b)} same as a) but plotted against $D$ and $\xi$ at $\theta=3.7^\circ$. \textbf{Blocks 1-12}: data points at $\theta=3.3,3.7,4.1^\circ$ at $D=0,10,20,30$~meV. In each block the panels show: \textbf{c)} the spectrum, \textbf{d)} the charge density and \textbf{e)} its Fourier transform. Panels  d) and e) are built from that same $S_z=2$ three-state manifold, a $\sqrt3\times\sqrt3$ superlattice is visible in the density wherever the gap is open.}
    \label{fig:one_third_full_page}
\end{figure}

\begin{figure}[H]
    \centering
    \includegraphics[width=0.8\columnwidth]{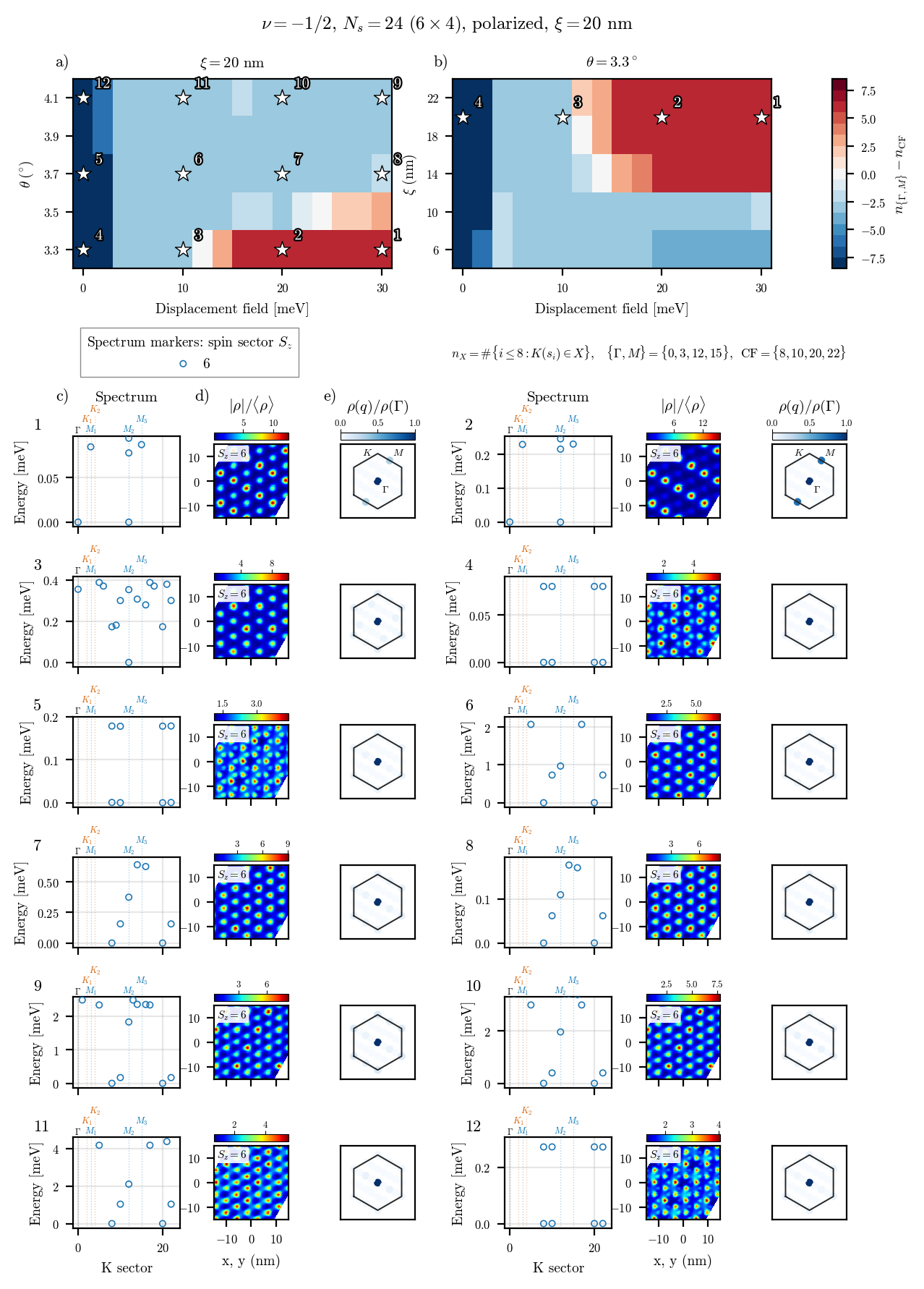}

     \caption{$\bm{\nu=-1/2}$ on $N_s=24$ ($6\times4$), \textbf{fully polarized} ($S_z=6$), $N_\mathrm{max}^1=0$, $\xi=20$~nm. \textbf{a)} Order score: among the $N=8$ lowest states, the number whose momenta lie in the manifold $\{\Gamma,M_1,M_2,M_3\}$ minus the number on the composite-Fermi-liquid momenta. Plotted against $D$ and $\theta$ at $\xi=20$~nm. \textbf{b)} same as a) but plotted against  against $D$ and $\xi$ at $\theta=3.3^\circ$. One colour block per integer: $-8$ means every one of the eight lowest states is liquid-like, $+6$ that the six-state crystal manifold is complete. \textbf{Blocks 1-12}: data points at $\theta=3.3,3.7,4.1^\circ$ at $D=0,10,20,30$~meV. In each block the panels show: \textbf{c)} the spectrum, \textbf{d)} the charge density and \textbf{e)} its Fourier transform. The crystal region corresponds to a single-$Q$ stripe with density maxima on alternate rows of moir\'e sites.}
    \label{fig:one_half_full_page}
\end{figure}

\begin{figure}[H]
    \centering
    \includegraphics[width=0.8\columnwidth]{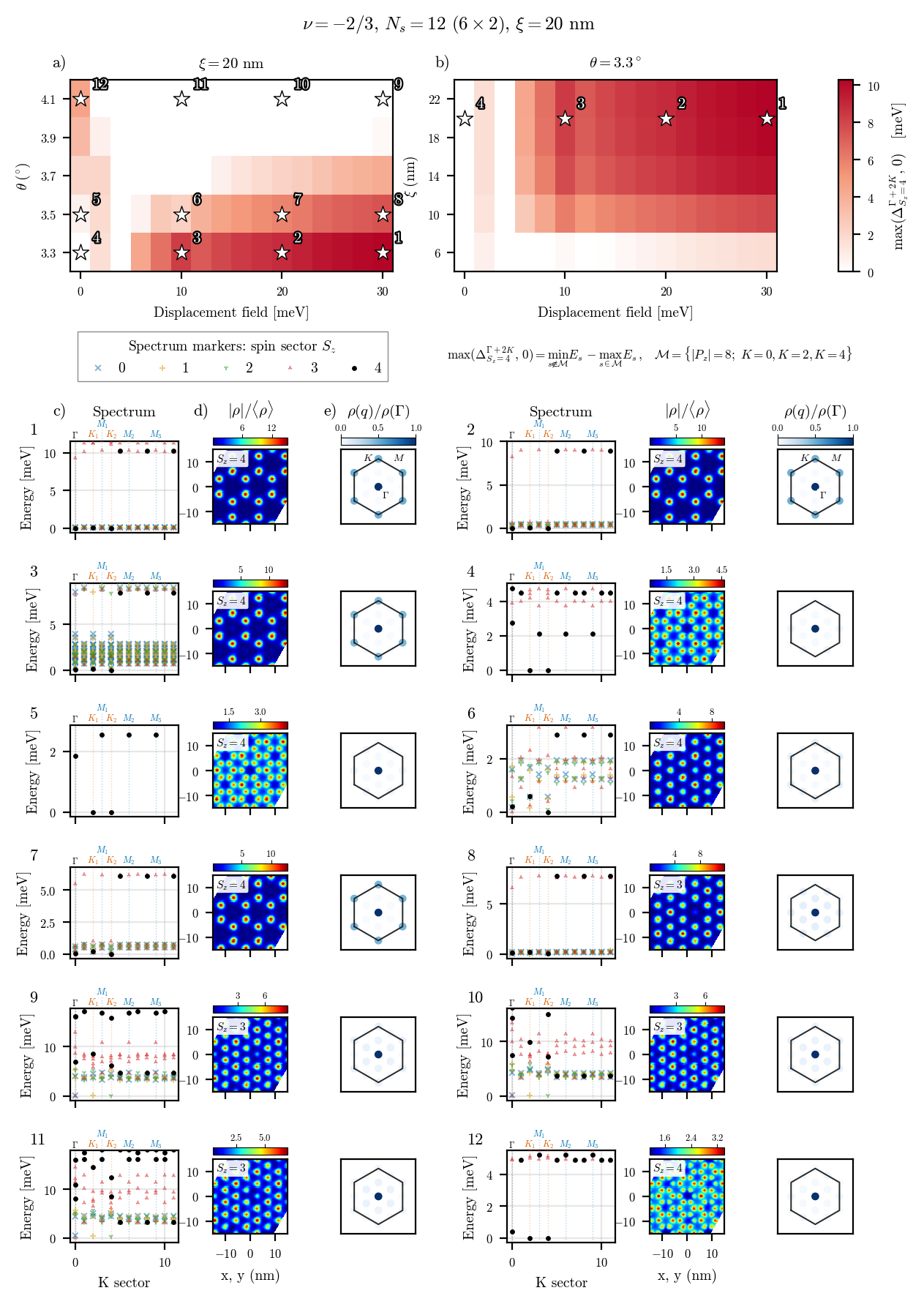}

    \caption{$\bm{\nu=-2/3}$ on $N_s=12$ ($6\times2$), $N_\mathrm{max}^1=2$, $\xi=20$~nm. \textbf{a)} $\max(\Delta^{\Gamma+2K}_{S_z=4},0)$ of the fully polarized three-state manifold at $\{\Gamma,K_1,K_2\}$ against $D$ and $\theta$ at $\xi=20$~nm. \textbf{b)} same as a) but plotted against $D$ and $\xi$ at $\theta=3.3^\circ$; white is where that manifold is no longer the lowest of its sector. \textbf{Blocks 1-12}: data points at $\theta=3.3,3.5,4.1^\circ$ at $D=0,10,20,30$~meV. In each block the panels show: \textbf{c)} the spectrum, \textbf{d)} the charge density and \textbf{e)} its Fourier transform. Panel (d) is built from the ground-state manifold, and its spin sector is \emph{not} constant across the page. A $\sqrt3\times\sqrt3$ superlattice is visible in the density for small $\theta$ and large $D$.}
    \label{fig:two_thirds_full_page}
\end{figure}

Figures \ref{fig:one_quarter_full_page},~\ref{fig:one_third_full_page},~\ref{fig:one_half_full_page}~and \ref{fig:two_thirds_full_page} share one layout. Panels a) are phase diagrams against $D$ and $\theta$ at fixed $\xi$, panels b) are phase diagrams against $D$ and $\xi$ at fixed $\theta$. Below them twelve data points are shown in blocks, each holding c) a spectrum , d) the charge density and e) its Fourier transform. The data points are numbered from 1 to 12 in the phase diagram and the corresponding number of each block is shown in its upper left corner. The letters c), d), e)  are only marked in the first block. 

For fillings $\nu=-1/4,\, -1/3$ and $-2/3$ the quantity mapped in the phase diagrams of panels a ) and b) is a gap defined in the following way. $\Delta^{X}_{S_z}$ is the lowest energy in spin sector $S_z$ outside the set of states $X$ minus the highest energy inside them; the definition, with the momentum sectors that make up $X$, is printed under panel b) of each page. In $max(\Delta,0)$ maps white marks where the manifold is not the lowest of its sector. 

For a cluster of size $N_x \times N_y$ the momentum sectors are numbered as $K=k_x+N_xk_y$, where $K_{x/y}\in \{ 0,1,\dots,N_{x/y}\}$. These numbers are used in the spectra c) as well as to to specify the momenta included in the manifolds of the quantities plotted in panels a) and b). 
\vfill
\newpage

\subsection{Real space profiles  at other spin sectors}
\label{app:other_spin_sectors}

\begin{figure}[H]
    \centering
    \includegraphics[width=1\columnwidth]{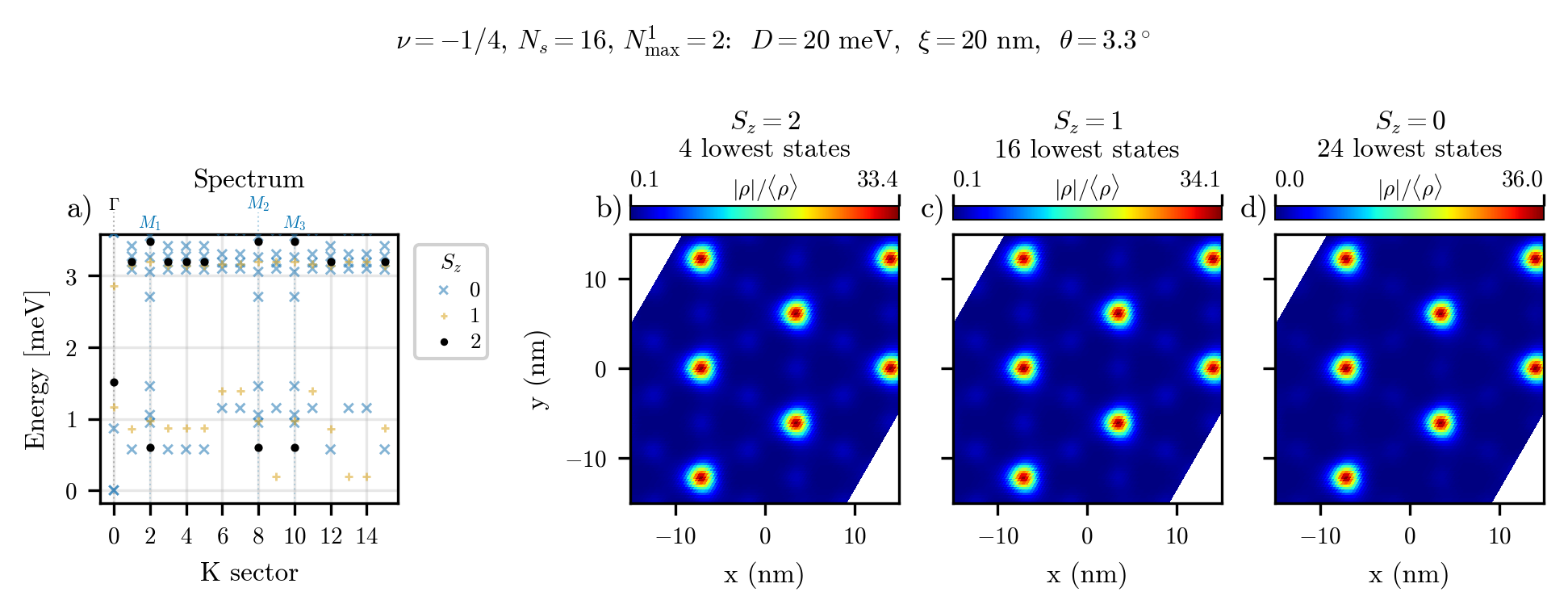}

    \caption{\textbf{$\bm{\nu=-1/4}$: the same $\bm{2\times2}$ order in every spin sector } ($N_s=16$, $N_\mathrm{max}^1=2$; $\theta=3.3^\circ$, $D=20$~meV, $\xi=20$~nm.  \textbf{a)} Spectrum. \textbf{b)}-\textbf{d)} charge density built from the lowest manifold of each spin sector: the four $S_z=2$ states at $\{\Gamma,M_1,M_2,M_3\}$, sixteen states of $S_z=1$ and twenty-four of $S_z=0$. Each manifold is separated from the rest of its sector, by $1.7$,$1.5$ and $1.2$~meV. All three give the same $2\times2$ superlattice.}
    \label{fig:one_quarter_other_spin_sectors}
\end{figure}

\begin{figure}[H]
    \centering
    \includegraphics[width=1\columnwidth]{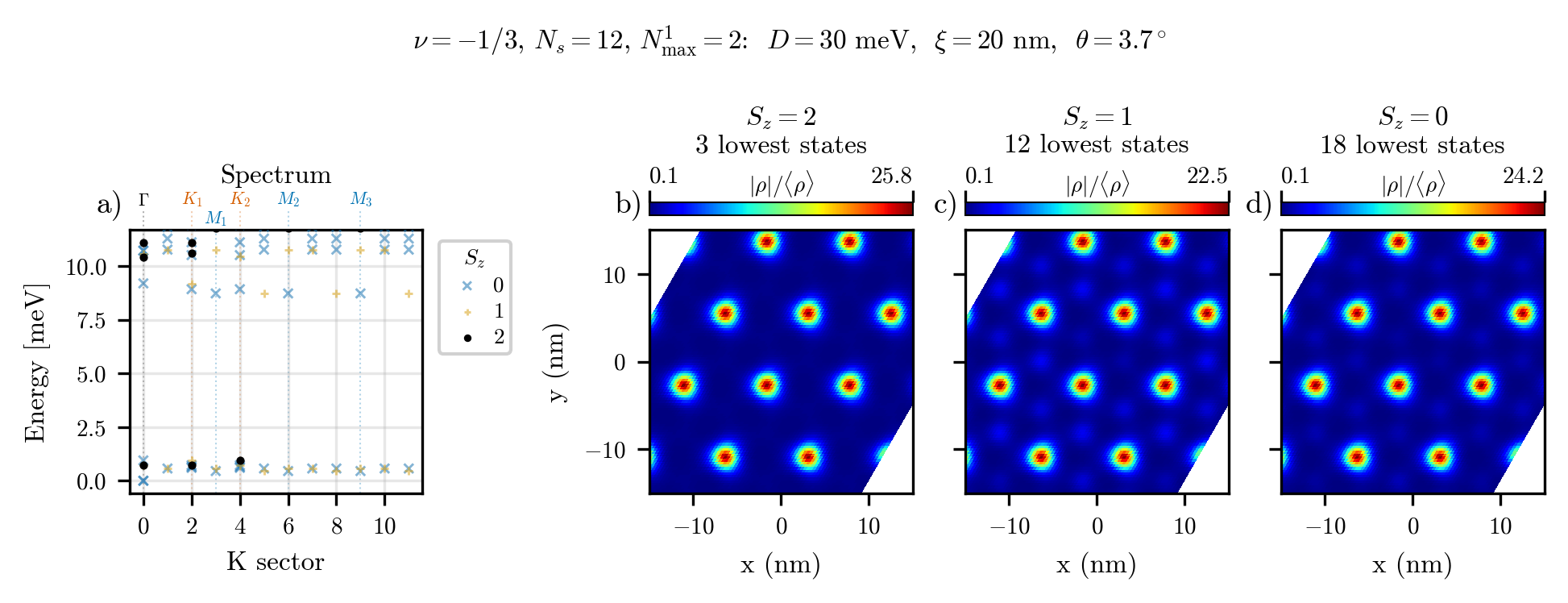}
    
     \caption{\textbf{$\bm{\nu=-1/3}$: the $\bm{\sqrt3\times\sqrt3}$ order belongs to the polarized sector} ($N_s=12$, $N_\mathrm{max}^1=2$; $\theta=3.7^\circ$, $D=30$~meV, $\xi=20$~nm. \textbf{a)} Spectrum. \textbf{b)}-\textbf{d)} charge density built from the lowest manifold of each spin sector: the three $S_z=2$ states at $\{\Gamma,K_1,K_2\}$, twelve states of $S_z=1$ and eighteen of $S_z=0$. All sectors give a $\sqrt3\times\sqrt3$ superlattice.}
    \label{fig:one_third_other_spin_sectors}
\end{figure}

\subsection{Results for different cluster sizes.}

\begin{figure}[H]
    \centering
    \includegraphics[width=1\columnwidth]{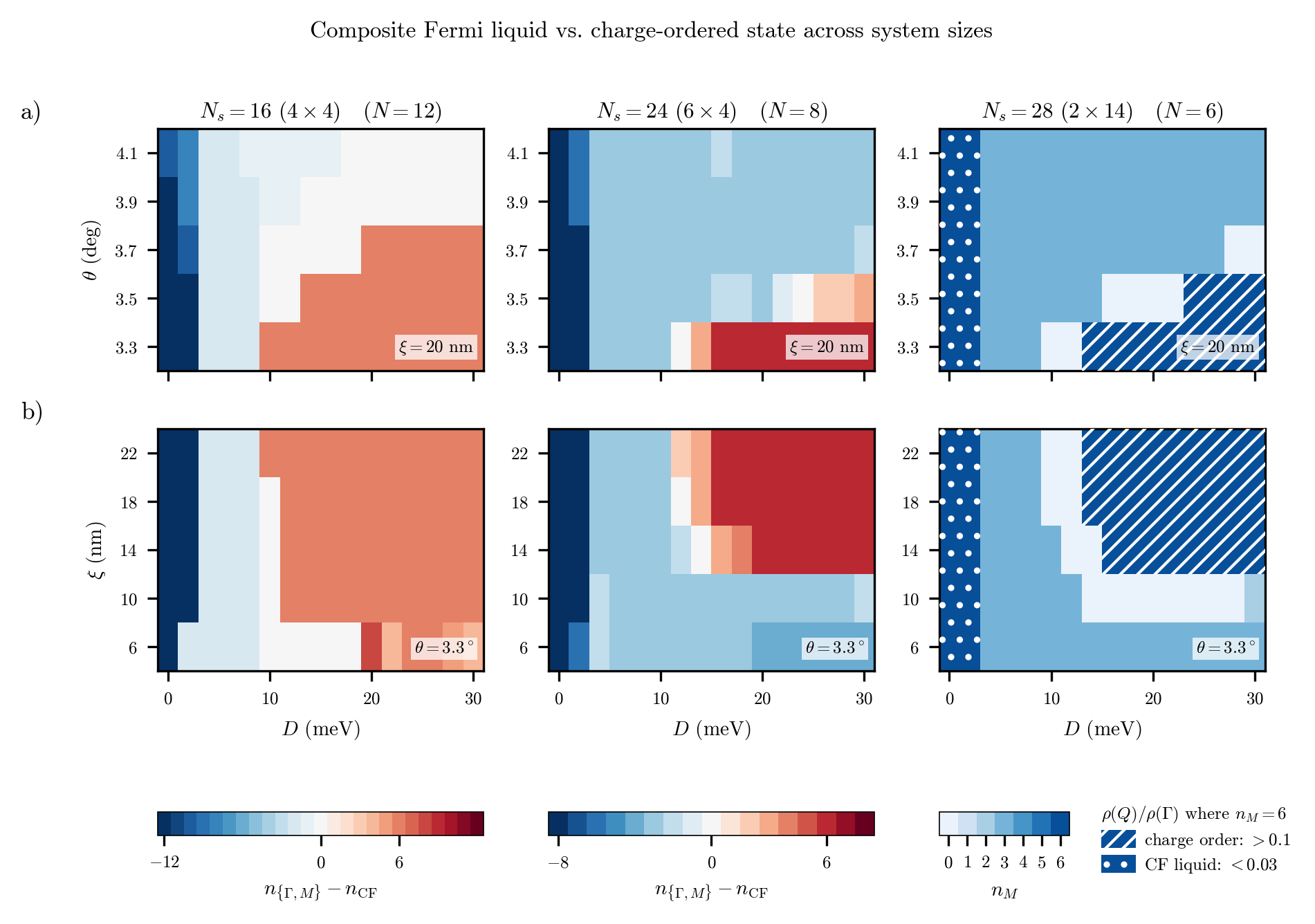}

     \caption{\textbf{Composite Fermi liquid and charge order at $\nu=-1/2$ on three clusters.} \textbf{a)} $N_s=16$ ($4\times4$), $24$ ($6\times4$) and $28$ ($2\times14$), all fully polarized, against $D$ and $\theta$ at $\xi=20$~nm. \textbf{b)} Same as a), against $D$ and $\xi$ at $\theta=3.3^\circ$. For $N_s=16$ and $24$ the color is the order score: among the $N$ lowest states, the number in $\{\Gamma,M_1,M_2,M_3\}$ minus the number on the ground momenta of the composite Fermi liquid. Each of these two columns uses its own $N$ and therefore its own color bar. For $N_s=28$ the ground states of the composite Fermi liquid and of the charge-ordered state lie in the same momentum sectors, the three $M$ points, so a score cannot separate them. This column shows instead $n_M$, the number of the six lowest states at $M$, and the cells with $n_M=6$ are hatched according to the density in momentum space: stripes for $\rho(Q)/\rho(\Gamma)\geq0.1$ (charge order) and dots for $\rho(Q)/\rho(\Gamma)<0.05$ (composite Fermi liquid), where $Q$ is the point in momentum space $\mathbf q\neq\Gamma$ where $\rho(Q)$ is alrgest.}
    \label{one_half_other_cluster_sizes}
\end{figure}

\vfill
\newpage
\begin{figure}[H]
    \centering
    \includegraphics[width=0.8\columnwidth]{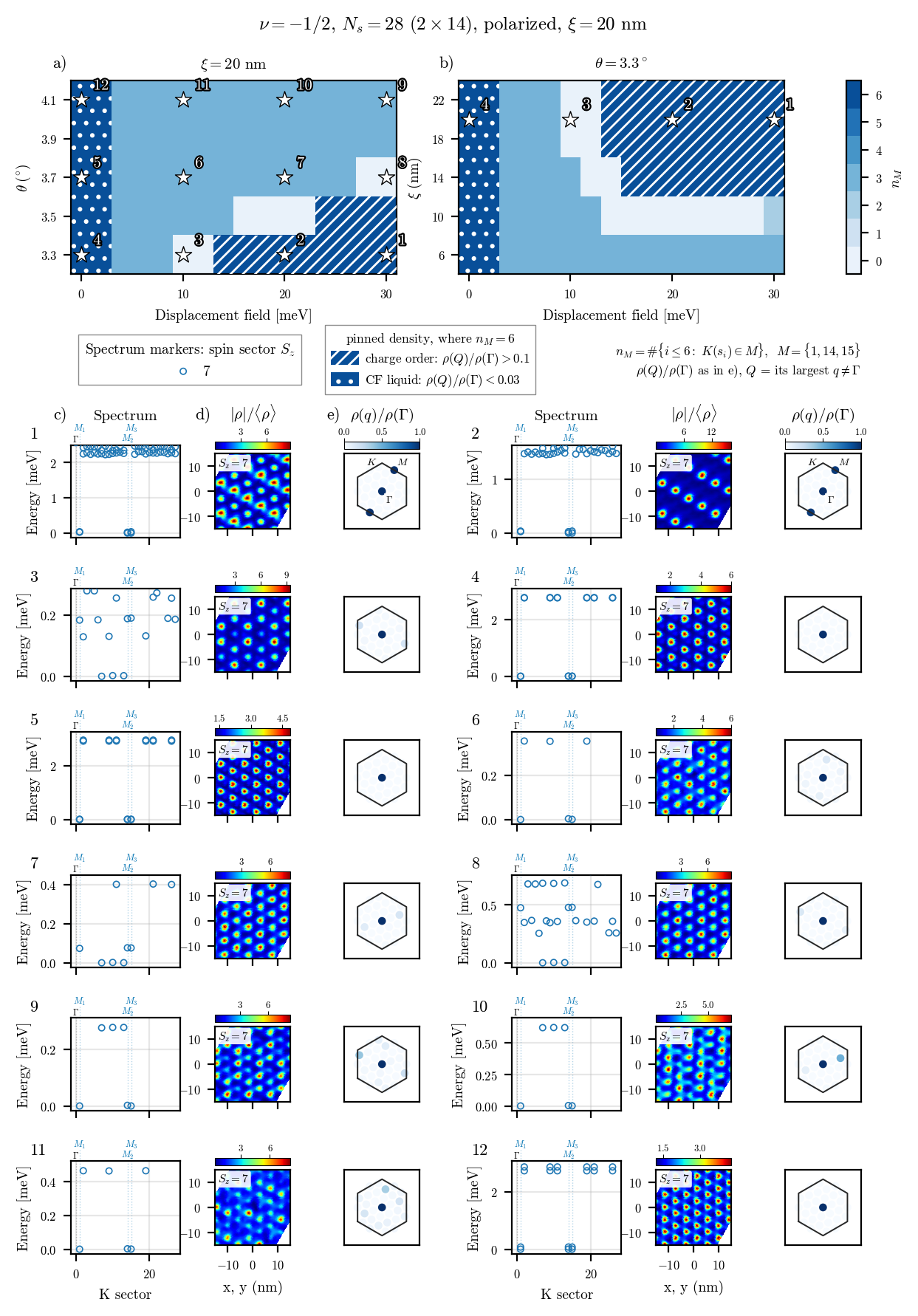}
    \caption{$\bm{\nu=-1/2}$ on $N_s=28$ ($2\times14$), \textbf{fully polarized} ($S_z=6$), $N_\mathrm{max}^1=0$, $\xi=20$~nm. \textbf{a)} Order score: among the $N=6$ lowest states, the number whose momenta lie in the manifold $\{M_1,M_2,M_3\}$. Plotted against $D$ and $\theta$ at $\xi=20$~nm. On this cluster the ground states of the composite Fermi liquid and of the charge-ordered state lie in the same momentum sectors, the three $M$ points, so the momenta alone cannot distinguish the two phases. Stripes mark $\rho(Q)/\rho(\Gamma)\geq0.1$ (charge order) and dots $\rho(Q)/\rho(\Gamma)<0.05$ (composite Fermi liquid), where $Q$ is the largest $\mathbf q\neq\Gamma$ of panel (e).   \textbf{b)} same as a) but plotted against  against $D$ and $\xi$ at $\theta=3.3^\circ$. \textbf{Blocks 1-12}: data points at $\theta=3.3,3.7,4.1^\circ$ at $D=0,10,20,30$~meV. In each block the panels show: \textbf{c)} the spectrum, \textbf{d)} the charge density and \textbf{e)} its Fourier transform. The crystal region corresponds to a single-$Q$ stripe with density maxima on alternate rows of moir\'e sites.}

    \label{fig:one_half_full_page_Ns_28}
\end{figure}

\section{Comparison among multiple \texorpdfstring{$N_\mathrm{max}^1$}{N\_max 1}}
\begin{figure}[H]
    \centering
    \includegraphics[width=0.5\columnwidth]{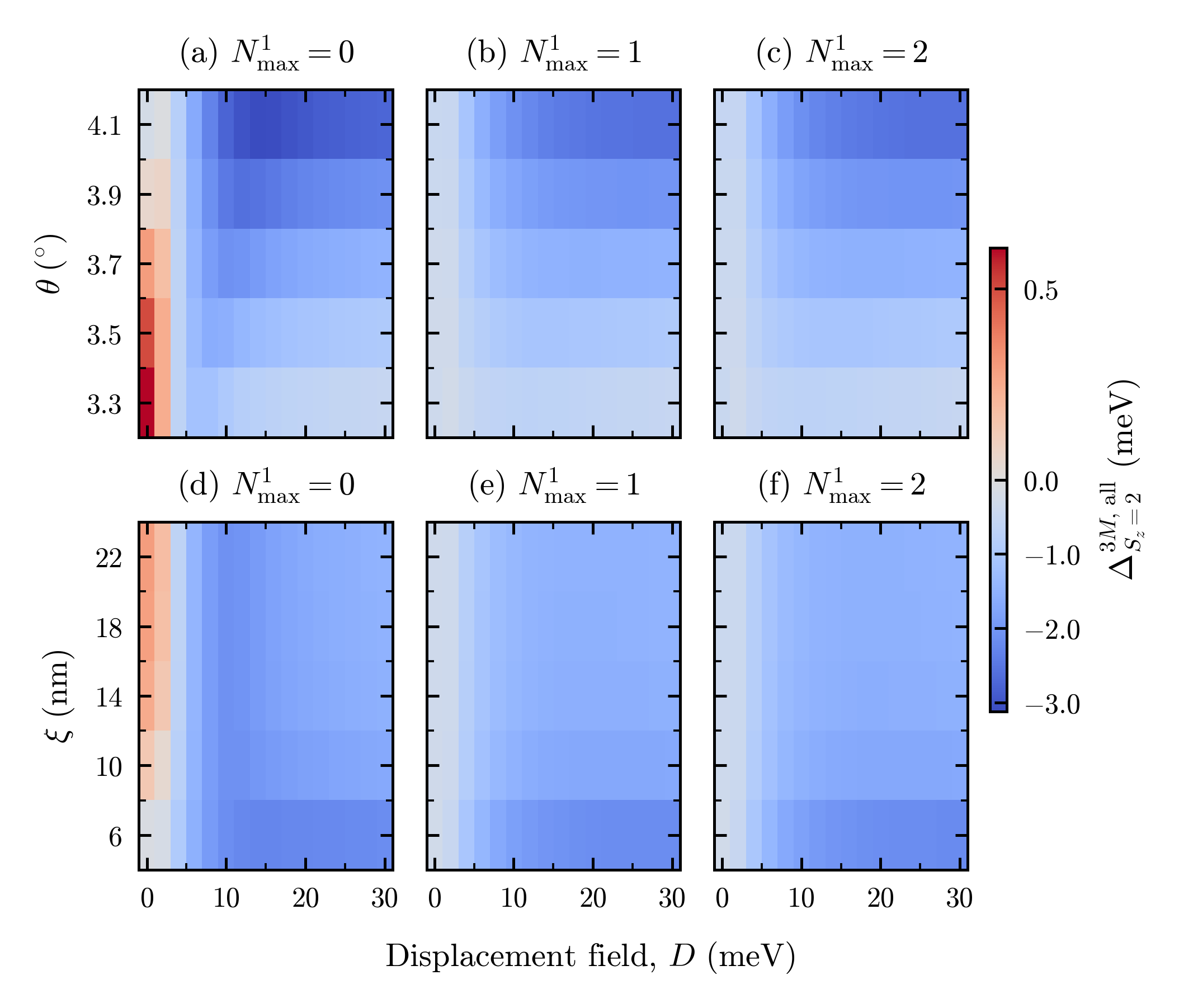}
      \caption{\textbf{Band mixing removes the $\bm{\nu=-1/4}$ Composite Fermi liquid} ($N_s=12$). The manifold $\mathcal{M}$ consists of the lowest fully polarized ($S_z=2$) state in each of the three momentum sectors $M_1$, $M_2$ and $M_3$. The color is its gap against the whole spectrum, $\Delta^{3M,\,\mathrm{all}}_{S_z=2}=\min_{s\notin\mathcal{M}}E_s-\max_{s\in\mathcal{M}}E_s ,$ where $s\notin\mathcal{M}$ runs over \emph{every} other eigenstate, in all spin sectors ($S_z=0,1,2$). Positive (red): the three states of $\mathcal{M}$ are the three lowest of the whole spectrum, separated from everything else by $\Delta$. Negative (blue): $\mathcal{M}$ is not the ground state manifold. \textbf{a)}-\textbf{c)}. Gap plotted for $N_\mathrm{max}^1=0,1,2$ (columns), against $D$ and $\theta$ at $\xi=20$ nm. \textbf{d)}-\textbf{f)}  Gap plotted against $D$ and $\xi$ at $\theta=3.7^\circ$. $\mathcal{M}$ It is the ground manifold only at $N_\mathrm{max}^1=0$. Allowing holes in the second band lowers the unpolarized states.}
     \label{CFL_and_band_mixing_one_quarter}
\end{figure}

\begin{figure}[H]
    \centering
    \includegraphics[width=0.5\columnwidth]{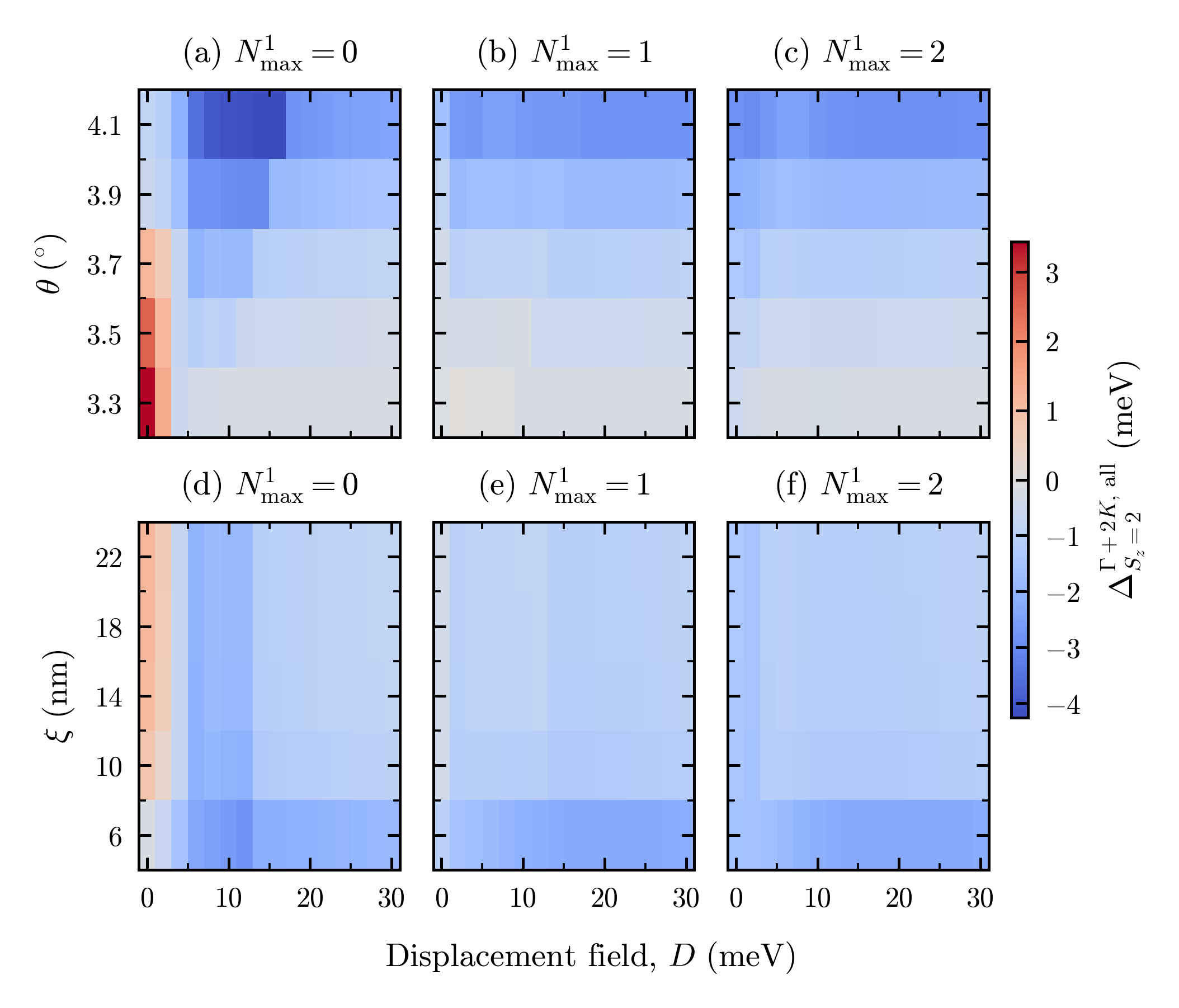}
     \caption{\textbf{Band mixing removes the $\bm{\nu=-1/3}$ fractional Chern insulator} ($N_s=12$). The manifold $\mathcal{M}$ consists of the lowest fully polarized ($S_z=2$) state in each of the three momentum sectors $\Gamma$, $K_1$ and $K_2$. The color is its gap against the whole spectrum, $\Delta^{\Gamma+2K,\,\mathrm{all}}_{S_z=2}\,=\,\min_{s\notin\mathcal{M}}E_s\,-\,\max_{s\in\mathcal{M}}E_s ,$ where $s\notin\mathcal{M}$ runs over \emph{every} other eigenstate, in all spin sectors ($S_z=0,1,2$). Positive (red): the three states of $\mathcal{M}$ are the three lowest of the whole spectrum, separated from everything else by $\Delta$. Negative (blue): $\mathcal{M}$ is not the ground state manifold. \textbf{a)}-\textbf{c)}. Gap plotted for $N_\mathrm{max}^1=0,1,2$ (columns), against $D$ and $\theta$ at $\xi=20$ nm. \textbf{d)}-\textbf{f)}  Gap plotted against $D$ and $\xi$ at $\theta=3.7^\circ$. $\mathcal{M}$ It is the ground manifold only at $N_\mathrm{max}^1=0$. Allowing holes in the second band lowers the unpolarized states.}
    \label{FCI_and_band_mixing_one_third}
\end{figure}

\begin{figure}[H]
    \centering
    \includegraphics[width=0.5\columnwidth]{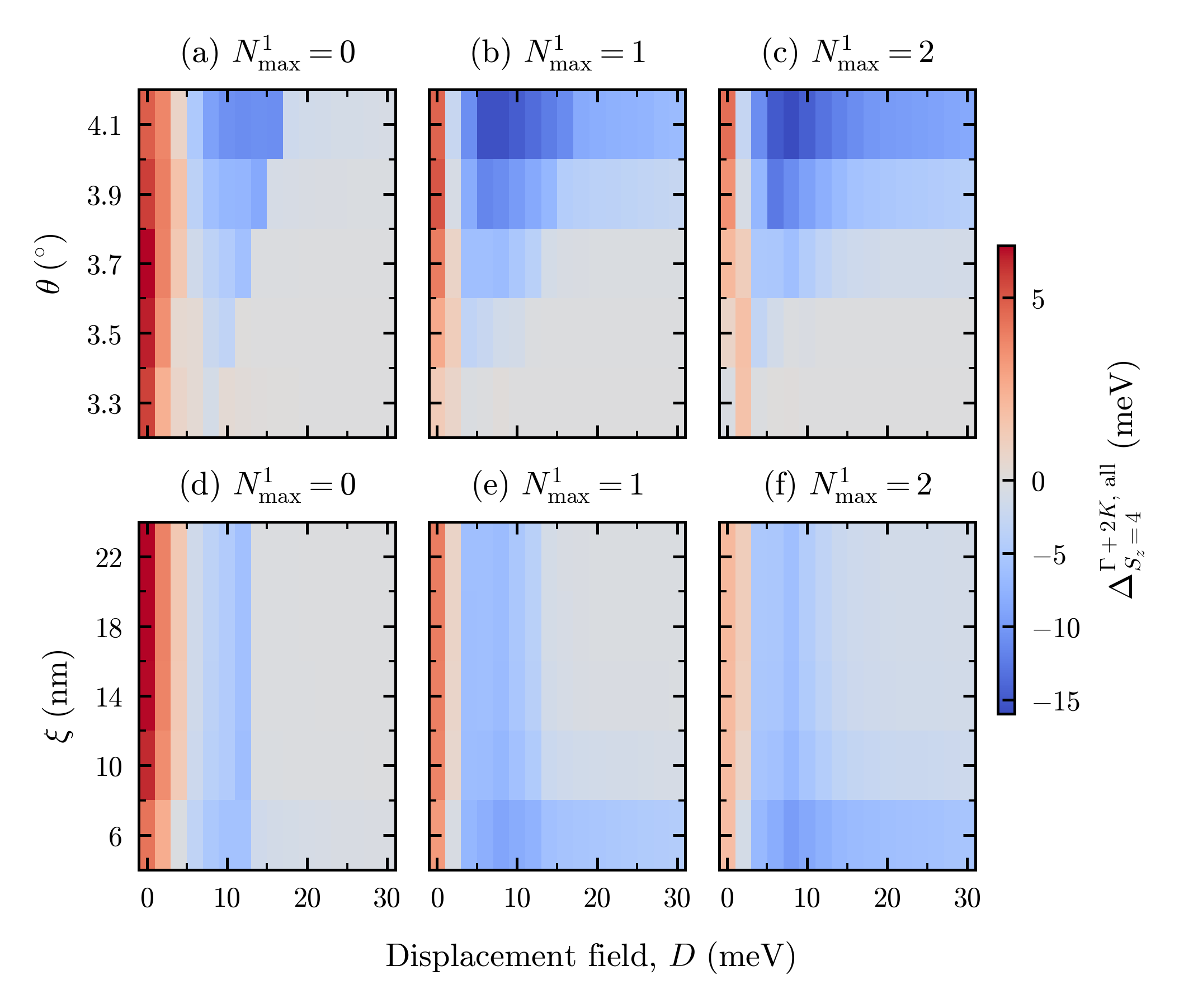}
     \caption{\textbf{Band mixing weakens the $\bm{\nu=-2/3}$ fractional Chern insulator} ($N_s=12$). The manifold $\mathcal{M}$ consists of the lowest fully polarized ($S_z=4$) state in each of the three momentum sectors $\Gamma$, $K_1$ and $K_2$. The color is its gap against the whole spectrum, $\Delta^{\Gamma+2K,\,\mathrm{all}}_{S_z=2}\,=\,\min_{s\notin\mathcal{M}}E_s\,-\,\max_{s\in\mathcal{M}}E_s ,$ where $s\notin\mathcal{M}$ runs over \emph{every} other eigenstate, in all spin sectors ($S_z=0,1,2$). Positive (red): the three states of $\mathcal{M}$ are the three lowest of the whole spectrum, separated from everything else by $\Delta$. Negative (blue): $\mathcal{M}$ is not the ground state manifold. \textbf{a)}-\textbf{c)}. Gap plotted for $N_\mathrm{max}^1=0,1,2$ (columns), against $D$ and $\theta$ at $\xi=20$ nm. \textbf{d)}-\textbf{f)}  Gap plotted against $D$ and $\xi$ at $\theta=3.7^\circ$.}
    \label{FCI_and_band_mixing_two_thirds}
\end{figure}

\end{appendix}

\end{document}